\documentclass[a4paper,11pt]{article}

\usepackage{jheppub} 

\usepackage[T1]{fontenc} 
\usepackage{float}
\usepackage{amsmath,amssymb,graphics,epsfig,subfigure}
\usepackage{color}
\usepackage{hyperref}
\usepackage{epstopdf}
\usepackage{graphicx}
\usepackage{braket} 
\usepackage{amsthm}
\usepackage{cases}

\begin{document}

\newcommand \e{\text{e}}
\newcommand \nn{\nonumber}
\newcommand \fc{\frac}
\newcommand \lt{\left}
\newcommand \rt{\right}
\newcommand \pd{\partial}
\newcommand \hmn{h_{\mu\nu}}
\newcommand \mn{{\mu\nu}}
\newcommand \tcb{\textcolor{blue}} 
\newcommand{\PR}[1]{\ensuremath{\left[#1\right]}} 
\newcommand{\PC}[1]{\ensuremath{\left(#1\right)}} 
\newcommand{\PX}[1]{\ensuremath{\left\lbrace#1\right\rbrace}} 
\newcommand{\BR}[1]{\ensuremath{\left\langle#1\right\vert}} 
\newcommand{\KT}[1]{\ensuremath{\left\vert#1\right\rangle}} 
\newcommand{\MD}[1]{\ensuremath{\left\vert#1\right\vert}} 

\title{Gravitational Echoes from Braneworlds}


\author{Chun-Chun Zhu,$^{a}$ $^{b}$}
\author{Jing Chen,$^{a}$ $^{b}$ $^{c}$}
\author{Wen-Di Guo,$^{a}$ $^{b}$}
\author{and Yu-Xiao Liu$^{a}$ $^{b}$ \footnote{corresponding author}}


\affiliation[a]{Lanzhou Center for Theoretical Physics, Key Laboratory for Quantum Theory and Applications of the Ministry of Education, Key Laboratory of Theoretical Physics of Gansu Province, School of Physical Science and Technology, Lanzhou University, Lanzhou 730000, China}
\affiliation[b]{Institute of Theoretical Physics and Research Center of Gravitation, School of Physical Science and Technology,  Lanzhou University, Lanzhou 730000, China}
\affiliation[c]{Science and Technology on Vacuum Technology and Physics Laboratory, Lanzhou Institute of Physics, Lanzhou, 73000, China}

\emailAdd{zhuchch21@lzu.edu.cn}
\emailAdd{chenj19@lzu.edu.cn}
\emailAdd{guowd@lzu.edu.cn}
\emailAdd{liuyx@lzu.edu.cn}

\abstract
   {Gravitational echoes can be used to probe the structure of spacetime. In this paper, we investigate the gravitational echoes in different  braneworld models in five-dimensional spacetime. We derive the gravitational perturbation equations of these models, and obtain the time-dependent evolution equations of the extra-dimensional and radial components. Using a Gaussian wave packet as initial data, we study the time evolution of the gravitational perturbation. By monitoring the evolution of the Gaussian wave packet, the gravitational echoes are observed whether the wave packet is generated from inside or outside the braneworld. Furthermore, we can restrict the parameters of the braneworld by calculating the strength of the first gravitational  echo and using the current gravitational wave data.}

\maketitle
\flushbottom

\section{Introduction}
In 2015, Laser Interferometer Gravitational-Wave Observatory (LIGO) detected the first gravitational wave signal GW150914, originating from the merger of a binary black hole system~\cite{LIGOScientific:2016aoc}. This groundbreaking event paved the way to do new researches on gravitational wave astronomy~\cite{LIGOScientific:2017vwq,LIGOScientific:2020iuh,Wang:2021srv,Guo:2022sss,Wang:2022nml,Yi:2023mbm,Cai:2023rta,Lin:2022huh}. The GW150914 signal was generated from the merger of two black holes with about 36 solar masses and 29 solar masses into a final black hole with about 62 solar masses ~\cite{LIGOScientific:2016aoc}. Binary black hole merger can be divided into three stages: inspiral, merger, and ringdown. Of particular interest to us is the final ringdown stage, as it contains crucial information about the structure of the final black hole.

The signal from the ringdown stage is generally composed of quasinormal modes of the black hole~\cite{Vishveshwara:1970zz,Chandrasekhar:1984siy,Kokkotas:1999bd}. Regge and Wheeler first introduced the concept~\cite{Regge:1957td} of quasinormal modes when they studied gravitational perturbations of the Schwarzschild black hole. A set of discrete quasinormal  frequencies was found in their pioneering work~\cite{Regge:1957td,Chandrasekhar:1984siy,Kokkotas:1999bd}. The presence of discrete frequencies is due to the radiation boundary conditions imposed at the black hole's event horizon and at infinity. In the classical theory, quasinormal modes could reflect the spacetime structure near the black hole horizon. Especially, we 
can use the late stage of the ringdown to distinguish black holes and compact objects. Although the early stage of the ringdown signal from compact objects is similar to that from black holes, modifications to the quantum structure near the horizon or alterations to the horizon itself may manifest in secondary pulses during the late stage of the ringdown~\cite{Cardoso:2016rao}. The reason is that the effective potential of gravitational perturbations in typical black hole models only has a single barrier, leading to a monotonic decay in the amplitude of signals during the ringdown~\cite{Regge:1957td,Chandrasekhar:1984siy,Leaver:1990zz,Guinn:1989bn,Berti:2003zu,Zhang:2023wwk}. On the other hand, quantum structures near the black hole horizon~\cite{Lunin:2001jy,Skenderis:2008qn,Almheiri:2012rt,Saravani:2012is}, such as firewalls~\cite{Almheiri:2012rt,Saravani:2012is},  may introduce new barriers to the effective potentials for gravitational perturbations near the horizon. Additionally, other horizonless compact objects, such as wormholes~\cite{Cardoso:2016rao,Bueno:2017hyj}, would exhibit multiple barriers in their effective potentials for gravitational perturbations. In such cases, although quasinormal modes still manifest as discrete quasinormal frequencies, some of these modes decay at a relatively slower rate, which could lead to the observation of multiple peaks in the signal. This occurs as the signal can be reflected multiple times by these barriers, generating a scattering resonance effect where the segment of signal with the resonance frequency will be quasi-localized between the barriers. This phenomenon results in multiple similar signal peaks, termed gravitational echoes~\cite{Cardoso:2017cqb,Cardoso:2016oxy,Cardoso:2019rvt,Conklin:2017lwb,Abedi:2016hgu,Capano:2022zqm,Abedi:2018npz,Zhong:2023pjz,Bronnikov:2019sbx}. Therefore, detecting gravitational echoes can provide strong evidence for the existence of  these previously undiscovered structures.

The extra-dimensional theories fundamentally alter our perspective on the concept of spacetime structure. These theories aim to either unify gravity and electromagnetism~\cite{kaluza:1921un,Klein:1926tv} or solve the gauge hierarchy problem (the fine-tuning problem) in the standard model of particle physics~\cite{Arkani-Hamed:1998jmv,Randall:1999ee,Randall:1999vf,Gremm:1999pj}. In the models addressing the hierarchy problem, Randall-Sundrum 1 (RS-I) model~\cite{Randall:1999ee} with a compact extra dimension stands out as more refined candidates and has successfully addressed the hierarchy problem. Based on it,  RS-II and thick brane models with an infinite extra dimension can restore the four-dimensional Newtonian potential~\cite{Randall:1999vf,Goldberger:1999uk,Gremm:1999pj,DeWolfe:1999cp,Bazeia:2008zx,Charmousis:2001hg,Arias:2002ew,Barcelo:2003wq,Bazeia:2004dh,Castillo-Felisola:2004omi,Kanno:2004nr,Barbosa-Cendejas:2005vog,Koerber:2008rx,Barbosa-Cendejas:2007cwl,Johnson:2008kc,Liu:2011wi,Chumbes:2011zt,Andrianov:2012ae,Kulaxizi:2014yxa,deSouzaDutra:2014ddw,Chakraborty:2015zxc,Karam:2018squ,Kanti:2005xa,Kanti:2006ua}.
In most extra-dimensional theories, we can observe that the effective potential felt by gravitational perturbations often features multiple barriers~\cite{Goldberger:1999uk,Gremm:1999pj,DeWolfe:1999cp,Bazeia:2008zx,Charmousis:2001hg,Arias:2002ew,Barcelo:2003wq,Bazeia:2004dh,Castillo-Felisola:2004omi,Kanno:2004nr,Barbosa-Cendejas:2005vog,Koerber:2008rx,Barbosa-Cendejas:2007cwl,Johnson:2008kc,Liu:2011wi,Chumbes:2011zt,Andrianov:2012ae,Kulaxizi:2014yxa,deSouzaDutra:2014ddw,Chakraborty:2015zxc,Karam:2018squ} and the gravitational waves can propagate along extra dimensions in the theory of infinite extra dimensions. Can gravitational echoes be generated in extra-dimensional models? And if so, are those gravitational echoes different from those generated from compact stars? Our research aims to  study the gravitational echoes in various thick brane models and the differences between gravitational echoes of compact stars and thick brane models. We use Gaussian wave packets as initial waveforms and employ numerical simulation to monitor their evolution and observe whether gravitational echoes occur. Furthermore, we simulate the waveform of Gaussian wave packets propagating along the brane, and calculate the time interval between the secondary pulse and the primary wave. We also calculate the velocity of the secondary pulse. In the process, it demonstrates the differences between the gravitational echoes generated from compact stars and those generated form thick brane. It is worth noting that LIGO claims to have no evidence for the existence of quasinormal modes or gravitational echoes~\cite{LIGOScientific:2016aoc,Westerweck:2017hus,Lo:2018sep,Ashton:2016xff,Nielsen:2018lkf,Uchikata:2019frs,Wang:2020ayy}. It indicates that either the secondary pulse or echoes must be sufficiently weak or the time interval between secondary pulse and the primary wave is sufficiently large, exceeding the detection ability of current detectors. Therefore, these observational outcomes impose certain restrictions on the parameter settings of extra-dimensional models.

The remaining part of this paper is organized as follows. In Sec.~\ref{5fieldeq}, we study the gravitational echoes from a thick brane generated by a canonical scalar field in five-dimensional general relativity framework. In Sec.~\ref{5FRT} and Sec.~\ref{5FDNDT}, we study the gravitational echoes in  $f(R)$ gravity and  the non-minimally derivative coupled scalar-tensor gravity in five-dimensional spacetime, respectively.  Finally, we give the discussions and conclusions in Sec.~\ref{conclusion}.

\section{Five-dimensional Thick Brane Model}~\label{5fieldeq}
In this section, we investigate the gravitational echoes from a flat thick brane that is generated by a canonical scalar field and has a flat 3-brane under the five-dimensional general relativity framework. It is worth noting that the infinite extra dimensions in thick brane models.

\subsection{Background}
\begin{figure*}[htbp]
	\centering
	\subfigure[~The scalar field]{\includegraphics[width=5cm]{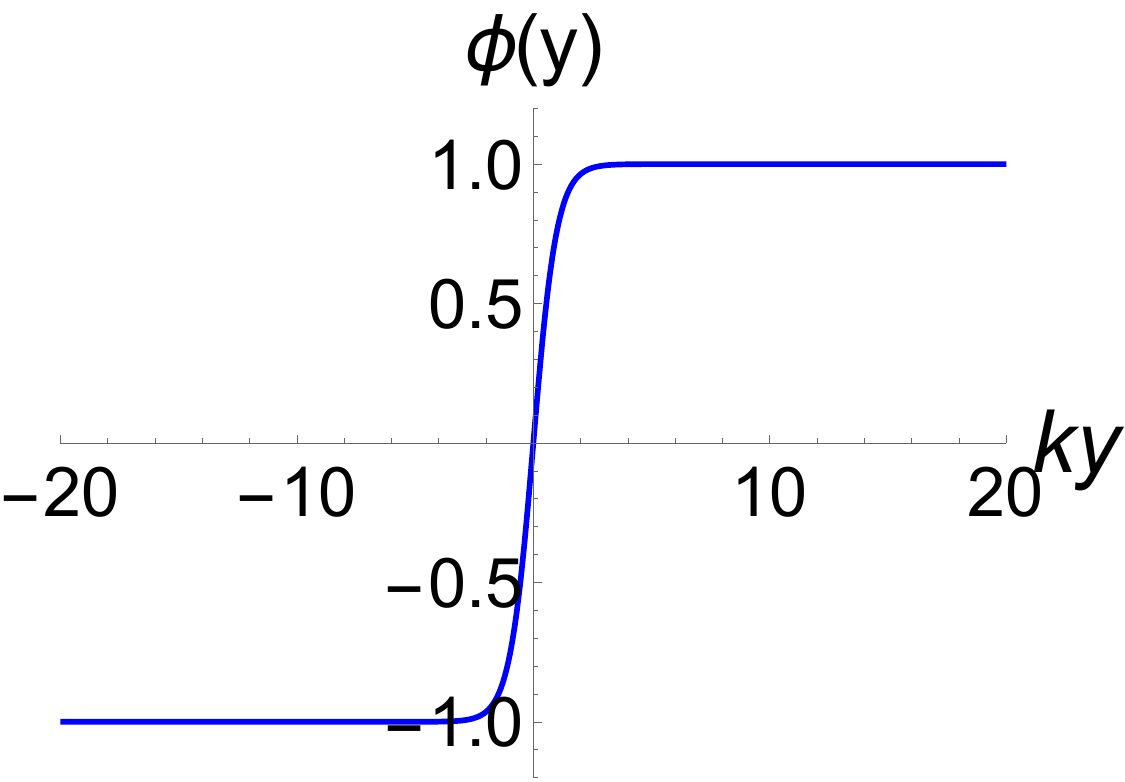}\label{figsinglescalar}}
	\subfigure[~The warp factor]{\includegraphics[width=5cm]{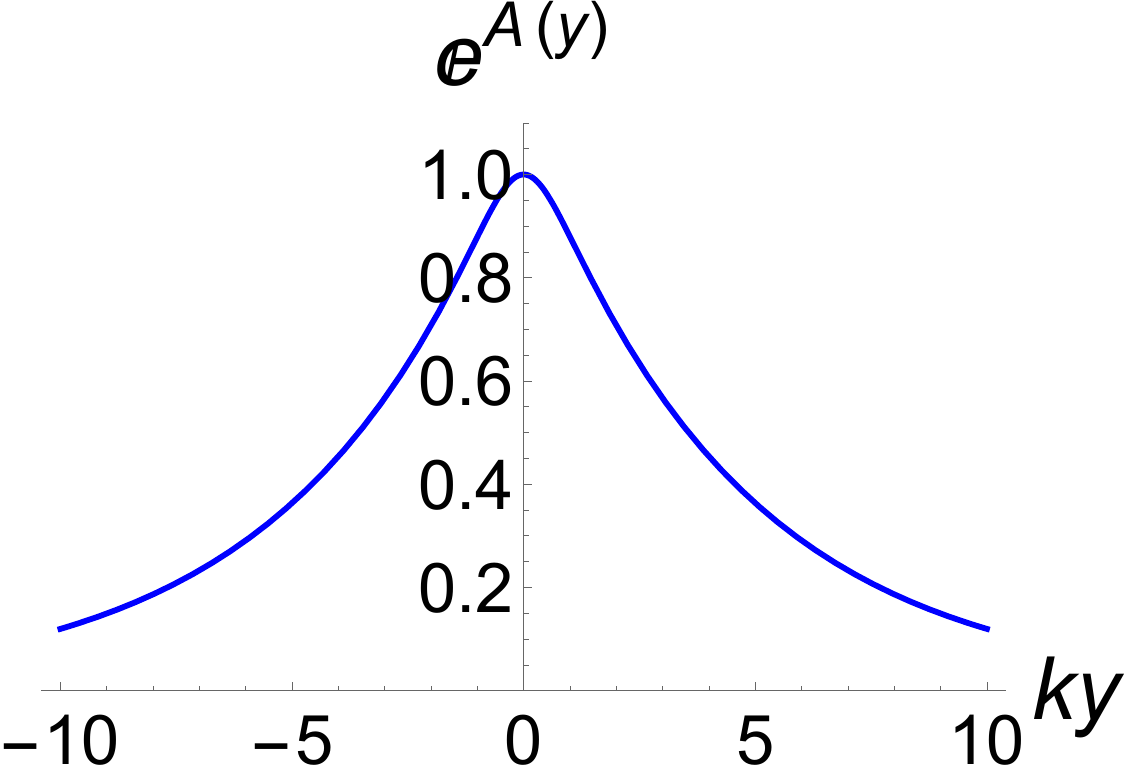}}
	\subfigure[~The scalar potential]{\includegraphics[width=5cm]{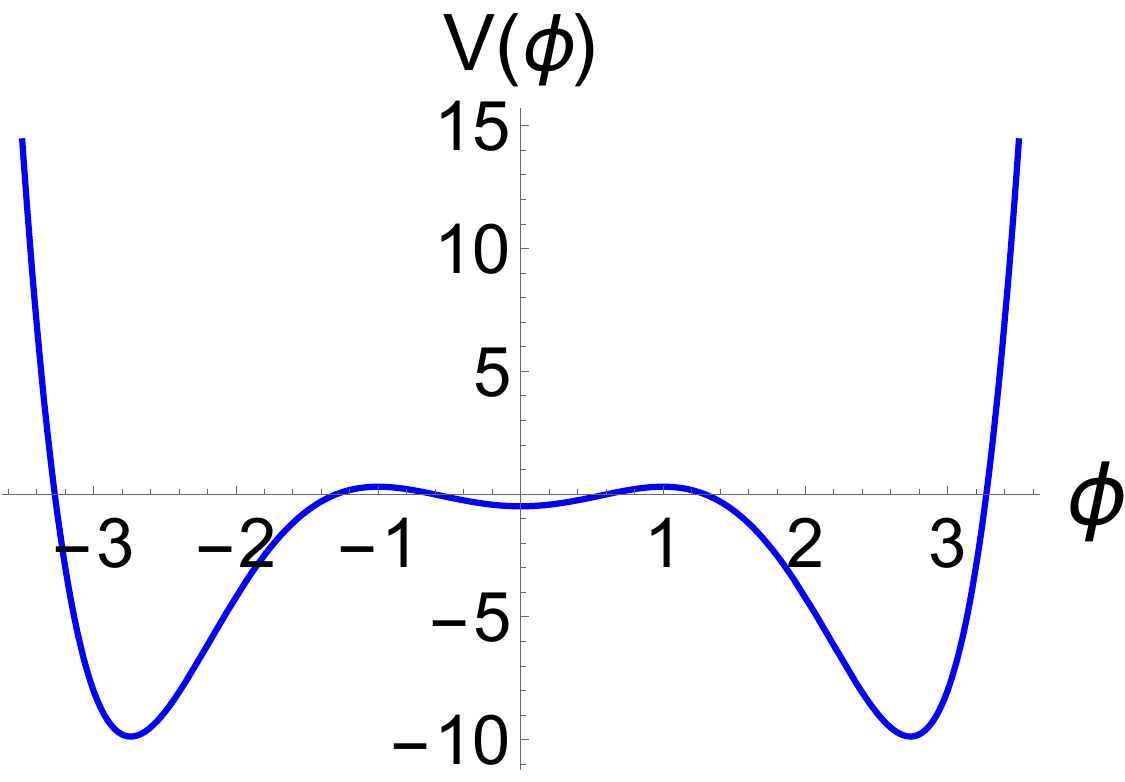}}
	\vskip -4mm \caption{Plots of the single kink scalar field,  the warp factor, and the scalar potential with $v=1$. }
	\label{Fig1kinkwarpedfactor}
\end{figure*}

First, we consider the five-dimensional gravity action with a real scalar field~\cite{Gremm:1999pj}:
\begin{eqnarray}
	S = 	\int{d^5x\sqrt{-g}~\left[\frac{M_5^3}{4}R-\frac{1}{2}\partial_M\phi\partial^M\phi-V(\phi)\right]}.\label{realscalaraction}\label{GRaction}
\end{eqnarray}
where the capital Latin letters $M, N, \dots=0,1,2,3,5$  label the five-dimensional spacetime indices. In this paper we set $M_5^3/4$=1.  By varying the action~\eqref{GRaction} with respect to the metric and the scalar field respectively, we can obtain the equations of motion:
\begin{eqnarray}
	G_{MN}&=&T^{(\phi)}_{MN},\\
	\Box\phi&=&\frac{d V(\phi)}{d\phi},
\end{eqnarray}
where $T^{(\phi)}_{MN}=\pd_M\phi\pd_N\phi-g_{MN}(\frac{1}{2}\partial_M\phi\partial^M\phi+V(\phi))$ is the energy-momentum tensor of the scalar field and $\Box=g^{MN}\bigtriangledown_M\bigtriangledown_N$. The five-dimensional metric describing a static flat brane is~\cite{Randall:1999vf,Randall:1999ee,Halyo:1999zx,Chamblin:1999cj,Liu:2008wd,Rubakov:1983bb}
\begin{eqnarray}
	ds^2&=&\text{e}^{2A(y)}\eta_{\mu\nu} dx^\mu dx^\nu+dy^2,\label{metric}
\end{eqnarray}
where $y\in(-\infty,\infty)$ is the extra-dimensional coordinate and $e^{2A(y)}$ is the warp factor. Here, Greek letters $\mu, \nu, \dots=0,1,2,3$  label the four-dimensional spacetime indices. With metric~\eqref{metric}, we can write the 
specific form of the field equations:
\begin{eqnarray}
	6 A^{\prime\prime}+12 A^{\prime 2}+2 V(\phi)+\phi^{\prime 2}&=&0,\\
	6 A^{\prime 2}+V(\phi)-\frac{1}{2} \phi^{\prime 2}&=&0,\label{motionGRbrane}\\
	\phi^{\prime\prime}+4A^{\prime}\phi^{\prime}-\frac{dV(\phi)}{d\phi}&=&0,\label{motionscalarGRbrane}
\end{eqnarray}
where prime denotes the derivative with respect to the extra-dimensional coordinate $y$. We can also derive Eq.~\eqref{motionscalarGRbrane} through $\bigtriangledown^MT^{(\phi)}_{MN}=0$. So only two of the above three equations are independent. Considering the following boundary conditions: $A(0)=0$, $A'(0)=0$, $\phi(y)=-\phi(-y)$, and $\phi(\pm\infty)\rightarrow \text{const}$, we can get the following solutions:
\begin{eqnarray}
	\phi(y)&=&v\,\text{tanh}\,(k y),\label{singlescalar}\\
	A(y)&=&-\frac{1}{18} v^2 \tanh ^2(k y)-\frac{2}{9} v^2 \ln (\cosh (k y)),\\
	V(\phi)&=& -\frac{2 k^2 \phi ^6}{27 v^2}+\frac{k^2 \left(24 v^2+27\right) \phi ^4}{54 v^2}+\frac{k^2 \left(-36 v^4-54 v^2\right) \phi ^2}{54 v^2}+\frac{k^2 v^2}{2},
	\label{1kinksolution}
\end{eqnarray}
where $k$ and $v$ are real parameters. The distributions of the warp factor, the scalar field, and the scalar potential are shown in Fig.~\ref{Fig1kinkwarpedfactor}.
The configuration of the scalar field~\eqref{singlescalar} is a single kink as shown in Fig.~\ref{figsinglescalar}. However, it should be pointed out that a double kink scalar field is also the solution of the field equations under the above boundary conditions. The solution of a double kink scalar field can be taken as 
\begin{eqnarray}
	\phi(y)=v~(\tanh(k y +b)+\tanh(k y-b)),
\end{eqnarray}
where $b$ is the dimensionless distance between the two kinks of the scalar field. Now, since there is no analytical solution for the warp factor and the scalar potential, they can only be depicted numerically. The range of $\phi(y)$ is  $(-2v,2v)$ when $y$ belongs to $(-\infty,\infty)$. 

\begin{figure*}[htbp]
	\centering
	\subfigure[~The scalar field]{\includegraphics[width=6cm]{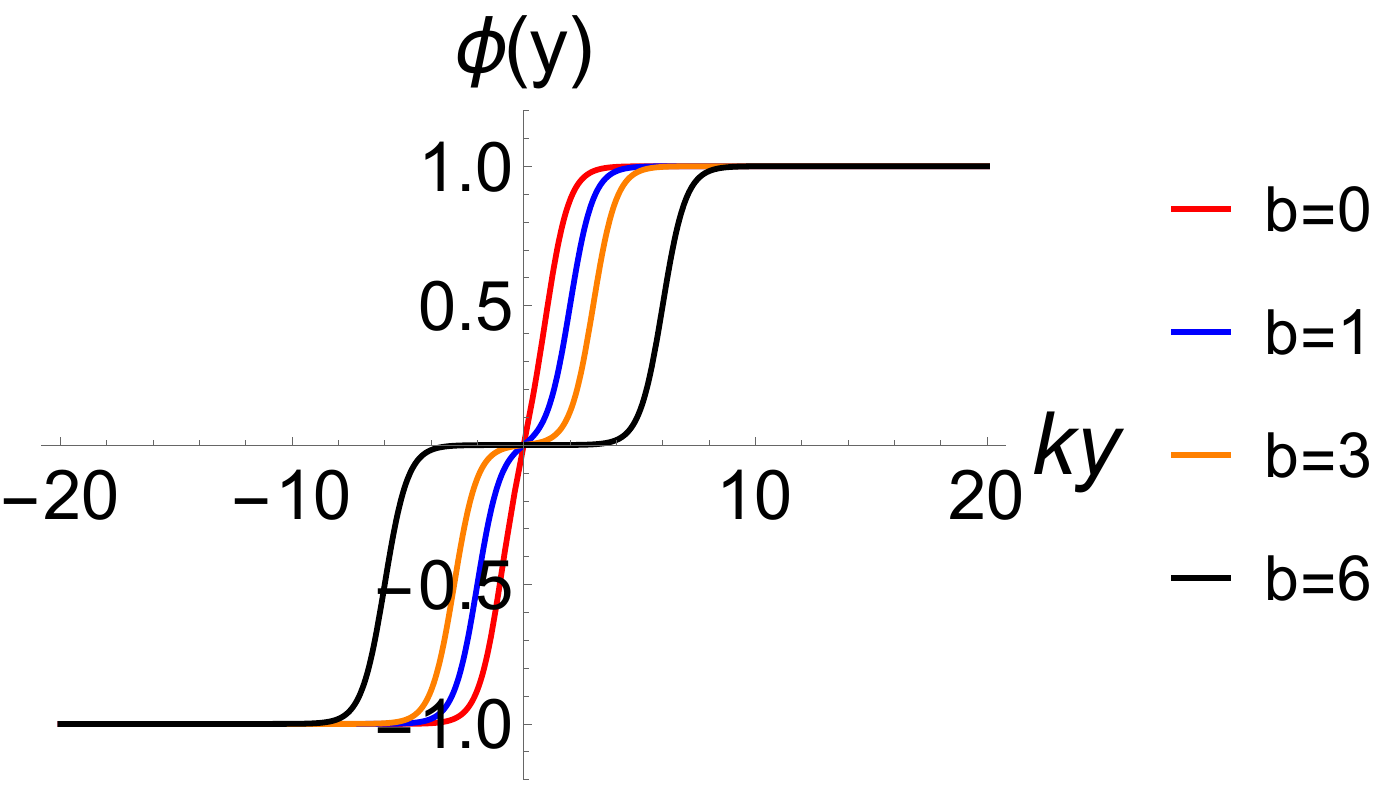}}
	\subfigure[~The warp factor]{\includegraphics[width=6cm]{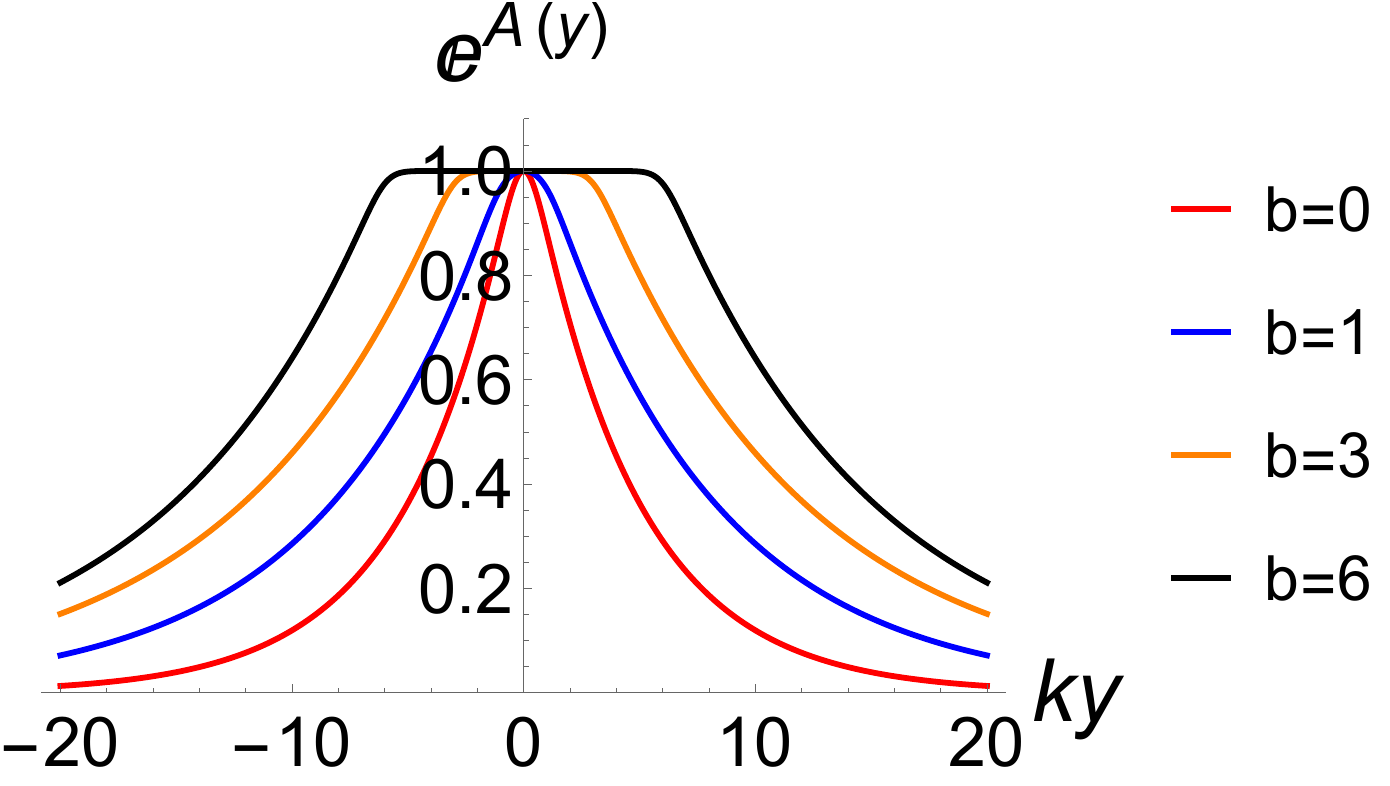}}
	\subfigure[~The scalar potential]{\includegraphics[width=6cm]{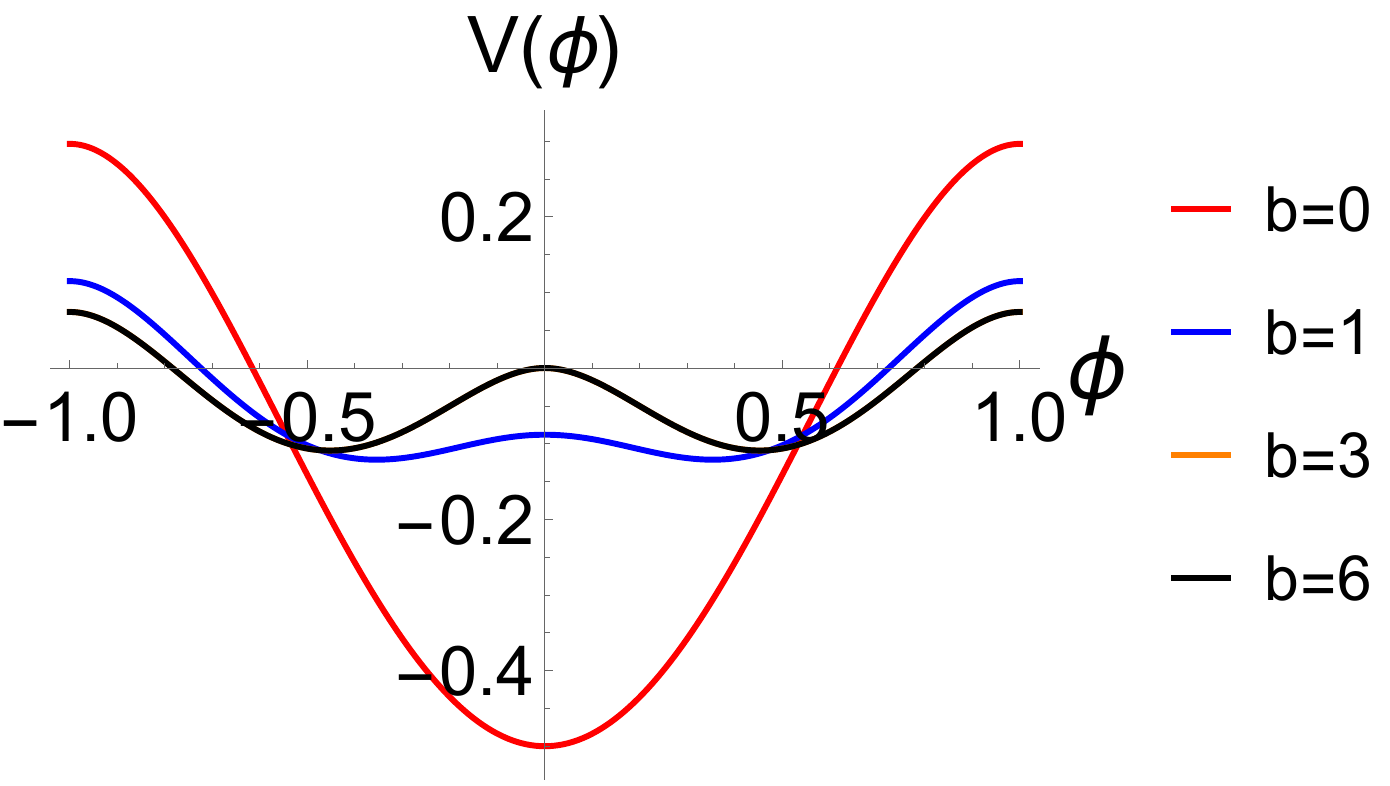}}
	\vskip -4mm \caption{Plots of the double  kink scalar field,  the warp factor, and the scalar potential  with  $v=0.5$ and $k=1$. }
	\label{Fig2kinkwarpedfactor}
\end{figure*}

The distributions of the warp factor, the scalar field, and the scalar potential are shown in Fig.~\ref{Fig2kinkwarpedfactor}. It can be seen that as the distance $b$ between the two kinks of the scalar field increases, the warp factor appears to plateau and its width also increases. The value of the scalar potential at $\phi=0$ increases with $b$. As the distance $b$ approaches infinity, the value of $V(0)$ approaches zero, and the values on both sides will decrease accordingly. 

Next we consider gravitational perturbations on the brane with the RS gauge $h_{(M0)}=0$. Since the scalar, vector, and tensor modes are decoupled from each other, we just focus on the tensor perturbation:
\begin{eqnarray}
	ds^2=\text{e}^{2A(y)}(\eta_{\mu\nu}+h_{\mu\nu})dx^{\mu}dx^{\nu}+dy^2. \label{5Dhmetric}
\end{eqnarray}
We consider the transverse and traceless (TT) gauge $\eta^{\mu\alpha}\pd_{\alpha}h_{\mu\nu}=\eta^{\mu\alpha}h_{\mu\nu}=0$ and make the transformation $dy=\e^{A}dz$.  Then, by redefining $h_{\mu\nu}=\e^{-3A(z)/2}\tilde{h}_{\mu\nu}$, the perturbation equation can be obtained:
\begin{eqnarray}
	\Box_{(4)} \tilde{h}_{\mu\nu}+\partial_z^2\tilde{h}_{\mu\nu}-V_{\text{eff}}(z)\tilde{h}_{\mu\nu}=0,\label{perturbation eq}
\end{eqnarray}
where $\Box_{(4)}=\eta^{\mu\nu}\partial_{\mu}\partial_{\nu}$ and the effective potential is
\begin{eqnarray}
	V_{\text{eff}}(z)=\frac{3}{2}\partial_z^2A+\frac{9}{4}(\partial_z A)^2.\label{effectivepotential} 
\end{eqnarray}
When $V_{\text{eff}}(z)\to 0$,   Then we do the following decomposition $\tilde{h}_{\mu\nu}(t,x^{i},z)=\varepsilon_{\mu\nu}(x^{i})H(t,z)$. The evolution equation of the extra-dimensional component is
\begin{eqnarray}
	\partial_t^2H(t,z)-\partial_z^2H(t,z)+V_{\text{eff}}(z)H(t,z)=-p^2H(t,z),\label{EDevolution eq}
\end{eqnarray}
where $p^2$ satisfies $	\partial_i^2\varepsilon_{\mu\nu}=-p^2\varepsilon_{\mu\nu}$. Then, by considering the further decomposition $H(t,z)=\psi(z)\e^{i\omega t}$, we can rewrite Eq.~\eqref{EDevolution eq} as
\begin{eqnarray}
	-\partial_z^2\psi(z)+V_{\text{eff}}(z)\psi(z)=m^2\psi(z),\label{EDperturbation eq}
\end{eqnarray}
where $m^2=\omega^2-p^2$ is the Kaluza-Klein (KK) mass of the KK mode on the brane. When $m=0$, the zero mass mode can be obtained:
\begin{eqnarray}
	\psi_0(z)=N_0\text{e}^{3A/2}.\label{zero mode}
\end{eqnarray}
For the single kink solution, when $y\rightarrow \pm\infty$, $A(y)\rightarrow \mp\frac{2kv^2}{9} y$. It is easy to show that $\int_{-\infty}^{+\infty}\psi_0^2(z)dz=N_0^2\int_{-\infty}^{+\infty}\e^{2A(y)}dy<+\infty$. Thus, the zero mode is localized near the brane. Unlike the zero mode, the massive KK modes are free states that can propagate along the extra dimension. The  zero mode localized on the brane means that the Newtonian gravitational potential on the brane can be restored, and the KK modes contribute to the higher-order correction to the Newtonian potential on the brane~\cite{Csaki:2000fc}.

\begin{figure*}[htbp]
	\centering
	\subfigure[~$v=0.5$]{\includegraphics[width=7cm]{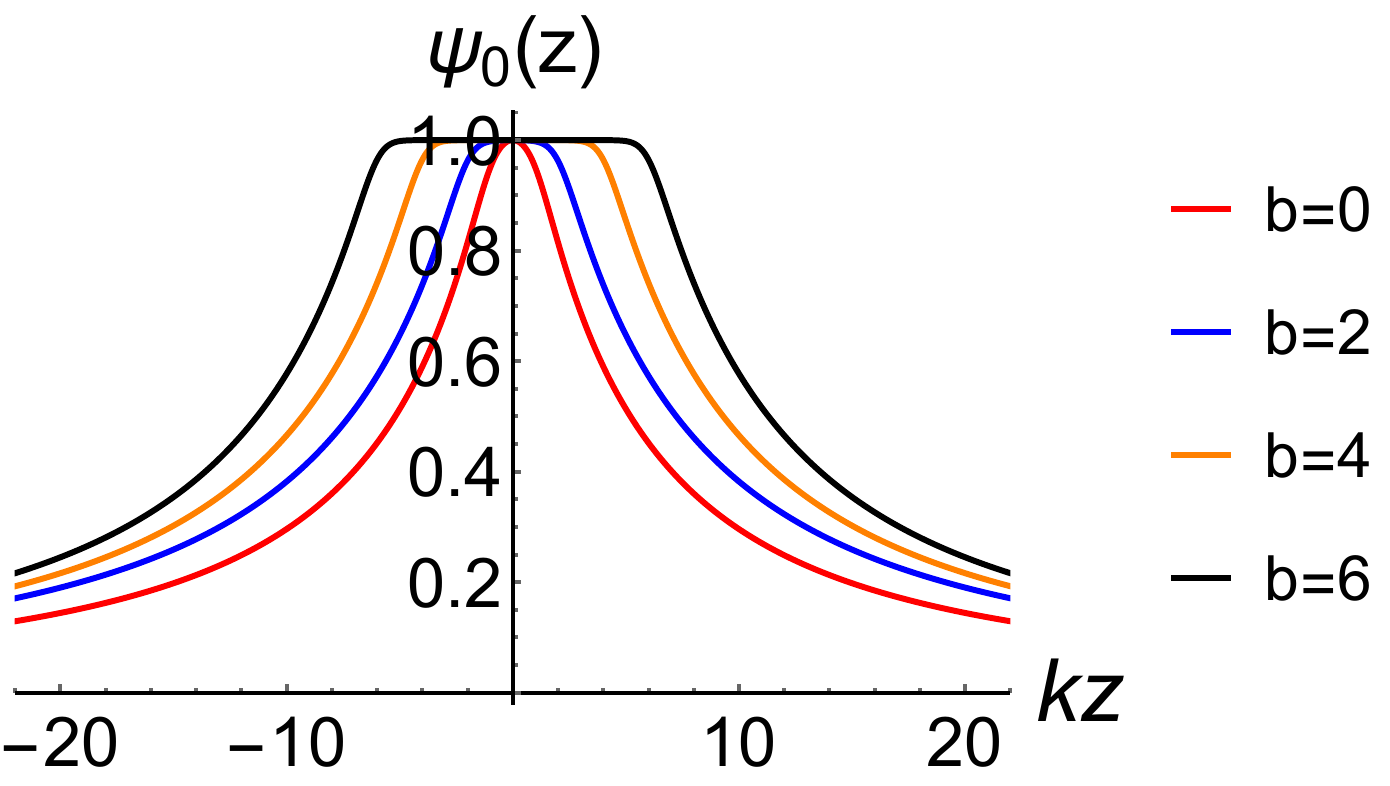}}
	\subfigure[~$b=6$]{\includegraphics[width=7cm]{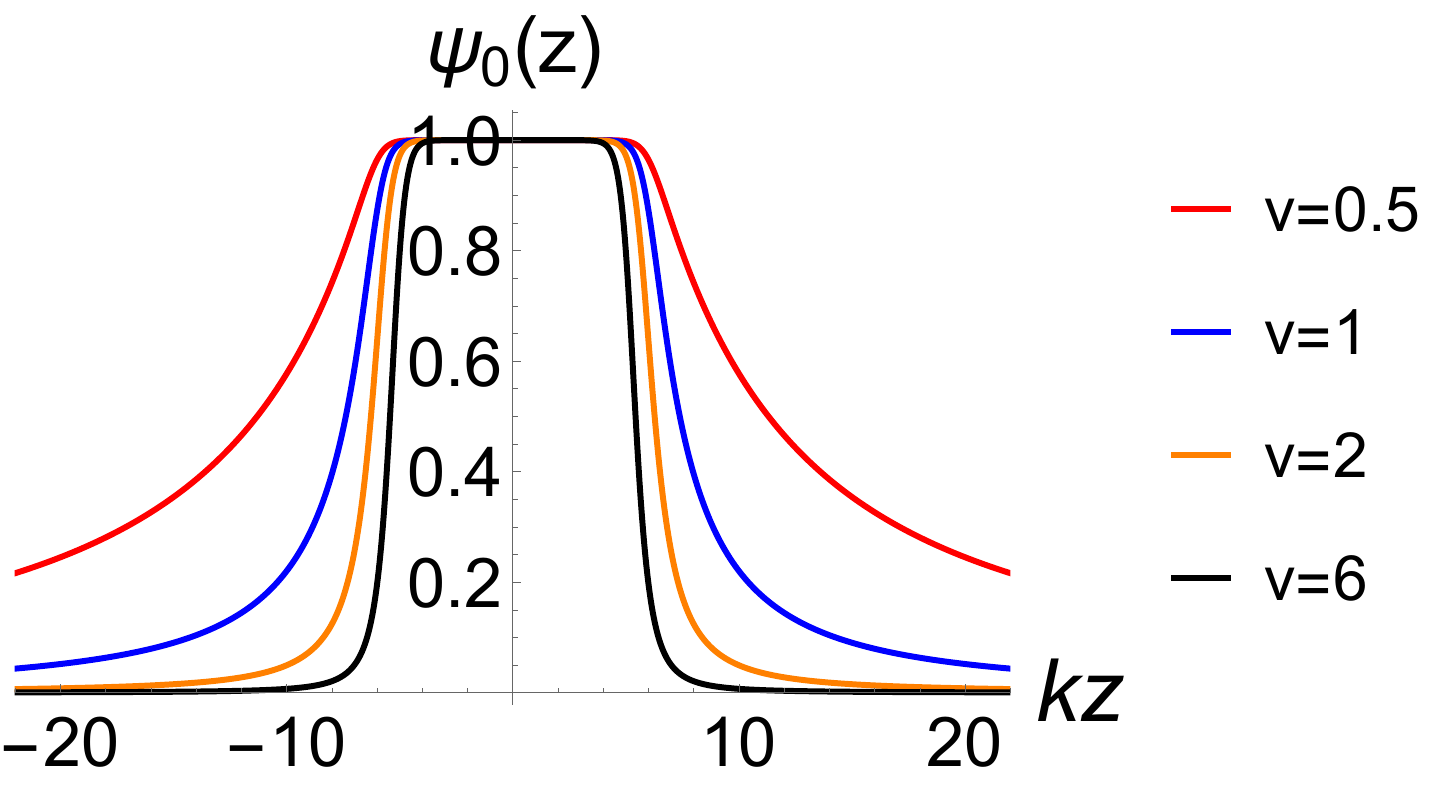}}
	\vskip -4mm \caption{Plots of the zero mode~\eqref{zero mode} with different values of $b$ and $v$ for the double kink scenario. }
	\label{Figzeromode}
\end{figure*}

\begin{figure*}[htbp]
	\centering
	\subfigure[~$v=0.5$]{\includegraphics[width=7cm]{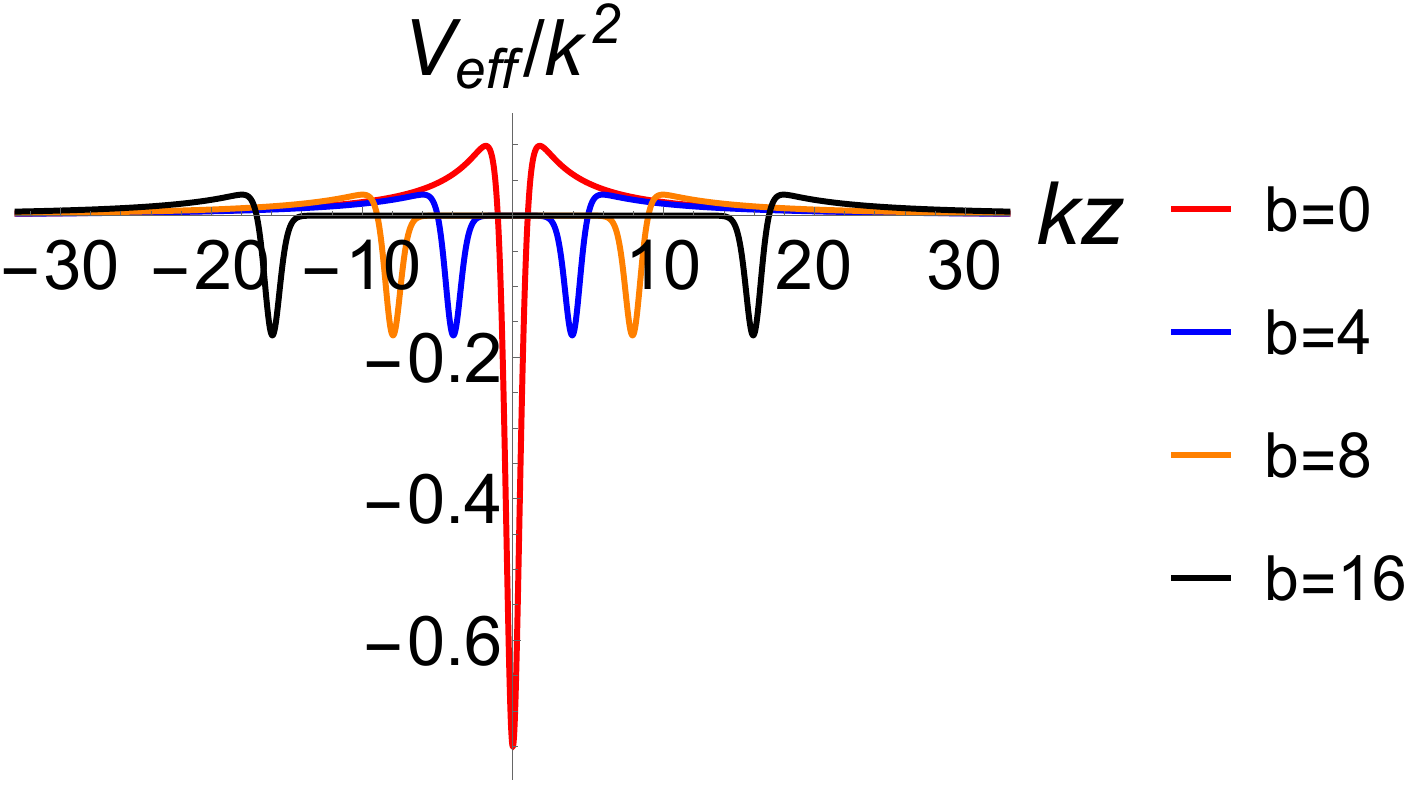}}
	\subfigure[~$b=6$]{\includegraphics[width=7cm]{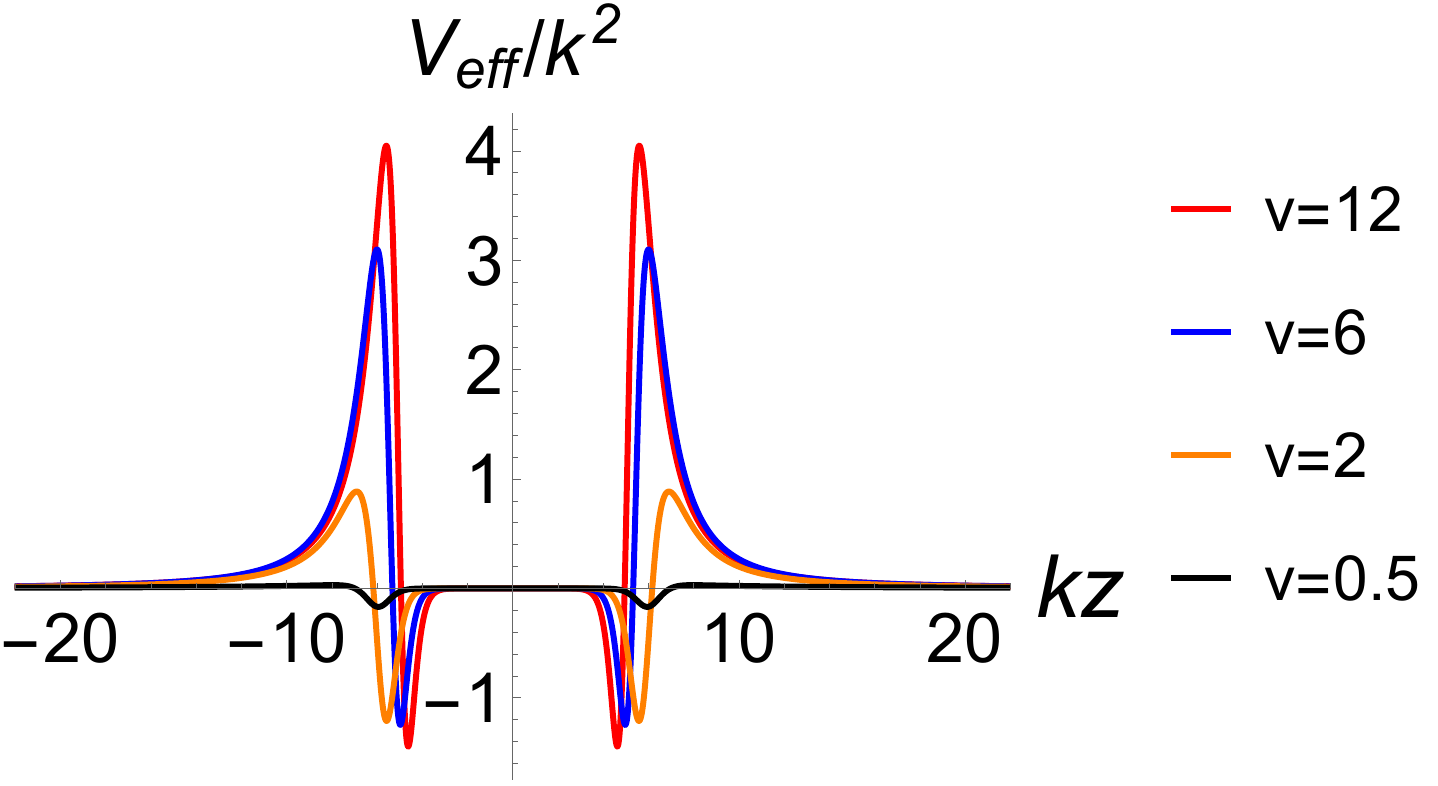}}
	\vskip -4mm \caption{Plots of the effective potential~\eqref{effectivepotential} with different values of $v$ and $b$ for the double kink scenario.}
	\label{FigVeff}
\end{figure*}

For the double kink solution, the distribution of the zero mode is shown in Fig.~\ref{Figzeromode}.
As $b$ increases, the trend on both sides of the zero mode remains the same, and the width of the platform near $y=0$ increases with $b$. The greater the vacuum expectation value $v$ of the scalar field, the steeper the peak of the zero mode.
The effective potential $V_{\text{eff}}$ with different values of  $b$ and $v$ is shown in Fig~\ref{FigVeff}. The distance between the two barriers increases with $b$ and the height of barriers increases with $v$.

\subsection{One-dimensional evolution}
\begin{figure*}[htbp]
	\centering
\subfigure[~$v=c$]{\includegraphics[width=5cm]{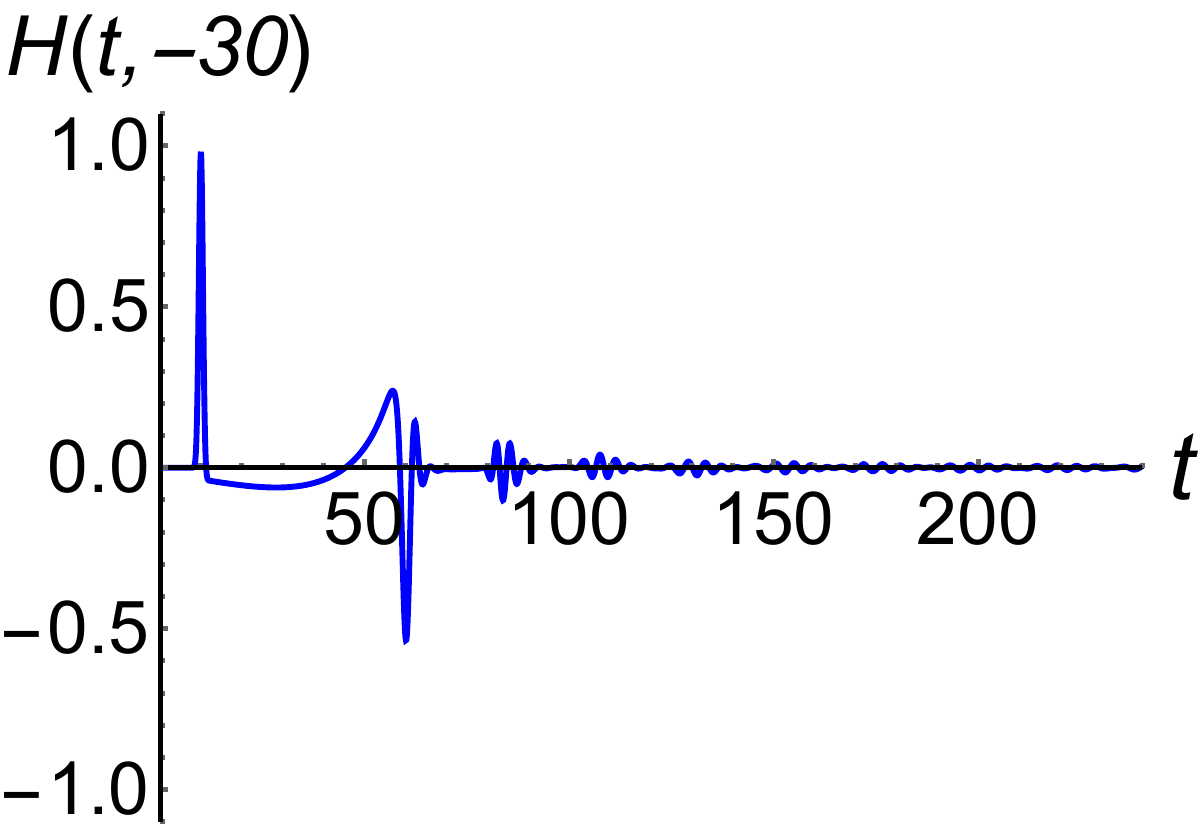}\label{FiWFevolutionc1}}
\subfigure[~$v=\frac{c}{2}$]{\includegraphics[width=5cm]{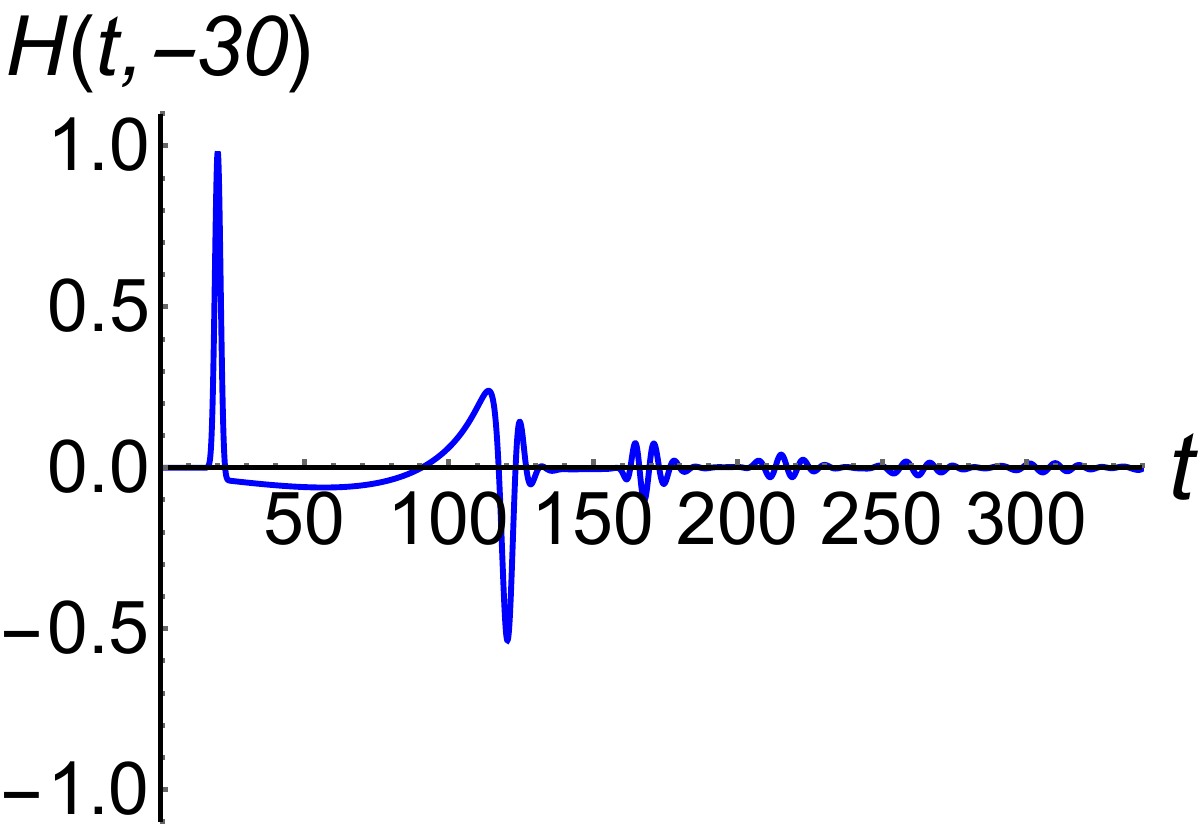}\label{FiWFevolutionc0.5}}
\subfigure[~$v=\frac{c}{3}$]{\includegraphics[width=5cm]{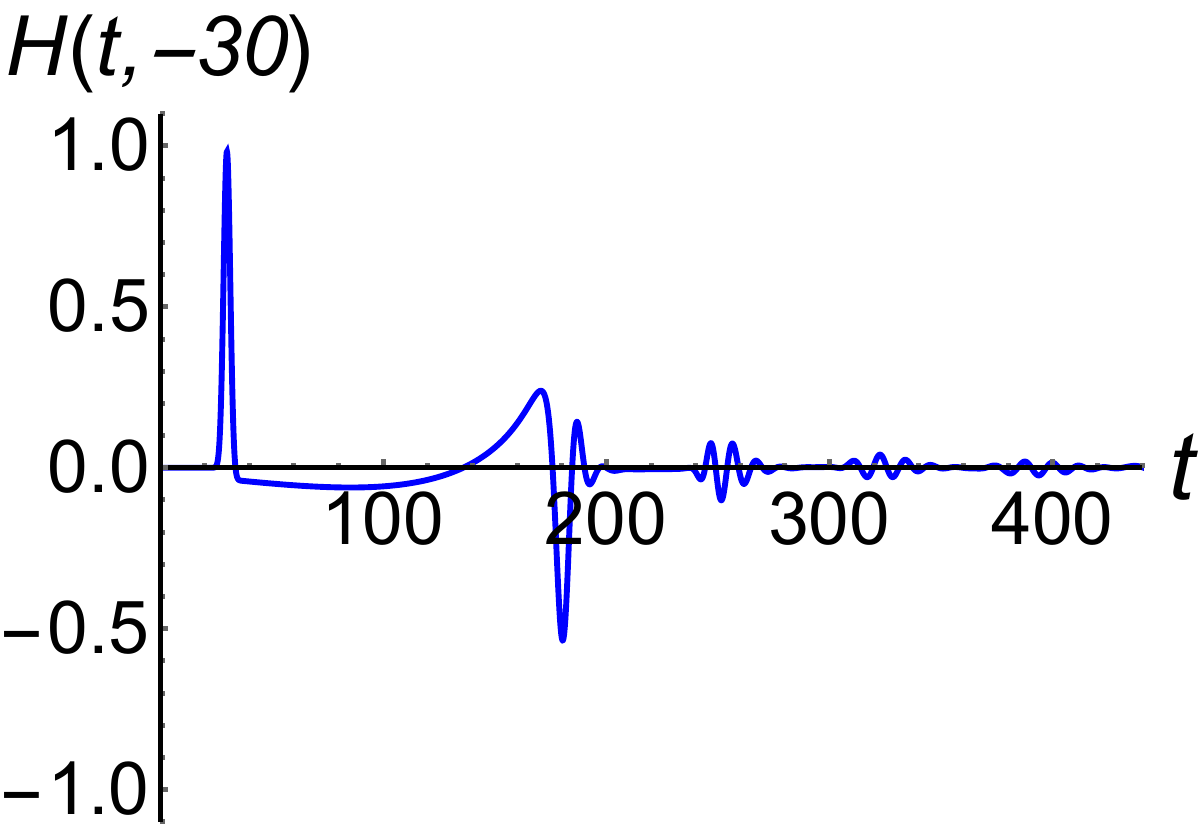}\label{FiWFevolutionc0.3}}
\subfigure[~$v=\frac{c}{4}$]{\includegraphics[width=5cm]{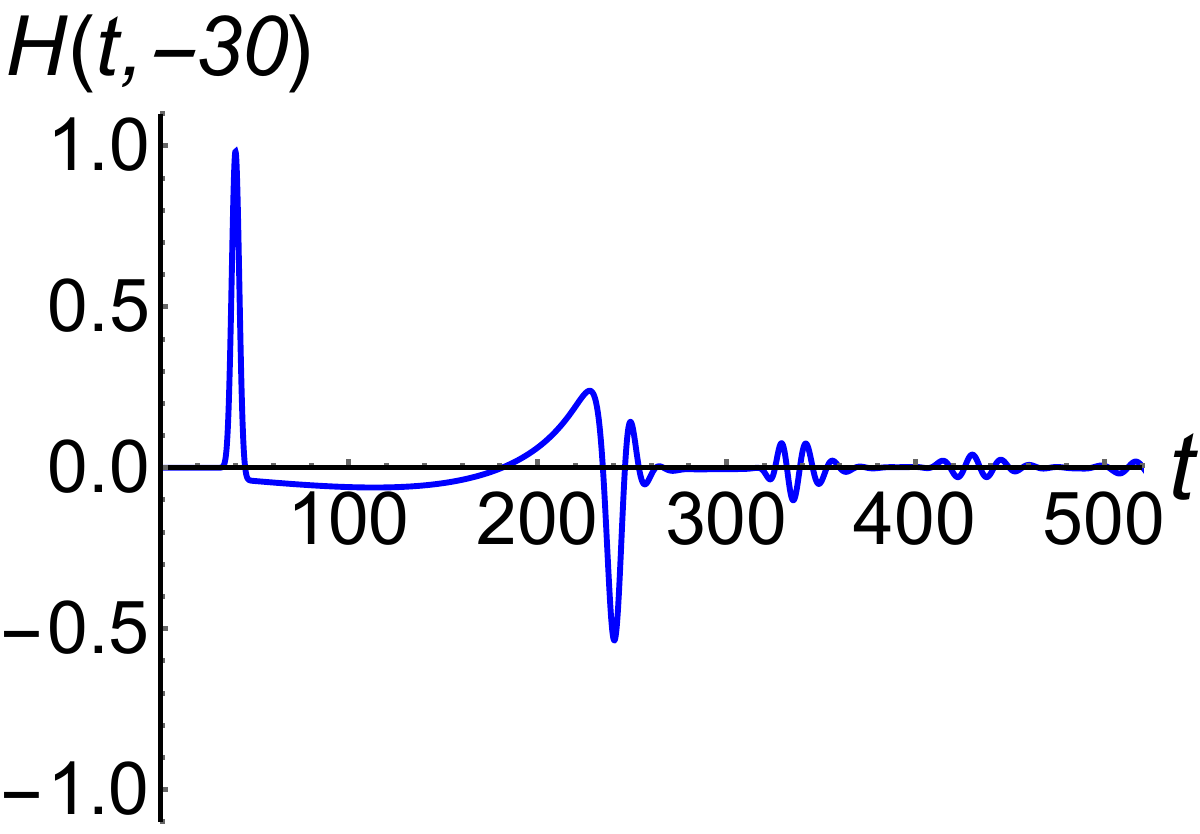}\label{FiWFevolutionc0.25}}
	\vskip -4mm \caption{Plots of the evolution $H(t, z)$ of different Gaussian wave packets located at fixed position $z_\text{ext}=-30$. The parameters of the double kink scalar and Gaussian wave packet are $b=6$, $v=6$ and $\sigma=0.5$.}
	\label{Figdifferentc}
\end{figure*}

Next, we investigate the evolution of a Gaussian packet. The evolution equation is given by Eq.~\eqref{EDevolution eq}, and the boundary conditions can be selected to be the radiative boundary conditions~\cite{Megevand:2007uy}. We initialize the system with a right-moving Gaussian wave packet $H(0,z)=\e^{-\frac{(z-z_0)^2}{\sigma}}$ propagating with different speeds speed as the initial condition. As shown in Fig.~\ref{Figdifferentc}, the whole evolution waveforms of Gaussian wave packets with different speeds are the same, only the time interval among the echoes will change with the different speeds. Therefore, the wave speed does not affect the production of echoes and other properties other than speed.

\begin{figure*}[htbp]
	\centering
	\subfigure[~$z_0=0, z_\text{ext}=0$]{\includegraphics[width=5cm]{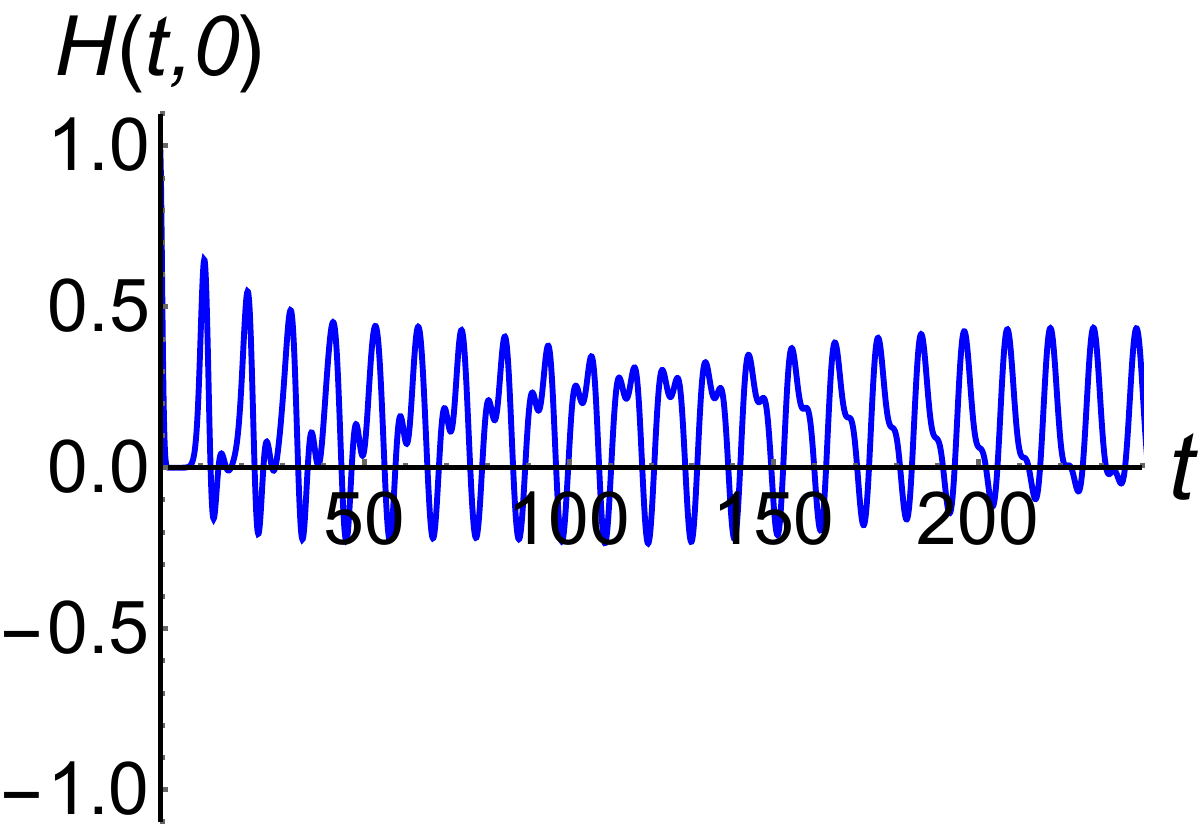}\label{FiWFevolution1}}
	\subfigure[~$z_0=0, z_\text{ext}=-30$]{\includegraphics[width=5cm]{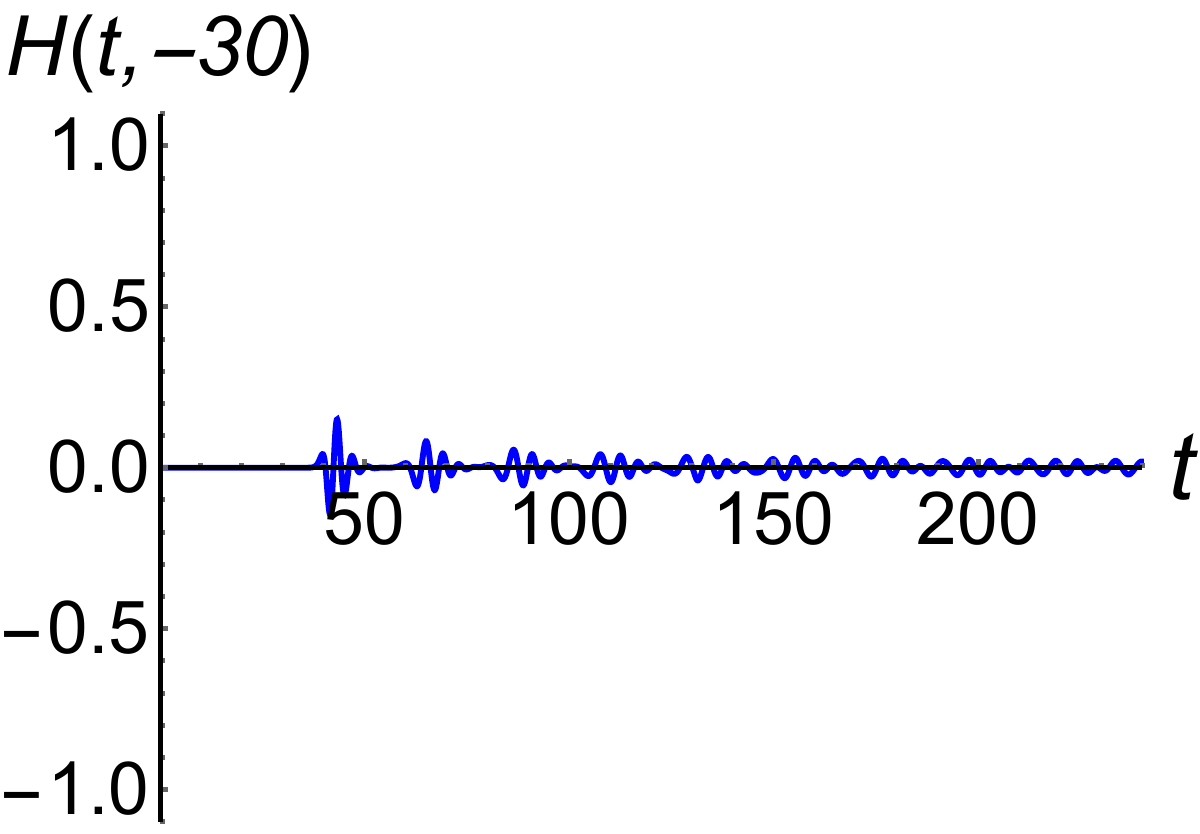}\label{FiWFevolution2}}
	\subfigure[~$z_0=0, z_\text{ext}=30$]{\includegraphics[width=5cm]{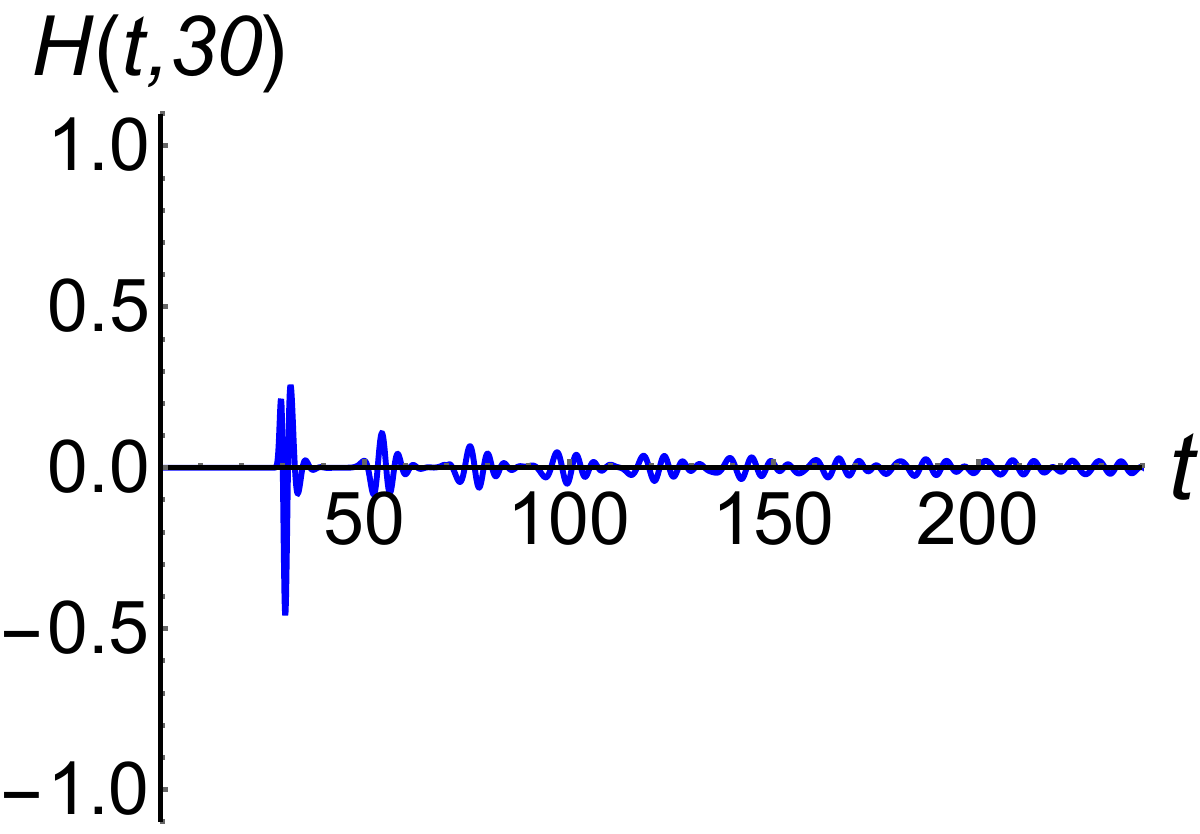}\label{FiWFevolution3}}
	\subfigure[~$z_0=-40, z_\text{ext}=0$]{\includegraphics[width=5cm]{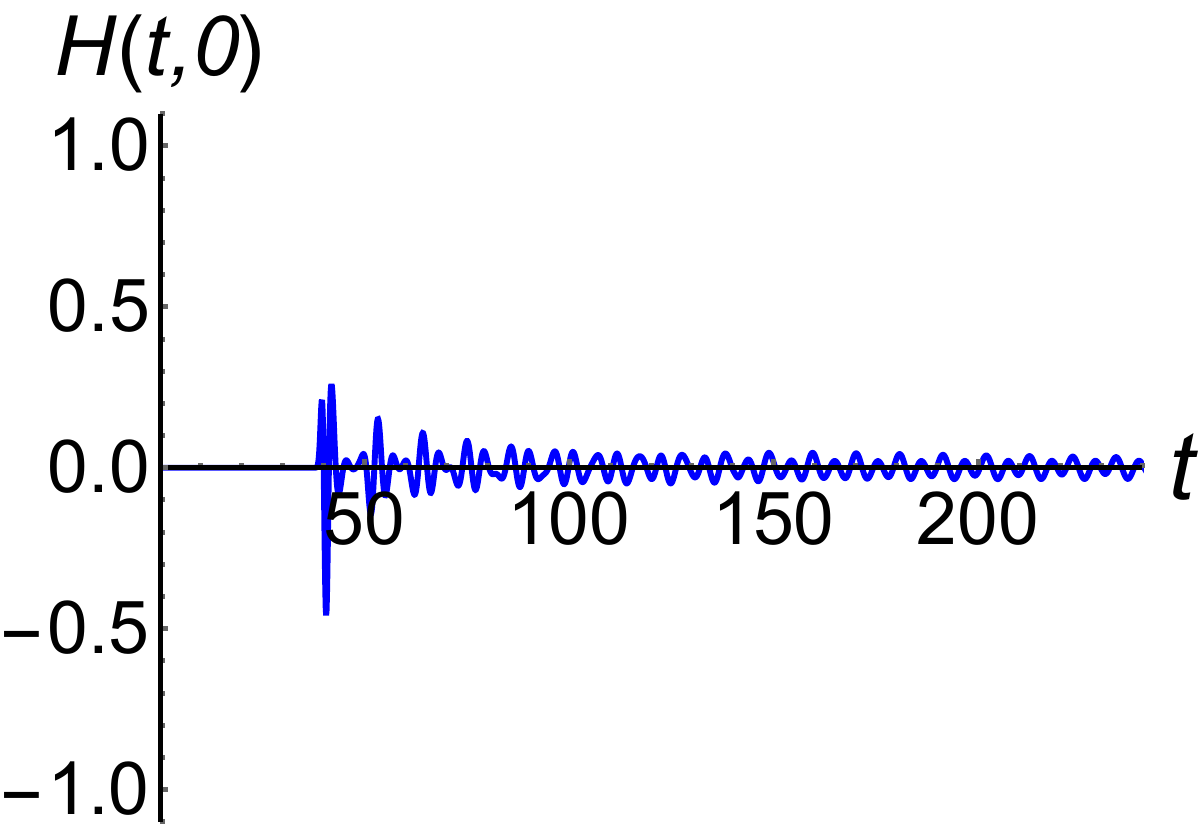}\label{FiWFevolution4}}
	\subfigure[~$z_0=-40, z_\text{ext}=-30$]{\includegraphics[width=5cm]{phi_tanhyWFy=40_30.pdf}\label{FiWFevolution5}}
	\subfigure[~$z_0=-40, z_\text{ext}=30$]{\includegraphics[width=5cm]{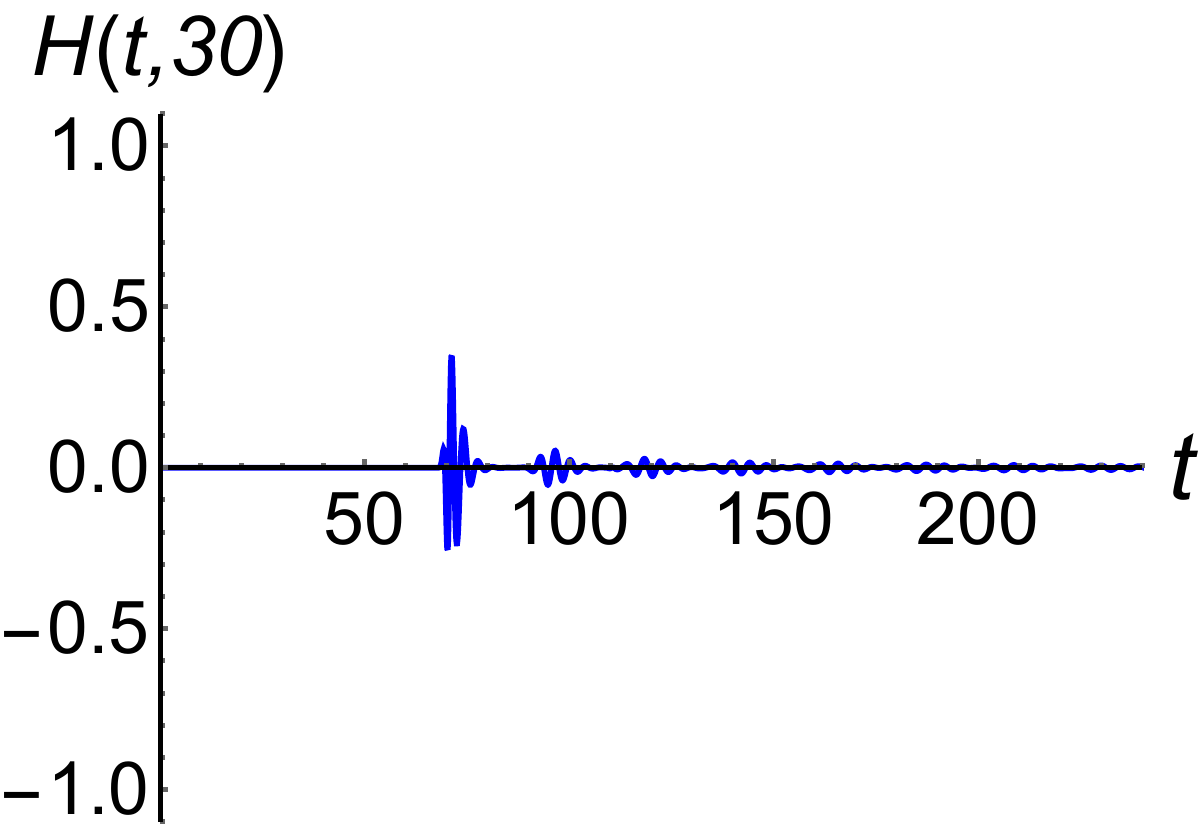}\label{FiWFevolution6}}
	\vskip -4mm \caption{Plots of the evolution $H(t, z)$ of different Gaussian wave packets located at different positions $z_\text{ext}$. The parameters of the double kink scalar and Gaussian wave packet are $b=6$, $v=6$ and $\sigma=0.5$.}
	\label{FiWFevolution}
\end{figure*}

\begin{figure*}[htbp]
	\centering
	\subfigure[~$b=4,v=6$]{\includegraphics[width=6cm]{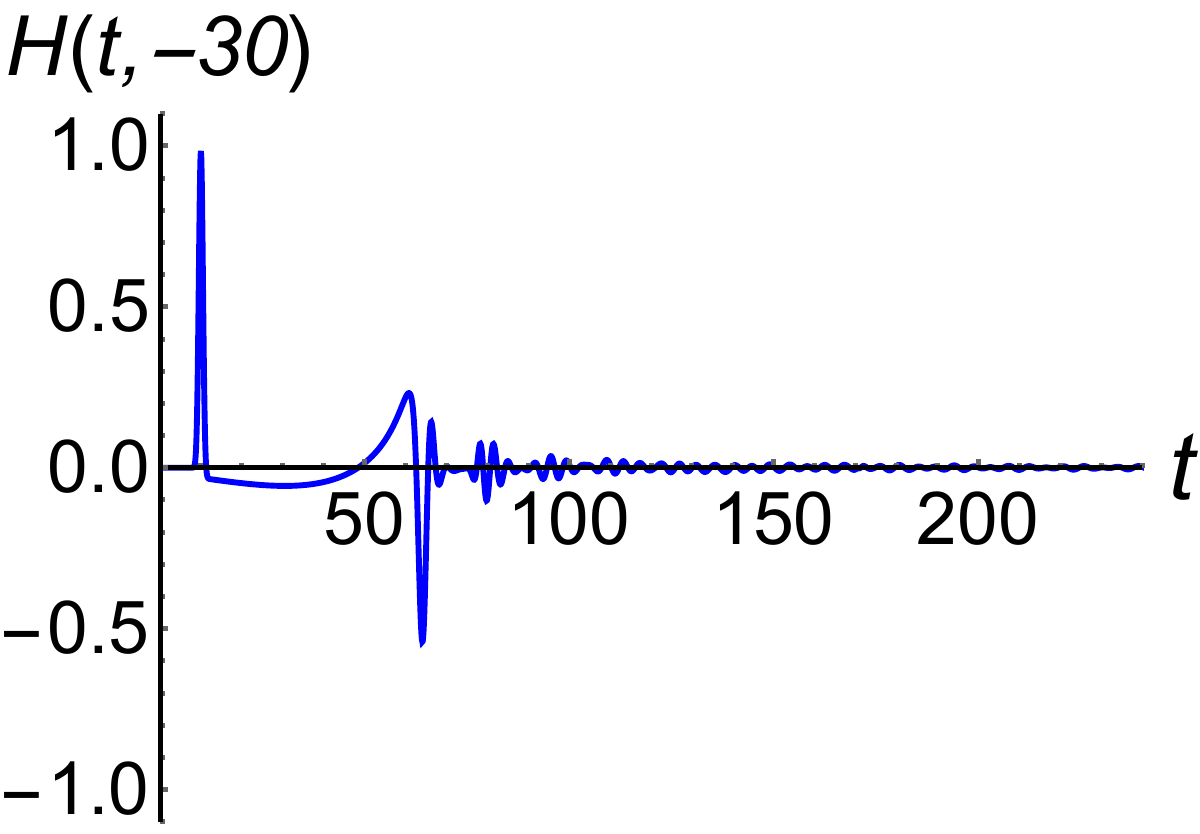}}
	\subfigure[~$b=8,v=6$]{\includegraphics[width=6cm]{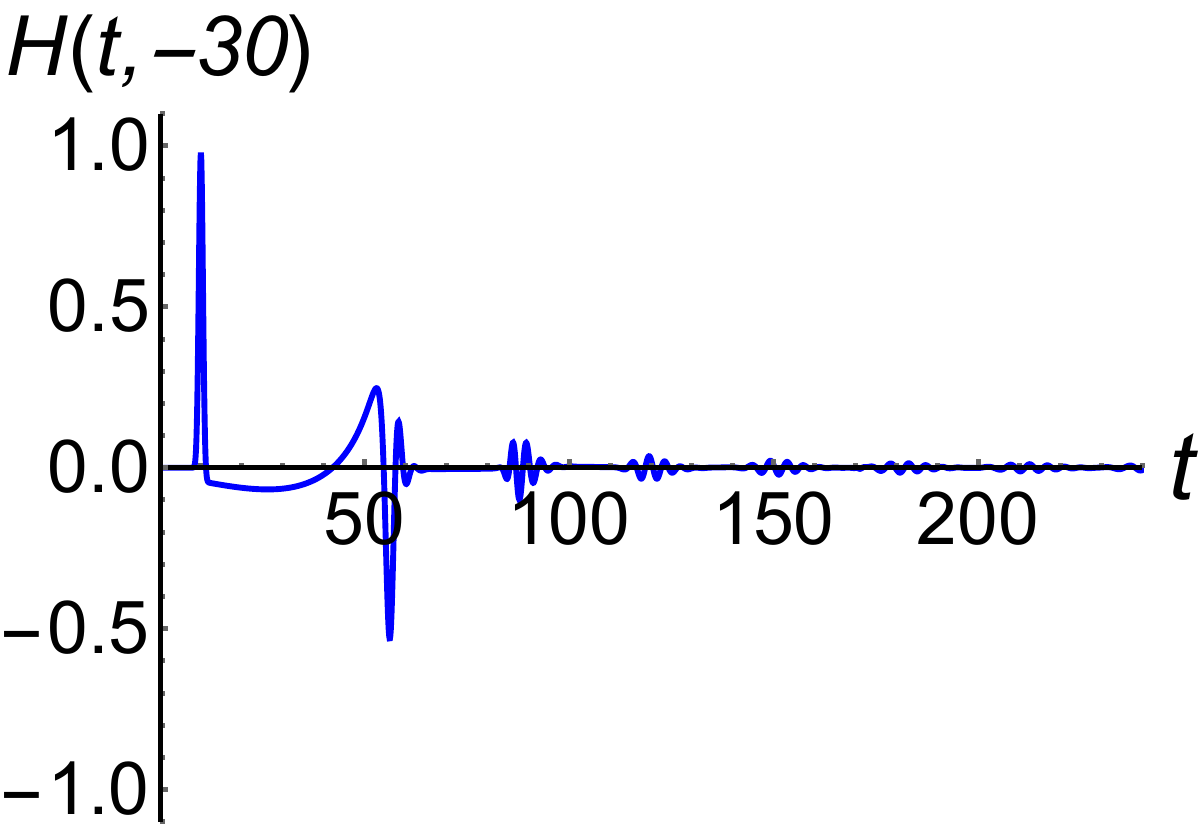}}	
	\subfigure[~$b=6,v=2$]{\includegraphics[width=6cm]{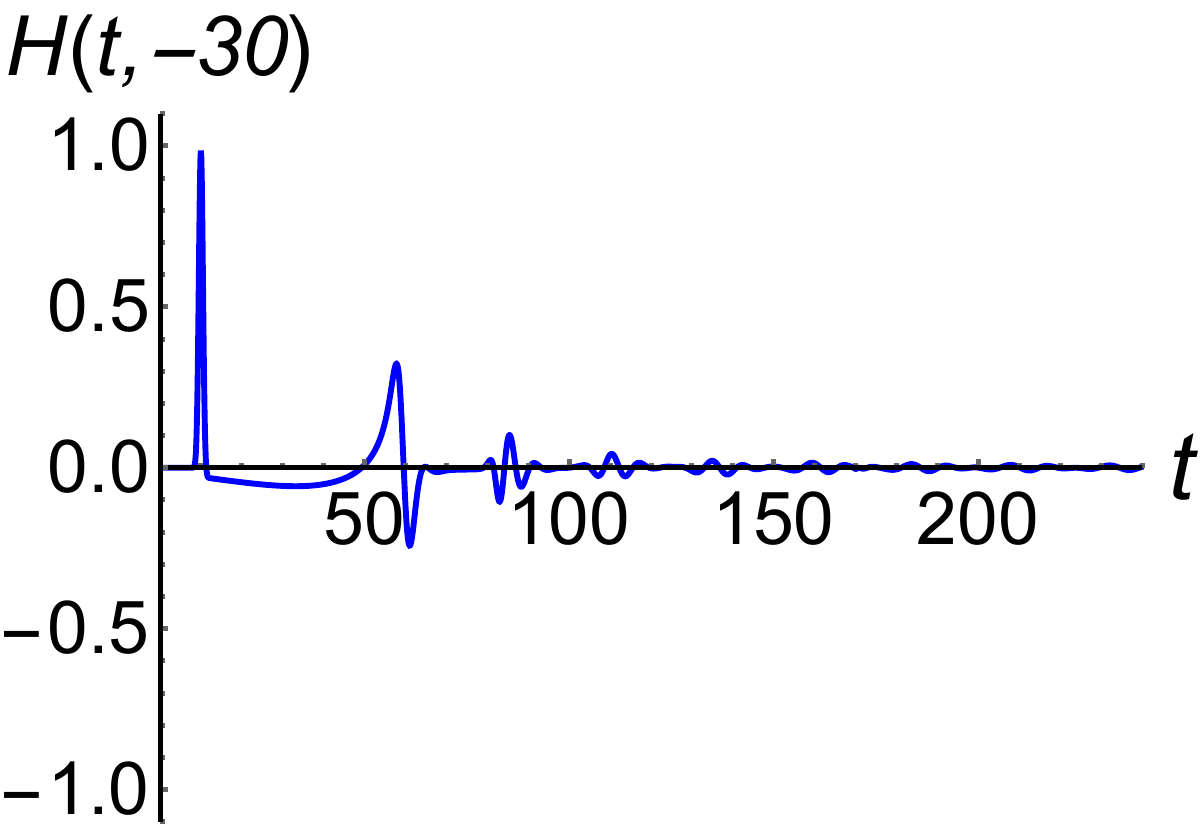}}
	\subfigure[~$b=6,v=12$]{\includegraphics[width=6cm]{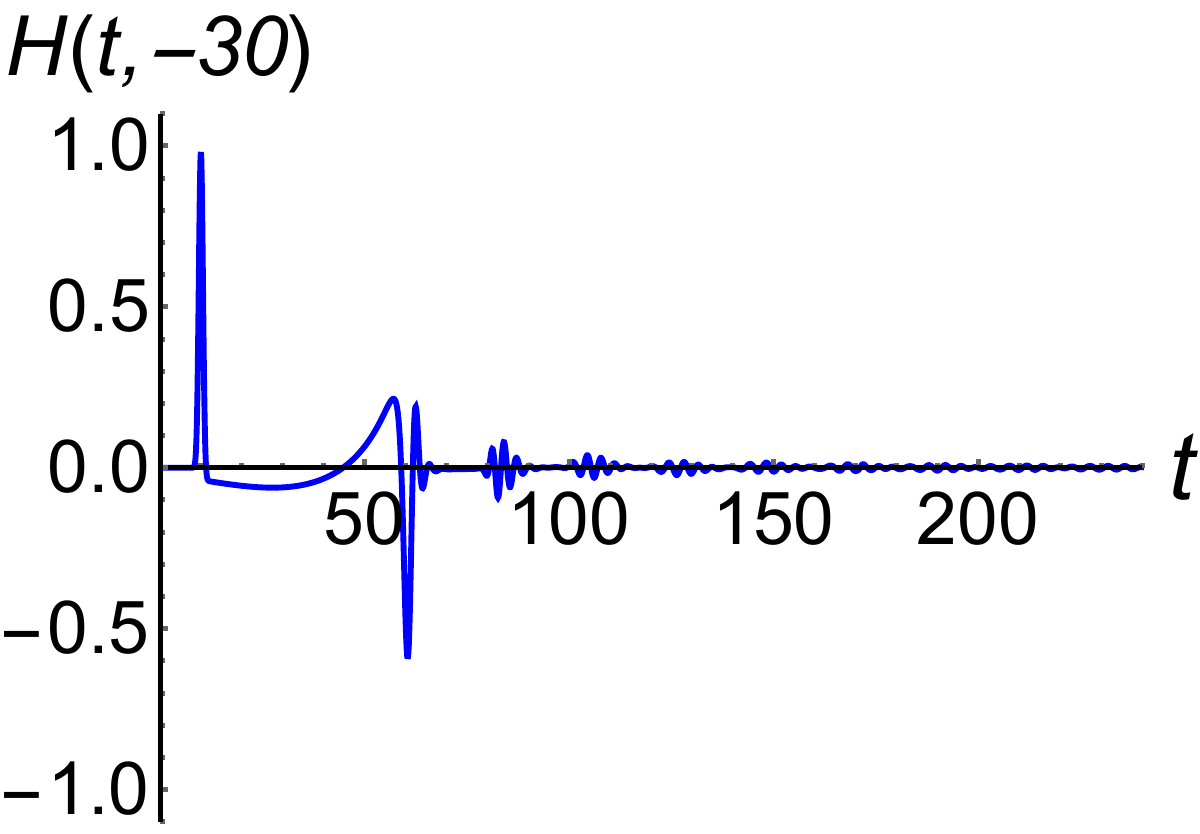}}
	\vskip -4mm \caption{Plots of the evolution $H(t, z)$  of the Gaussian wave packets for different values of the parameters of the double kink scalar field. The parameters of the Gaussian wave packet are set to $z_0=-40$ and $\sigma=0.5$. }
	\label{FiWFdvevolution}
\end{figure*}

  For convenience, we only consider the case where the gravitational waves propagate only along the extra dimension (no momentum along the brane). In this case the waves propagate with the speed of light. The reason is that in the five-dimensional general relativity framework, the gravitational waves propagate at the speed of light from Eq.~\eqref{perturbation eq}. The evolution behaviors of different Gaussian wave packets at a fixed point of extra dimension are shown in Figs.~\ref{FiWFevolution} and \ref{FiWFdvevolution}.

Figure~\ref{FiWFevolution}  illustrates the evolutionary behavior of Gaussian wave packets positioned at various locations: $z_0=0$ (inside the potential well) and $z_0=-40$ (outside the potential well).  In Figs.~\ref{FiWFevolution1} and \ref{FiWFevolution4}, it can be seen that the echoes are not distinct for Gaussian wave packets whether they are located inside or outside the potential well. The reason is that, for the signal within the potential well, the wave packet encounters reflections from both barriers. Additionally, the difference between the widths of the wave packet and the potential well is small.  Consequently, the time intervals between the reflected pulses are small. The echo phenomenon in other subgraphs is relatively obvious. Figure~\ref{FiWFdvevolution} shows the effect of the vacuum expectation  value $v$ and the distance $b$ between the two kinks of the scalar field on the echoes. The larger the distance between the two kinks, the greater the time interval among the echoes. The larger vacuum expectation value $v$, the larger the amplitude of the first echo, and the smaller the amplitude of the subsequent echoes. However, it results in smaller amplitude for the subsequent echoes, though with a slower rate of attenuation. The echoes occur because a wave with the resonant frequency is reflected back and forth within the cavity, with a portion being transmitted during each reflection. The amplitude of the primary wave is the largest, while the amplitudes of the subsequent echo pulses gradually decrease. Due to the high reflectivity of the barrier, the amplitude decay in the quasi-potential well is slow, as evidenced in the left panel of Fig.~\ref{FiWFevolution}.  We can perform spectral analysis of the waveform to demonstrate this point,
\begin{eqnarray}
	F\big[ \tilde{H}(f,z_\text{ext})\big] =\left|A\sum_{p}H(t_p,z_\text{ext})~\e^{-i2\pi ft_p}\right|,\label{discrete Fourier transform}
\end{eqnarray}
where $A=1/\text{max}(	F\left[\tilde{H}(f,z_\text{ext})\right] )$ is a normalized constant, and $z_\text{ext}$ is a fixed point. We use the transfer matrix to calculate the transmittance spectra.

\begin{figure*}[htbp]
	\centering
	\subfigure[~$t=0\sim2000$]{\includegraphics[width=6cm]{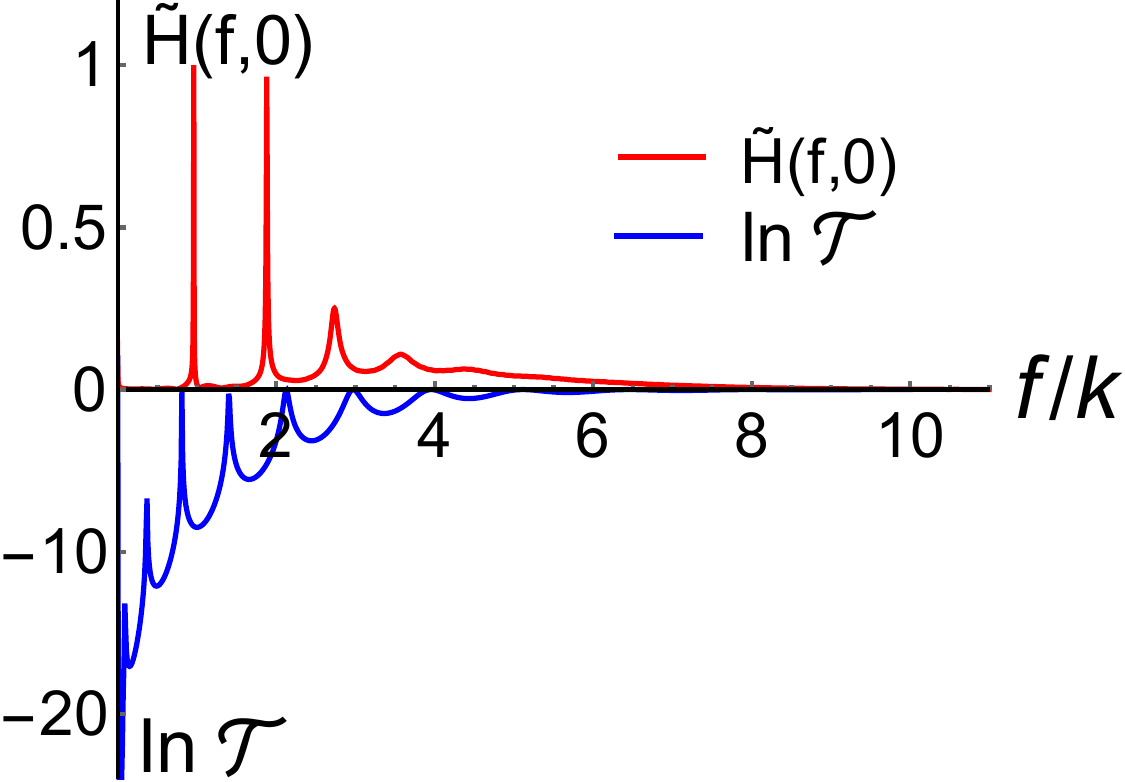}\label{Figfourier1}}
	\subfigure[~$t=0\sim2000$]{\includegraphics[width=6cm]{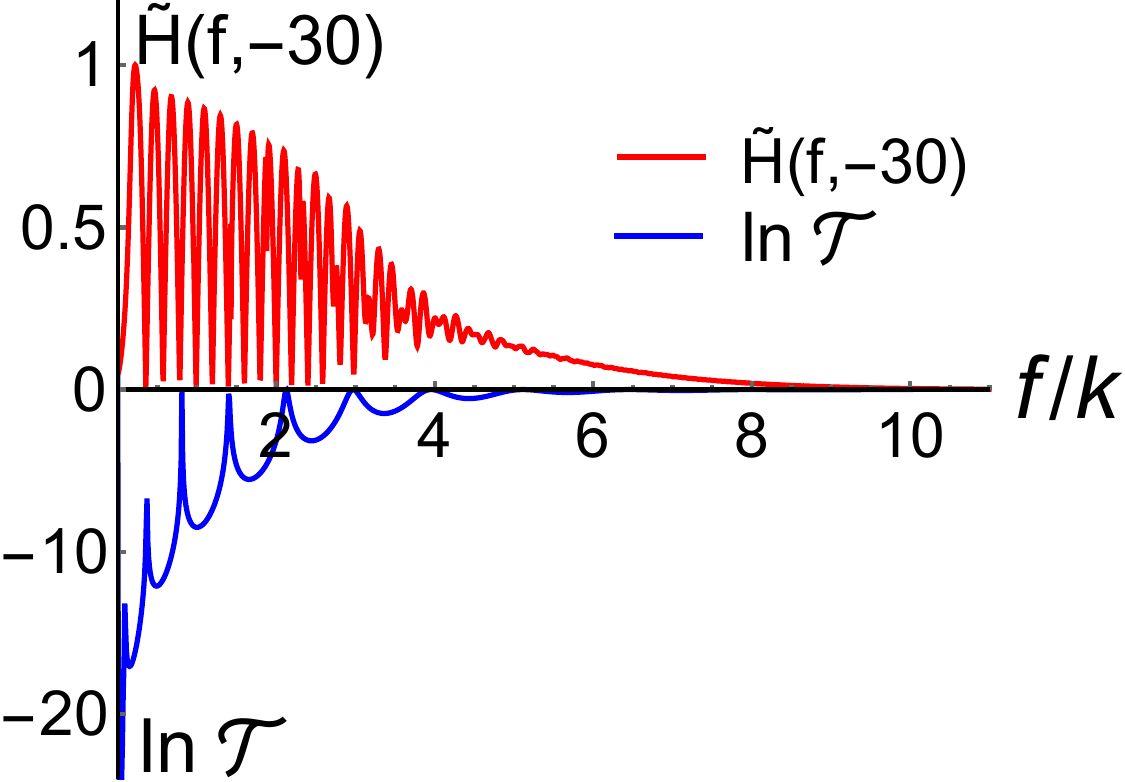}\label{Figfourier2}}
	\subfigure[~$t=0\sim100$]{\includegraphics[width=6cm]{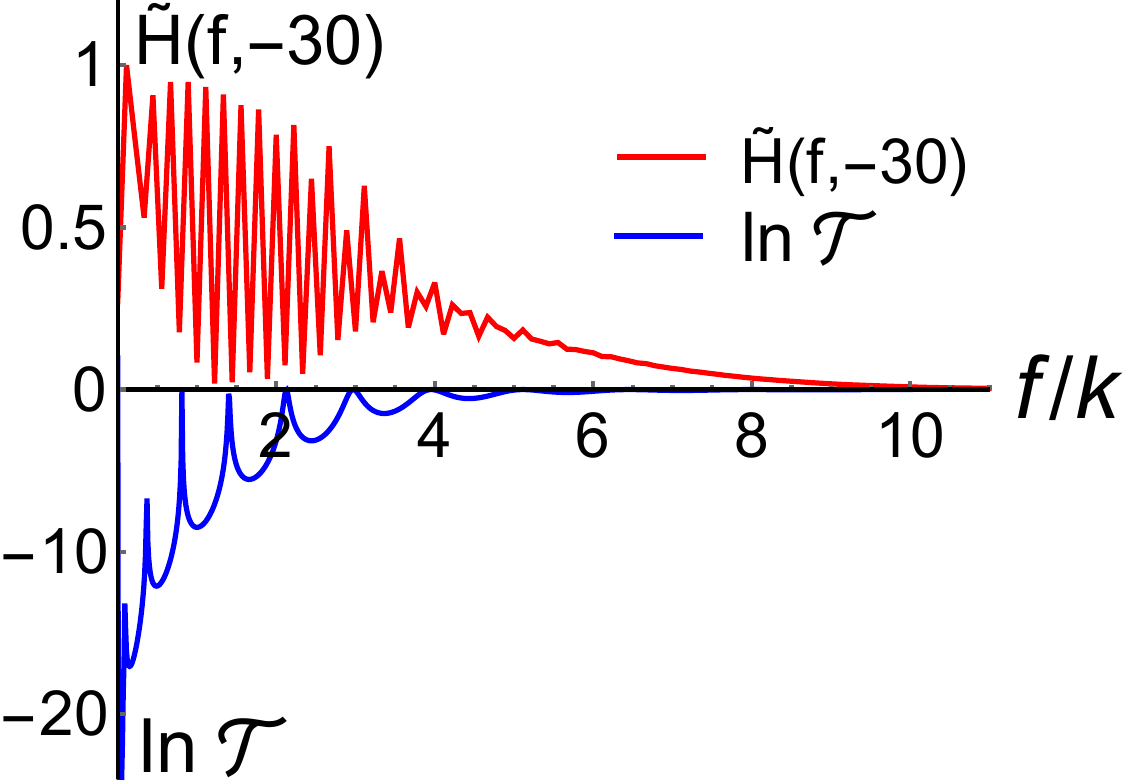}\label{Figfourier3}}
	\subfigure[~$t=100\sim500$]{\includegraphics[width=6cm]{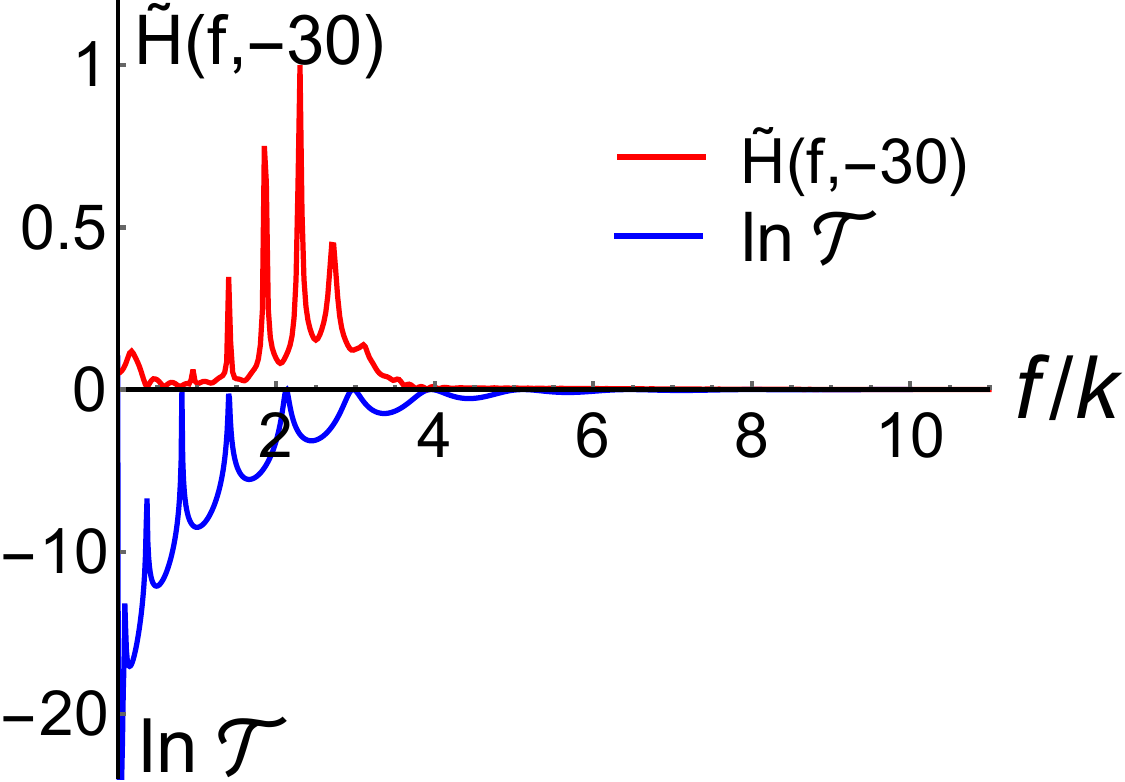}\label{Figfourier4}}
	\subfigure[~$t=500\sim1000$]{\includegraphics[width=6cm]{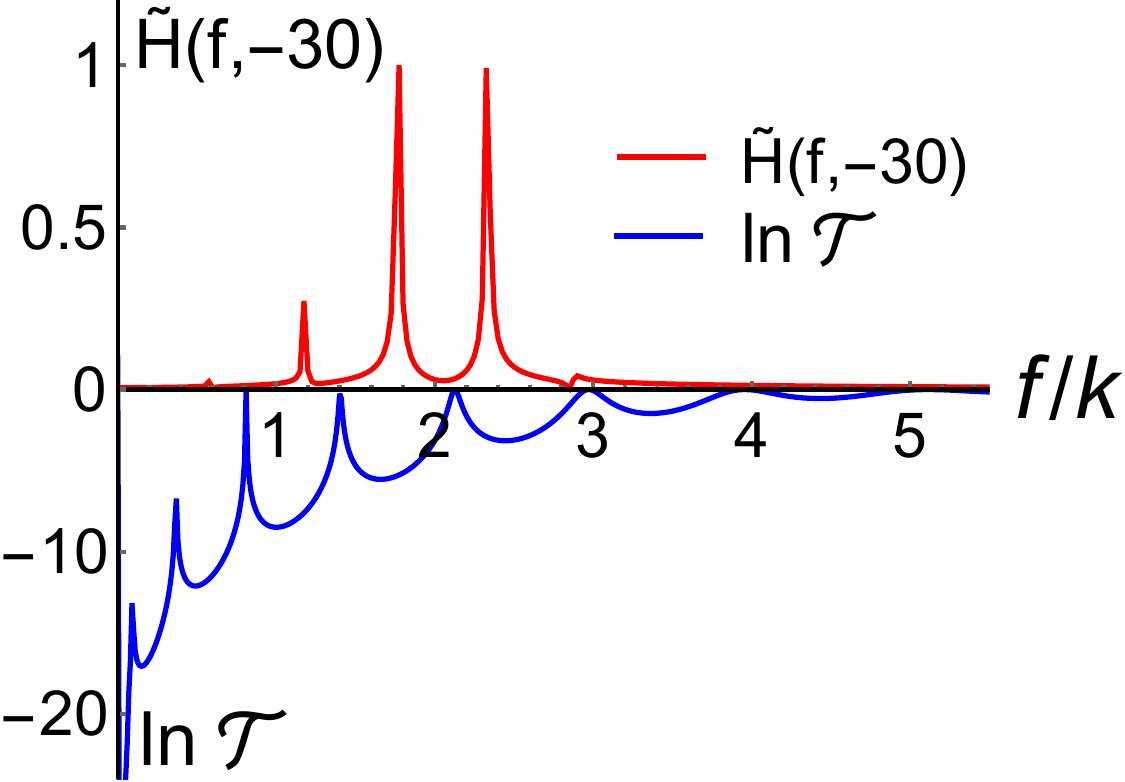}\label{Figfourier5}}
	\subfigure[~$t=1000\sim2000$]{\includegraphics[width=6cm]{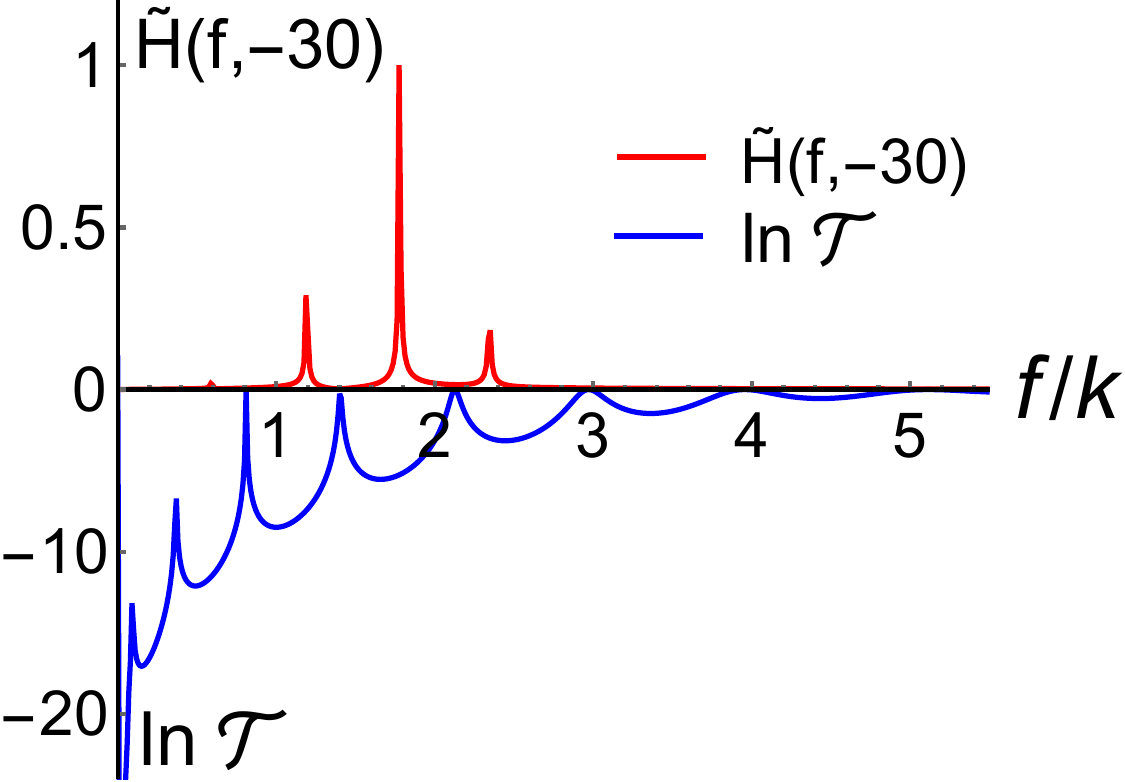}\label{Figfourier6}}
	\vskip -4mm \caption{Plots of the Fourier transform spectra of the waveforms in Figs.~\ref{FiWFevolution4} and \ref{FiWFevolution5} within different time intervals, and the transmittance $\mathcal{T}$ under the effective potential~\eqref{effectivepotential} with parameters $b=6$ and $v=6$.  }
	\label{Figfourier}
\end{figure*}

Figure~\ref{Figfourier} shows the frequency spectra of the waveforms illustrated in Figs.~\ref{FiWFevolution4} and \ref{FiWFevolution5} and the transmittance of the gravitational perturbations under the effective potential~\eqref{effectivepotential}. The frequency spectrum of the waveform at an internal point $z_0=0$ of the potential well is shown in Fig.~\ref{Figfourier1}. Due to the formation of a resonant cavity by the two barriers of the potential, the internal waves undergo multiple reflections from these barriers. Therefore, the wave in the resonant cavity has low transmittance (and consequently high reflectivity). The frequency spectrum of the waveform at an external point $z_0=-30$ is shown in Fig.~\ref{Figfourier2}. Its frequencies have no obvious pattern. To analysis this complex spectrum, we calculate and present frequency spectra for different time intervals in Figs.~\ref{Figfourier3}-\ref{Figfourier6}. The frequency spectrum for the early time interval $t=0\sim100$ is shown in Fig.~\ref{Figfourier3}. This waveform  contains the first three pulses, that can be seen from Fig.~\ref{FiWFevolution5}. The frequency distribution of this waveform is roughly similar to Fig.~\ref{Figfourier2}. Later signals are mainly the echo pulses. The later the echo appears, the lower its transmittance, which can be seen in Figs.~\ref{Figfourier4}-\ref{Figfourier6}.  The reason is that, echoes are the part of the wave that is transmitted from the potential well. The more times the wave is reflected back and forth by the two barriers, the greater the reflection coefficient of the echoes. The frequencies of the high reflectivity echoe in the late-time stages can be utilized as characteristic frequencies to comprehend the spacetime structure of extra dimensions.

The expression for the effective stress-momentum tensor of gravitational waves is given by~\cite{Isaacson:1968hbi,Isaacson:1968zza}
\begin{equation}
	T_{MN}= \frac{1}{32\pi} \langle\partial_M h^{\mu\nu}\partial_N h_{\mu\nu}\rangle,
	\label{energy-momentum}           
\end{equation}
where the angular brackets $\langle...\rangle$ denote a spatial average over several wavelengths. By using the stress-energy tensor $T_{MN}$ and the time-like killing vector $k^N=(\e^{-A},0,0,0,0)$, the conserved current is given by
\begin{equation}
	J_{M}=T_{MN}k^{N}.
	\label{current}
\end{equation}
Then we can define the energy of the gravitational perturbation~\cite{Pavlidou:2000cs}:
\begin{equation}
	E=\int J^0 \sqrt{-g} d^3xdz.
	\label{abstract energy}
\end{equation}
Assuming the gravitational perturbation is a plane wave on the brane, we solely focus on the evolution of the extra dimensional component. Consequently, we separate the energy into two parts as follows,
\begin{equation}
	E=\varepsilon_b\times \varepsilon_e,
\end{equation}
where
\begin{eqnarray}
	\varepsilon_b&=& \frac{1}{32\pi}\int \epsilon_{\mu\nu}\epsilon^{\mu\nu} d^3x, \\  
	\varepsilon_e&=& \int_{-z_b}^{z_b} (\partial_t H(t,z))^2 dz.
	\label{energycomponent}
\end{eqnarray}
In the following part, we only focus on the extra-dimensional part $\epsilon_e$.

\begin{figure*}[htbp]
	\centering
	\subfigure[~$z_0=0$]{\includegraphics[width=6cm]{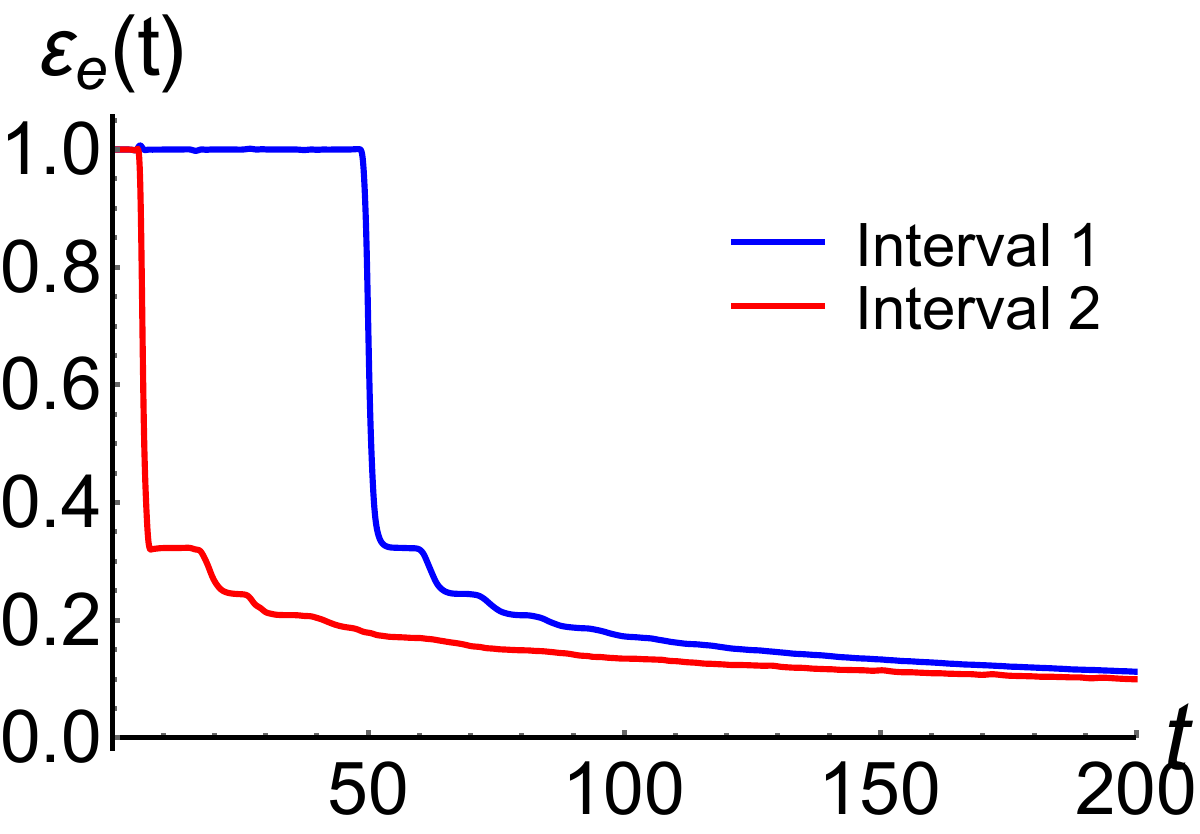}\label{Fienergyevolution1}}
	\subfigure[~$z_0=-40$]{\includegraphics[width=6cm]{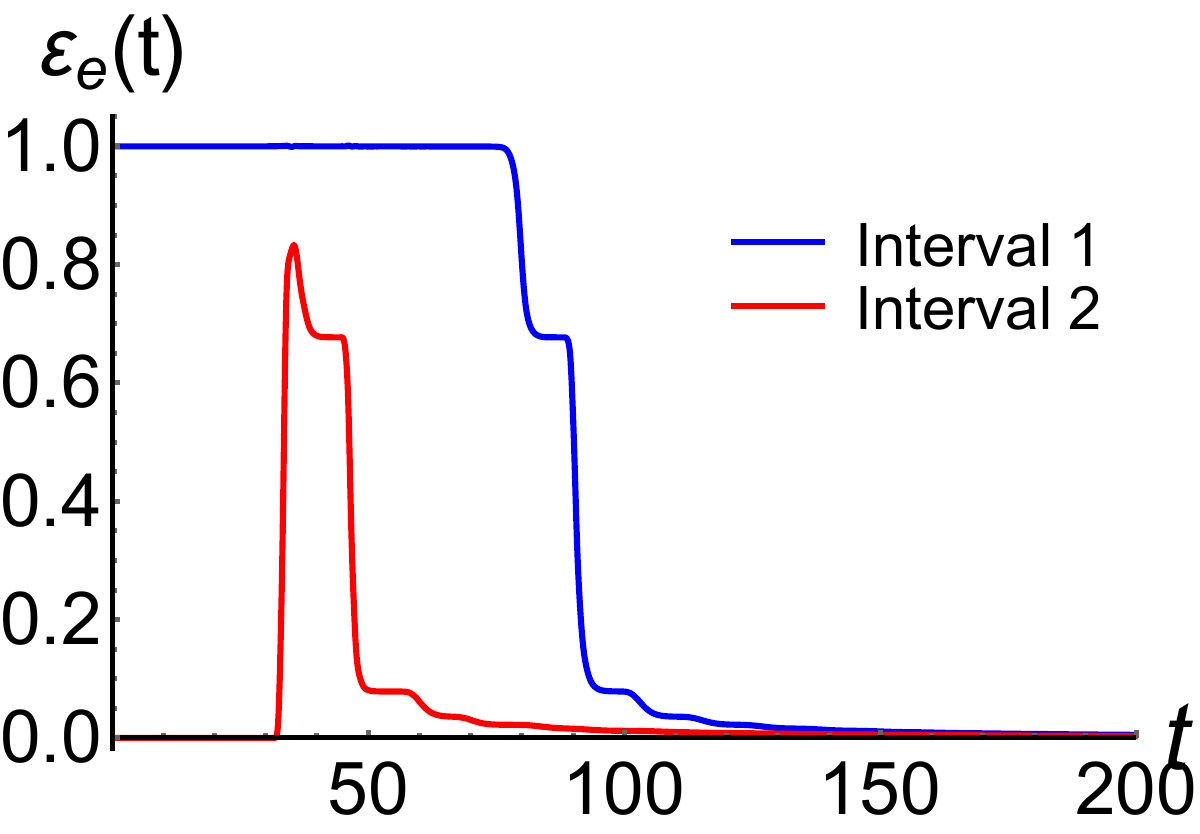}\label{Fienergyevolution2}}
	
	\vskip -4mm \caption{Plots of the evolution of energy with different Gaussian wave packets. Interval 1 represents $(-z_b,z_b)=(-50,50)$ (the whole interval of evolution) and interval 2 represents $(-z_b,z_b)=(-6,6)$ (inside the brane). The parameters of the double kink scalar and the Gaussian wave packet  are set to $b=8$, $v=6$, and $\sigma=0.5$. }
	\label{Fienergyevolution}
\end{figure*}

Figure \ref{Fienergyevolution} illustrates the energy evolution of an initial wave packet both inside and outside the potential well. Interval 1 represents the whole interval of evolution and interval 2 represents the region of the brane or the quasi well.  Figure~\ref{Fienergyevolution1} shows the energy evolution of the wave packet initially located in the interval 2. The initial energy of the wave packet is localized inside the interval 2.  However, as a part of the wave propagates out of the interval 2, the energy of the signal on the brane will decrease. Figure~\ref{Fienergyevolution2} illustrates the energy evolution of the wave packet initially located outside the interval 2 but in the interval 1. After some time, a part of the wave passes through the barrier and enters the potential well, thus the energy of the wave packet on the brane will increase. Then as the wave propagates out the interval 2, the energy inside the interval 2 will decrease. It can be seen that, both the energy in the interval 1 and interval 2 decrease in a step-like manner. This is because echo pulses are generated one after another, and whenever an echo pulse propagates out of the interval 1 or 2, the energy of the wave rapidly decreases. In later stages, the energies of the two intervals will be almost the same, as only the waves quasi-localized on the brane slowly leak out.

\begin{figure*}[htbp]
	\centering
	\includegraphics[width=6cm]{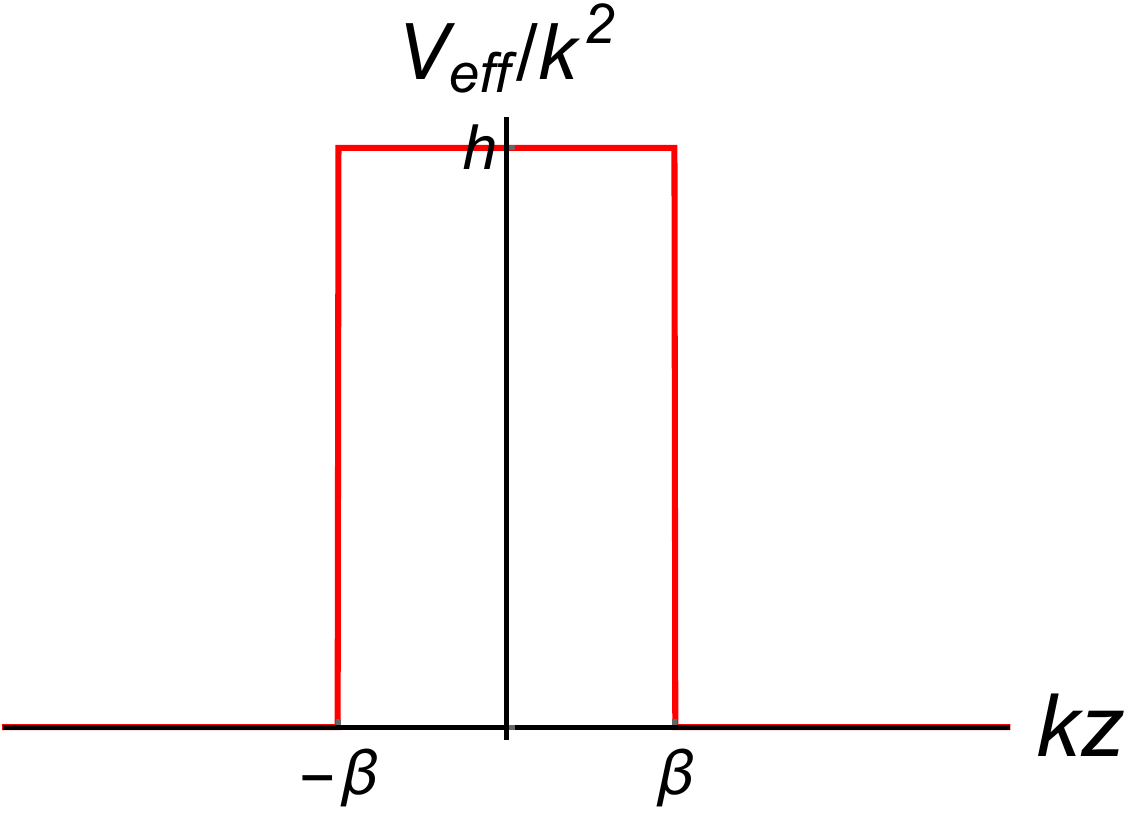}
	\includegraphics[width=6cm]{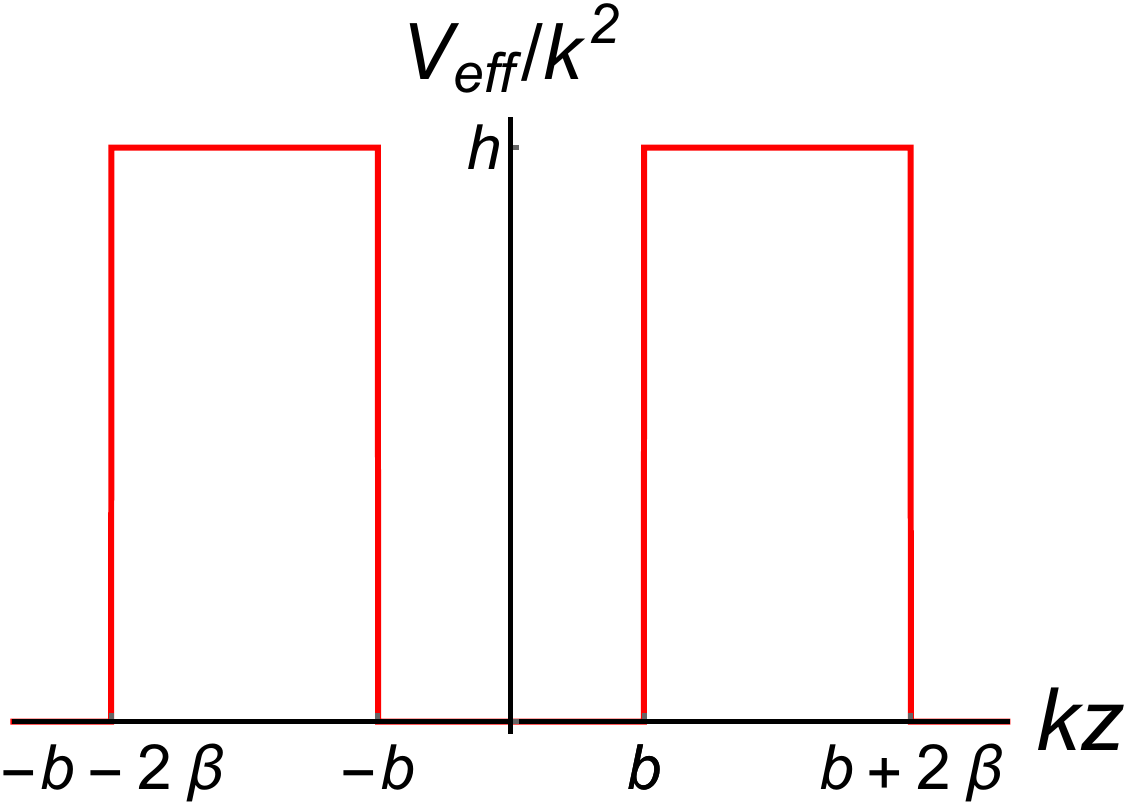}
	\vskip -4mm \caption{Plots of the single and the double barrier potentials.}
	\label{Fig1squarebarrier}
\end{figure*}

To explain the problem of the number of peaks per pulse in subsequent pulses, we use the finite high square barrier as a toy model. We also choose the Gaussian wave packet $H(0,z)=\e^{-\frac{(z-z_0)^2}{\sigma}}$ as the initial data with the parameters $z_0=-40$ and $\sigma=1$. The square barrier function is
\begin{equation}
	V_{\text{eff}}(z)=\left\{
	\begin{aligned}
		h&,& \quad  -\beta \leqslant z  \leqslant \beta,\\
		0&,&\quad   \text{otherwise},
	\end{aligned}	\label{singlebarrier}	
	\right.
\end{equation}
where $2\beta$ and $h$ are the width and the height of the barrier, respectively. The shape of the
square barrier is shown in the left panel of Fig.~\ref{Fig1squarebarrier}.

\begin{figure*}[htbp]
	\centering
	\subfigure[~$\beta=10,h=3$]{\includegraphics[width=6cm]{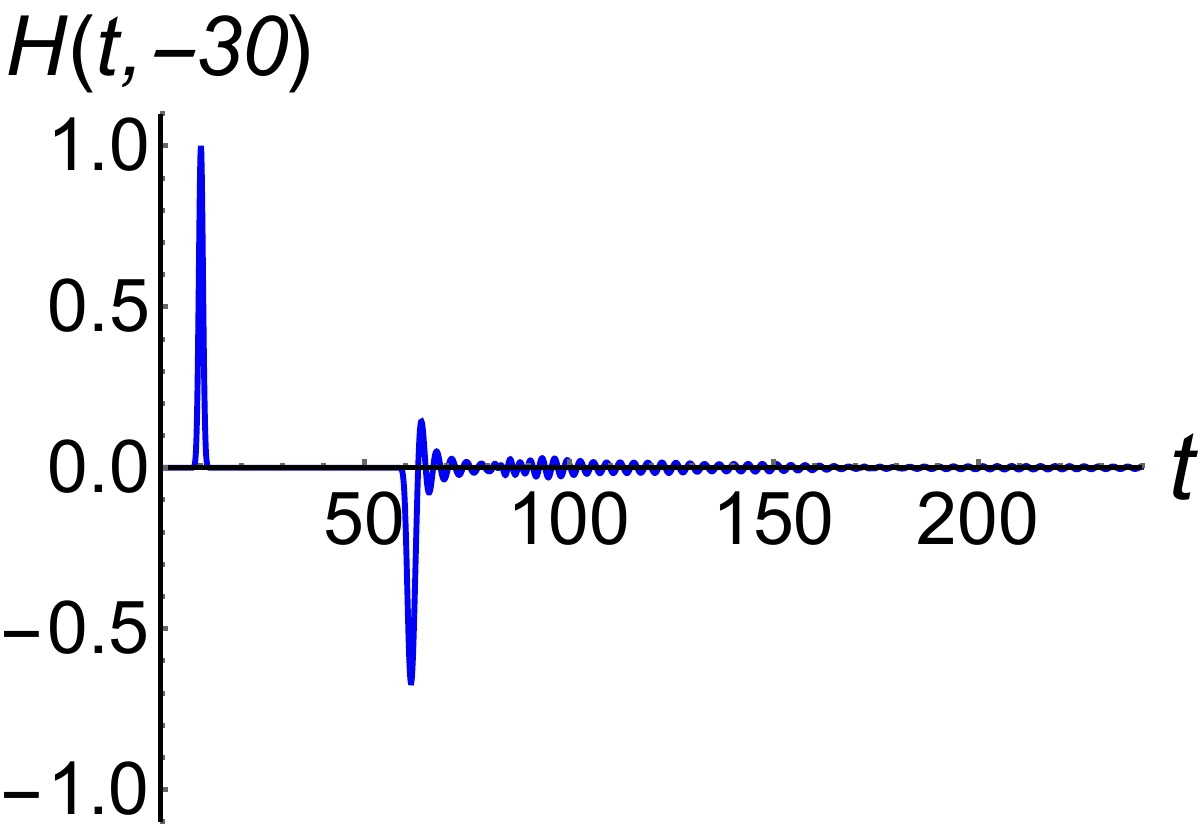}}
	\subfigure[~$\beta=2,h=3$]{\includegraphics[width=6cm]{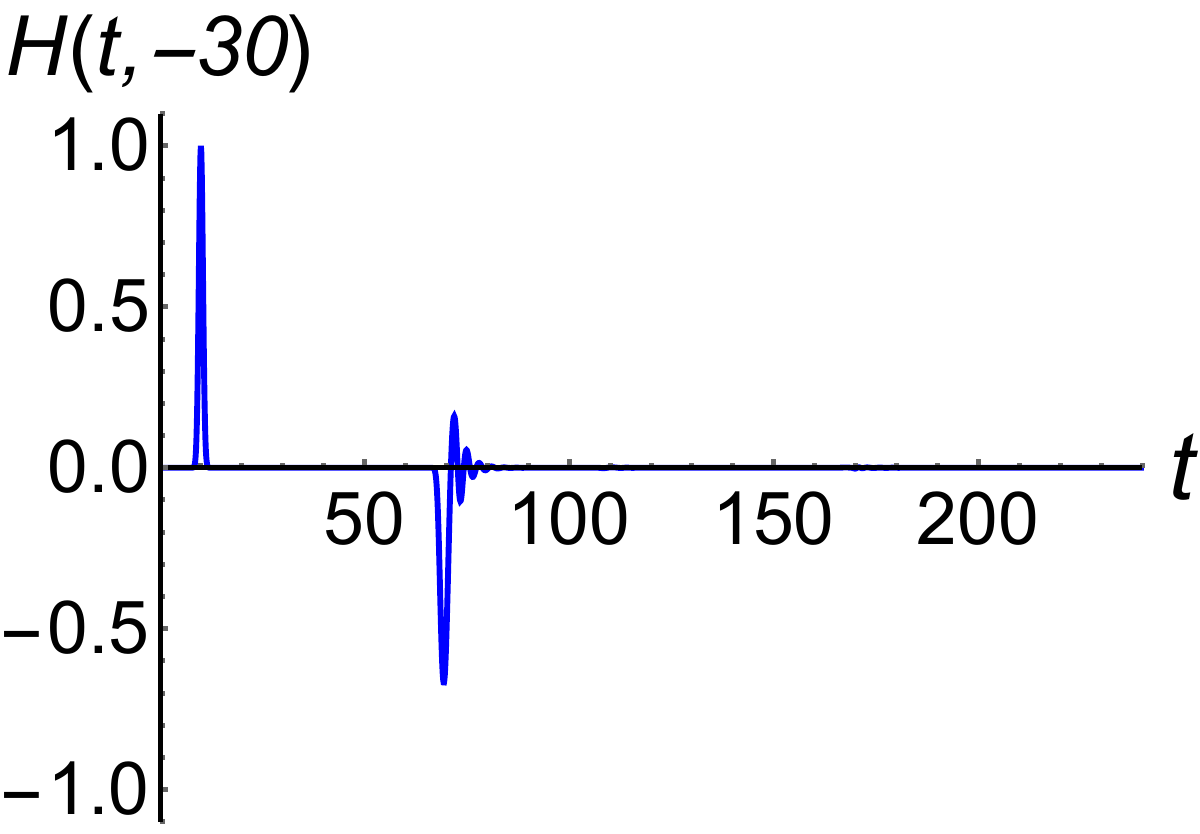}}
	\subfigure[~$\beta=10,h=0.5$]{\includegraphics[width=6cm]{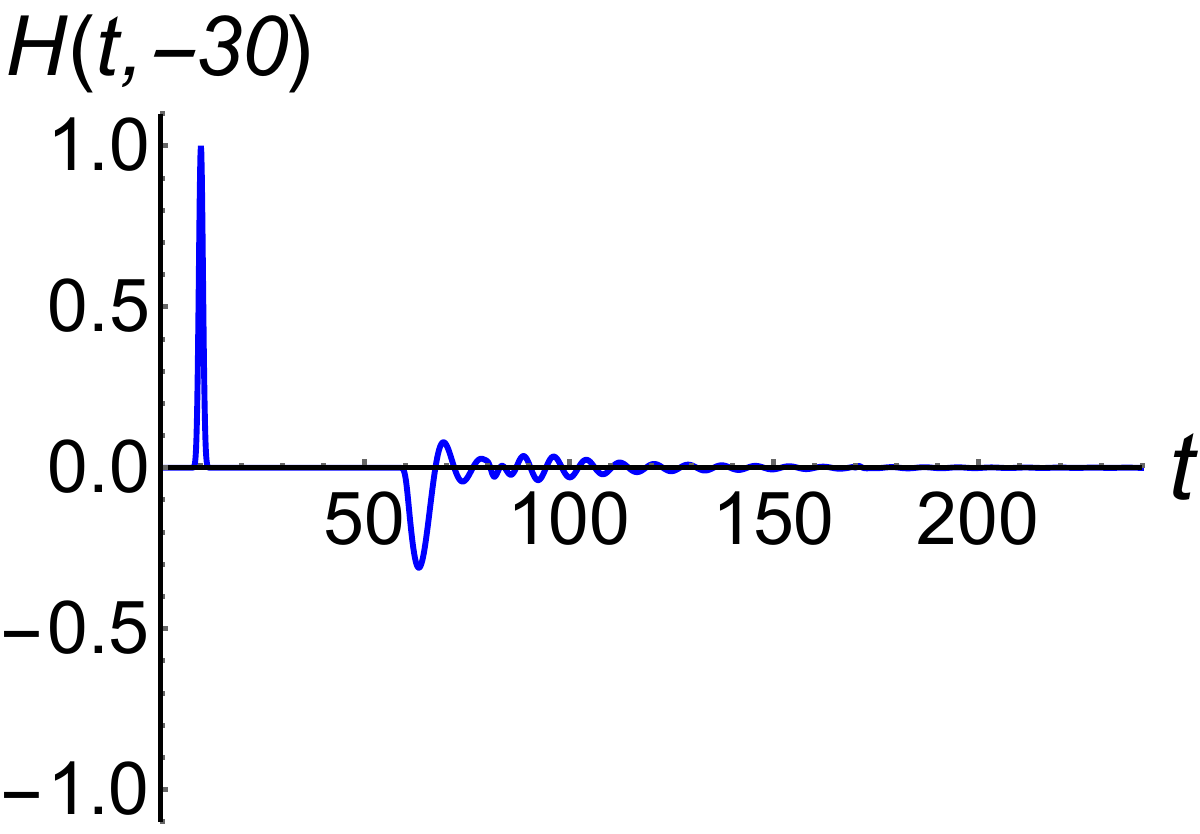}}
	\subfigure[~$\beta=10,h=10$]{\includegraphics[width=6cm]{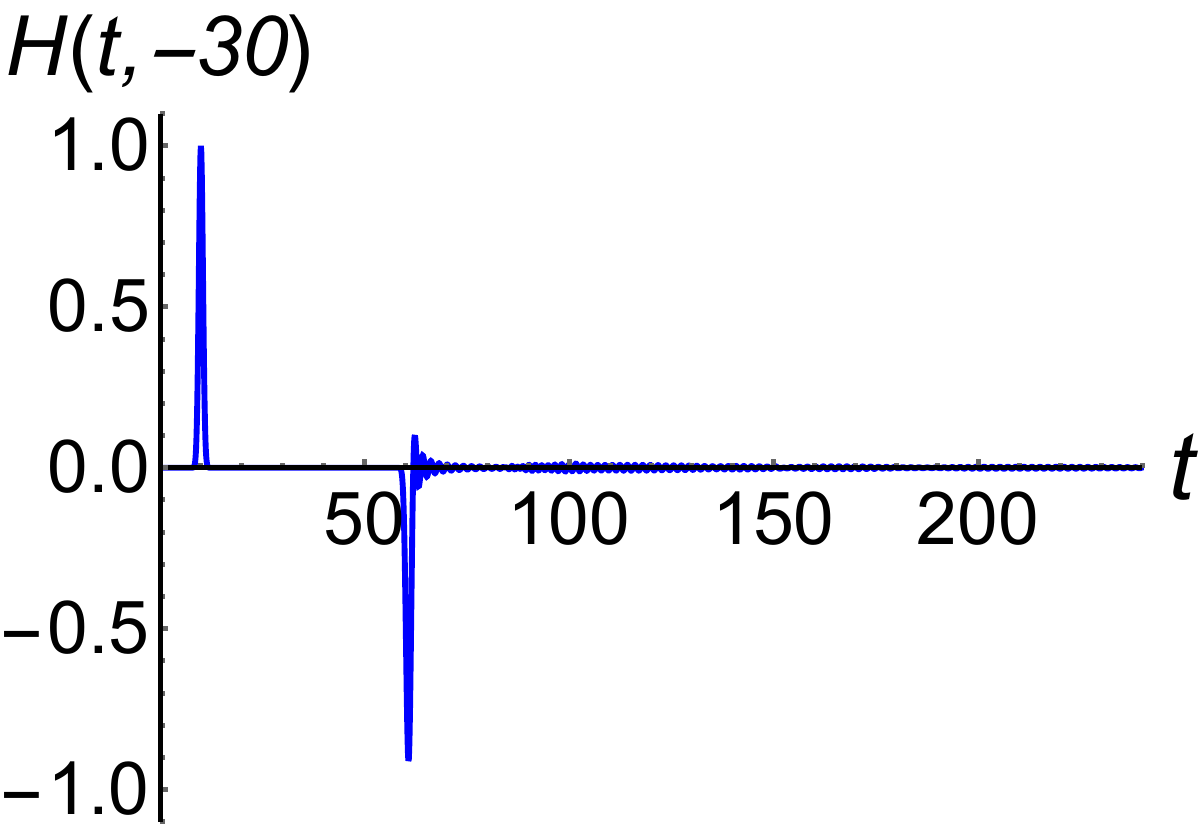}}
	
	\vskip -4mm \caption{Plots of the evolution $H(t, z)$ of a Gaussian wave packet under the single  barrier. The parameters of the Gaussian wave packet  are set to $\sigma=0.5$ and $z_0=-40$.}
	\label{Fig1squarebarrierevolution}
\end{figure*}

Figure~\ref{Fig1squarebarrierevolution} illustrates the impact of the width and height of the square barrier on the evolution of wave packets. It can be seen that the number of peaks of the subsequent pulse is actually infinite. However, the greater the barrier width, the more pronounced the peaks of the subsequent pulses. The higher the height of the barrier, the larger the height of the first reflected peak, and conversely the weaker the subsequent peaks. We also simulate the wave packet evolution under the double barriers. The barrier function is

\begin{equation}
	V_{\text{eff}}=\left\{
	\begin{aligned}
		h&,&\quad b < \rvert z \rvert <b+2\beta,\\
		0&,&\quad  \text{otherwise},
	\end{aligned}	\label{doublebarrier}	
	\right.
\end{equation}
which can be seen in the right panel of Fig.~\ref{Fig1squarebarrier}.
\begin{figure*}[htbp]
	\centering
	\subfigure[~$\beta=10,h=3$]{\includegraphics[width=6cm]{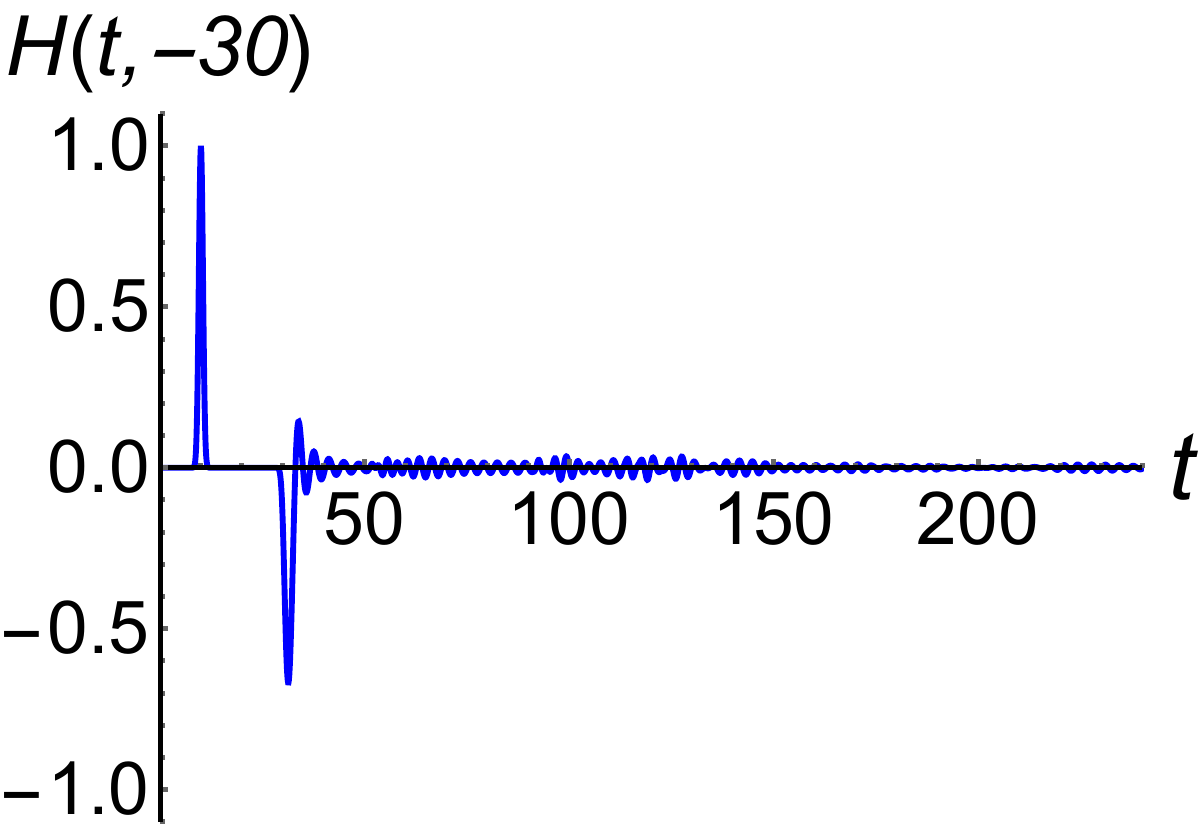}}
	\subfigure[~$\beta=2,h=3$]{\includegraphics[width=6cm]{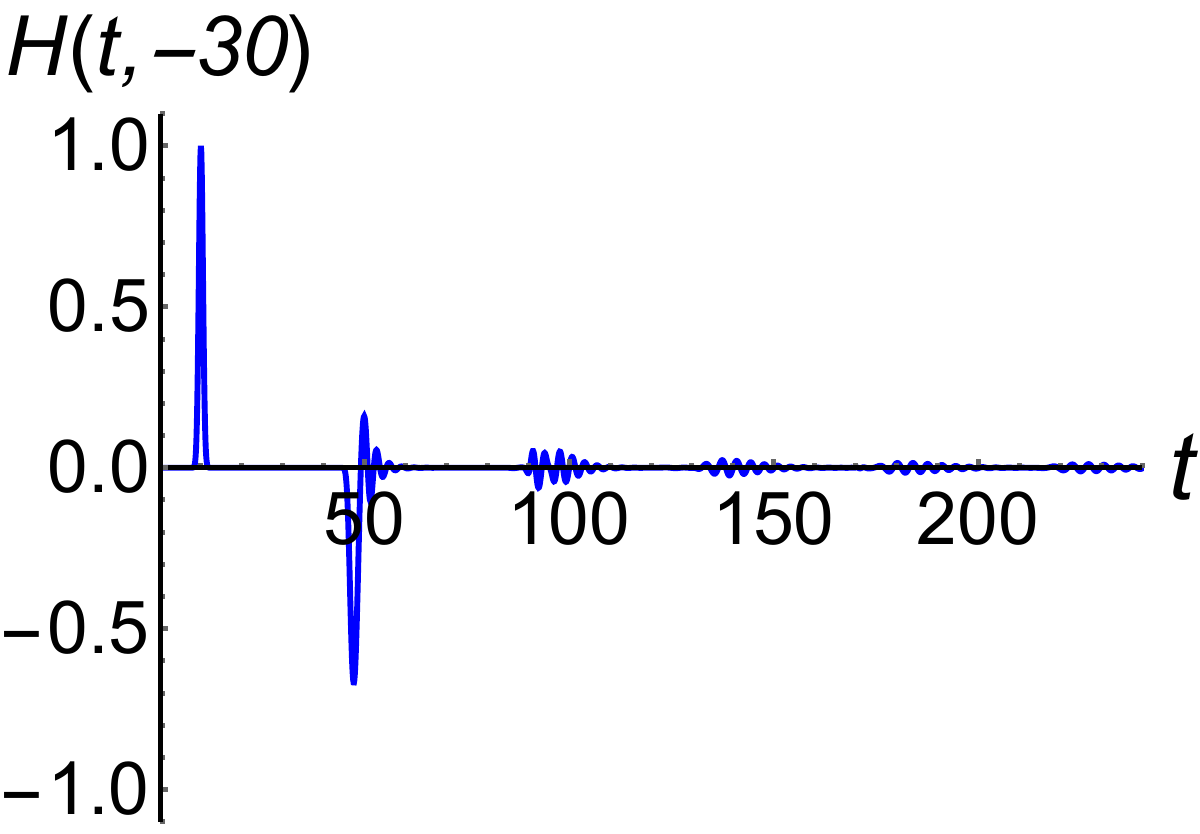}}
	\vskip -4mm \caption{Plots of the evolution $H(t, z)$ of a Gaussian wave packet under the double barriers. The parameters of the Gaussian wave packet  are set to $\sigma=0.5$ and $z_0=-40$.}
	\label{Fig2squarebarrierevolution}
\end{figure*}	

As can be seen in Fig.~\ref{Fig2squarebarrierevolution}, the signals have comparatively obvious echoes in the case of double barriers. The smaller the width of the barriers, the shorter the pulse signal, and the more pronounced the echoes. 

\subsection{Other scalar potentials}
In the previous part, we analyzed the impact of the double kink scalar field's properties on gravitational echoes. The properties of a kink is closely related to the scalar potential. Therefore in this subsection, we will investigate whether gravitational echoes exist for different types of scalar potentials.

First, we consider the following typical scalar potential with two minima (vacua):
\begin{eqnarray}
V(\phi) &=& -\frac{2 k^2 }{27 v^2}\phi ^6 + k^2 \left(\frac{4}{9}+\frac{1}{2v^2}\right)\phi^4-k^2 \left(\frac{2}{3}v^2+1 \right)\phi ^2+\frac{k^2 v^2}{2},
\end{eqnarray}
for which we have
\begin{eqnarray}
\frac{dV(\phi)}{d\phi}&=&-\frac{4 k^2 }{9 v^2}\phi  (\phi -v) (\phi+v ) \left(\phi+\sqrt{\frac{3}{2}} \sqrt{2 v^2+3} \right) \left(\phi -\sqrt{\frac{3}{2}} \sqrt{2 v^2+3}\right).
\label{singlekinkscalarpotential}
\end{eqnarray}
It can be seen that the scalar potential have two minima and three maxima, and the two minima $V=-8 k^2 v^4/27$ are located at $\phi= \pm v$. Therefore, the vacuum expectation value $v$ increases with the absolute values of the minima. For the above scalar potential, we can get an analytic single kink solution for the scalar field: 
\begin{eqnarray}
\phi(y)=v \tanh (k y).
\end{eqnarray}

\begin{figure*}[htbp]
	\centering
	\subfigure[the scalar field]{\includegraphics[width=6cm]{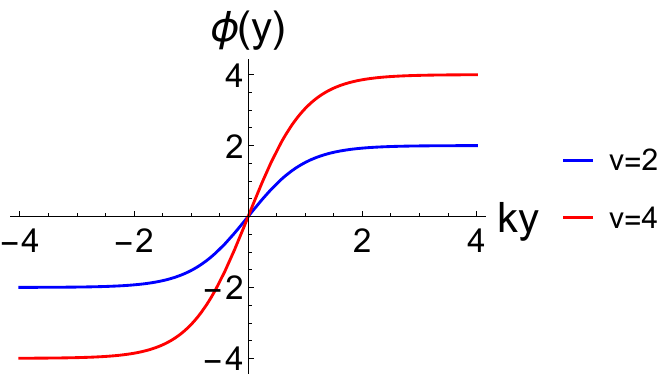}}
	\subfigure[the scalar potential]{\includegraphics[width=6cm]{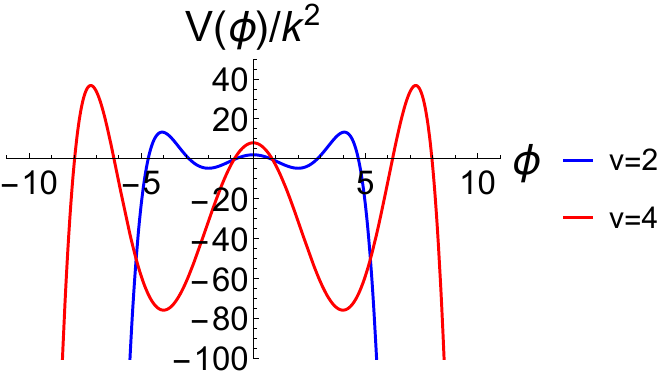}\label{13b}}
	\subfigure[the effective potential]{\includegraphics[width=6cm]{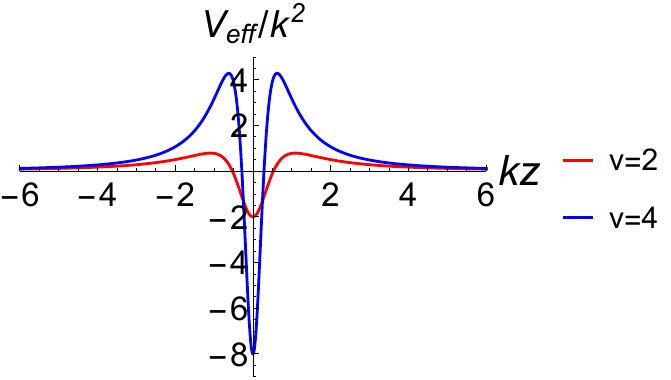}}
	\subfigure[the supersymmetric potential]{\includegraphics[width=6cm]{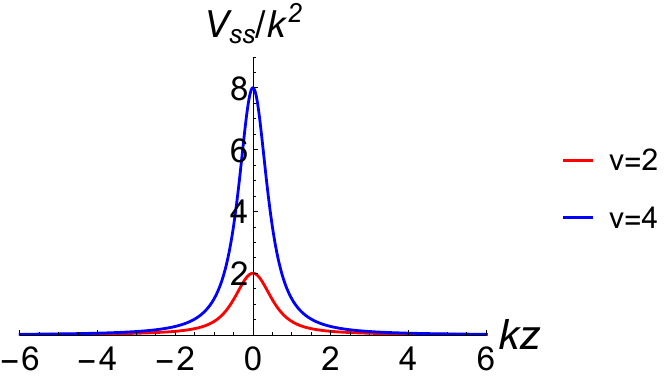}\label{13d}}
	\subfigure[the evolution of a Gaussian wave packet ($v=2$)]{\includegraphics[width=5cm]{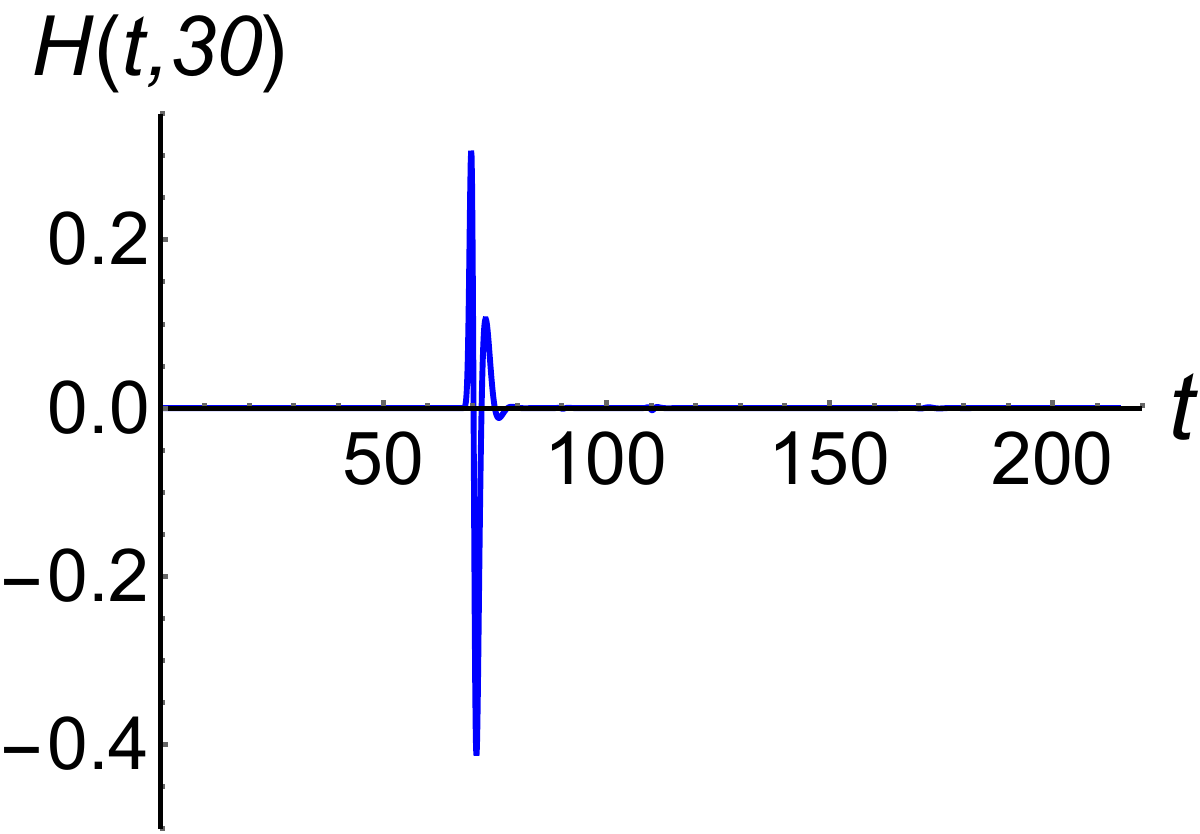}\label{13e}}
	\subfigure[the evolution of a Gaussian wave packet ($v=4$)]{\includegraphics[width=5cm]{1kinkv2GWE.pdf}\label{13f}}
	\vskip -4mm \caption{Plots of the single kink scalar $\phi(y)$, the scalar potential $V(\phi)$, the effective potential $V_{\text{eff}}(z)$ of gravitational perturbation, the supersymmetric potential $V_{\text{ss}}(z)$, and the evolution $H(t, 30)$ of a Gaussian wave packet under corresponding spacetime with different vacuum expectation value $v$.}
	\label{Fig1kinkscalarsystem}
\end{figure*}

The single kink scalar $\phi(y)$, the scalar potential $V(\phi)$, the effective potential $V_{\text{eff}}(z)$ of the gravitational perturbation, the supersymmetric potential $V_{\text{ss}}(z)$, and the evolution $H(t, 30)$ of a Gaussian wave packet are shown in Fig.~\ref{Fig1kinkscalarsystem}. The vacuum expectation value $v$ increases with the absolute values of the two minima in the scalar potential which can be seen in Eq.~\eqref{singlekinkscalarpotential} and Fig.~\ref{13b}. For the single kink solution, as shown in Fig.~\ref{13d}, the supersymmetric potential of the gravitational perturbations does not exhibit multiple barriers. Therefore, gravitational echoes are not observed during the evolution of a Gaussian wave packet, which can be seen in Figs.~\ref{13e} and \ref{13f}.

\begin{figure*}[htbp]
	\centering
	\subfigure[the scalar field ($b$=2)]{\includegraphics[width=6cm]{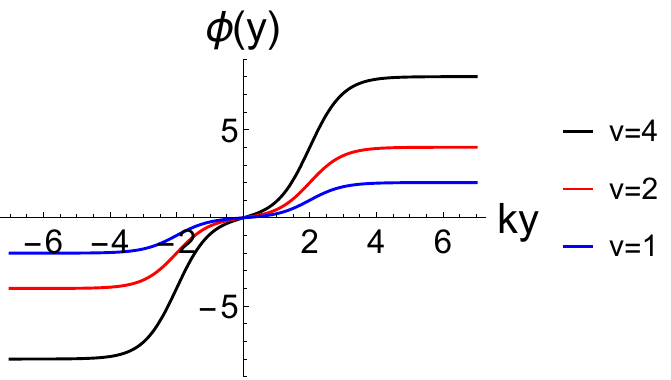}}
	\subfigure[the scalar field ($v$=1)]{\includegraphics[width=6cm]{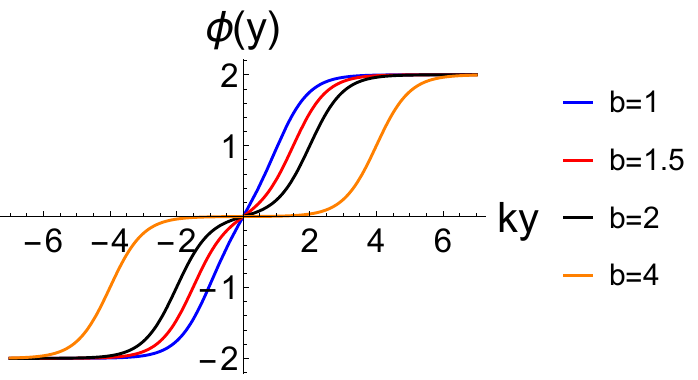}}
	\subfigure[the scalar potential ($b$=2)]{\includegraphics[width=6cm]{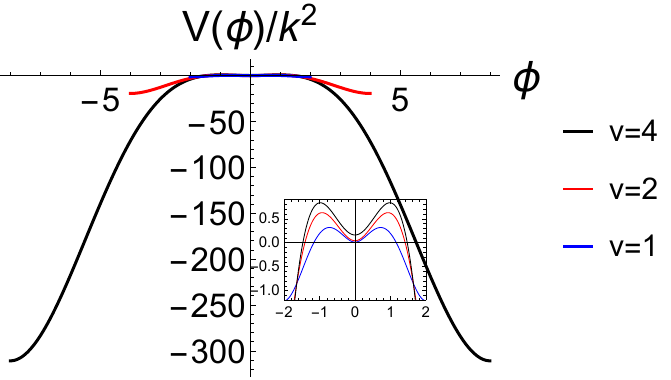}}
	\subfigure[the scalar potential ($v$=1)]{\includegraphics[width=6cm]{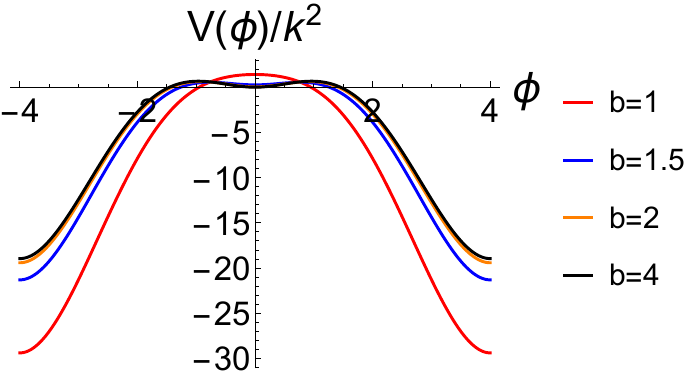}}
	\subfigure[the effective potential ($b$=2)]{\includegraphics[width=6cm]{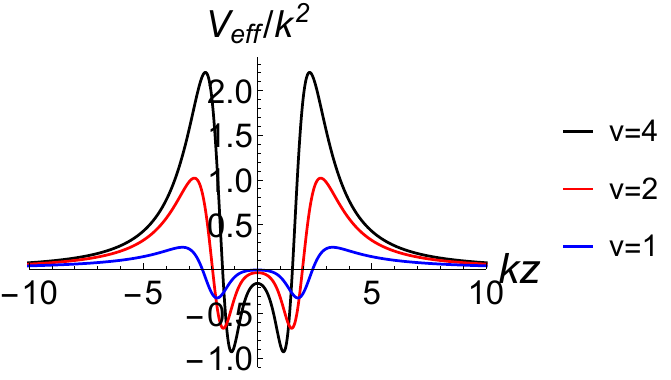}}
	\subfigure[the effective potential ($v$=1)]{\includegraphics[width=6cm]{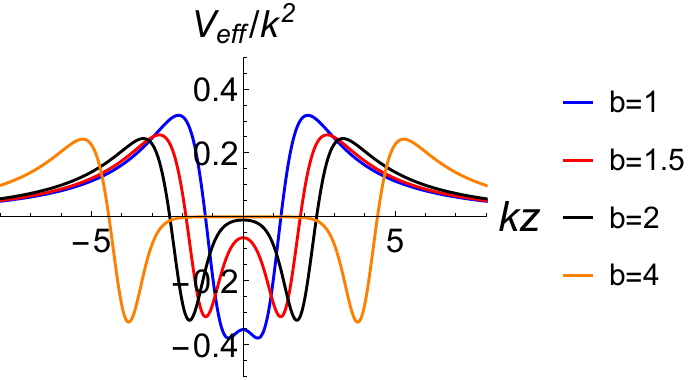}}
	\vskip -4mm \caption{Plots of the  double kinks scalar $\phi(y)$, the scalar potential $V(\phi)$, the effective potentials $V_{\text{eff}}(z)$ of gravitational perturbation with different $v$ and $b$.}
	\label{Fig2kinkscalarsystem}
\end{figure*}	

Second, we consider the scalar potential with there minima, one at $\phi=0$ and the others at both sides. It is worth noting that  such a potential can support a double kink solution if the minimum in the middle is greater than the ones on both sides. Here, we give some numerical solutions.  As shown in Fig.~\ref{Fig2kinkscalarsystem}, the greater the distance between the two vacua in the scalar potential, the greater the vacuum expectation value of the scalar field, and the higher the barrier corresponding to the effective potential of gravitational fluctuation, the greater the reflectivity of the high frequency, and the greater the proportion of the high frequency when the gravitational echo becomes obvious. The closer the value of the scalar potential at $\phi=0$ is to zero, the larger the distance between the two kinks, the larger the distance between the two barriers of the effective potential,  the larger the reflectivity of the low frequency wave, and the larger the proportion of low frequency in the echoes.

Third, we consider the following typical scalar potential with a negative minimum value (vacuum) at zero point:
\begin{eqnarray}
	V(\phi)& =& \frac{2 k^2}{27 v^2} \phi ^6-k^2\left(\frac{2}{9}+\frac{1}{2v^2}\right) \phi ^4-k^2\left(\frac{2}{9}+\frac{1}{2v^2}\right)\phi ^2-\frac{2 k^2 v^4}{27},
\end{eqnarray}
for which we can get an analytic soliton solution for the scalar field: 
\begin{eqnarray}
	\phi (y)&=&v\, \text{sech}(k y).
\end{eqnarray}

\begin{figure*}[htbp]
	\centering
	\subfigure[the scalar field]{\includegraphics[width=6cm]{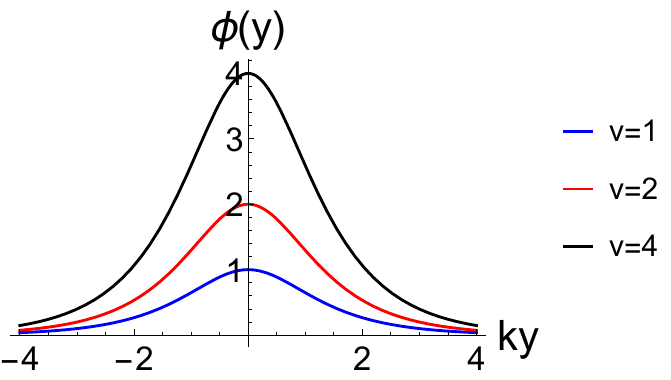}}
	\subfigure[the scalar potential]{\includegraphics[width=6cm]{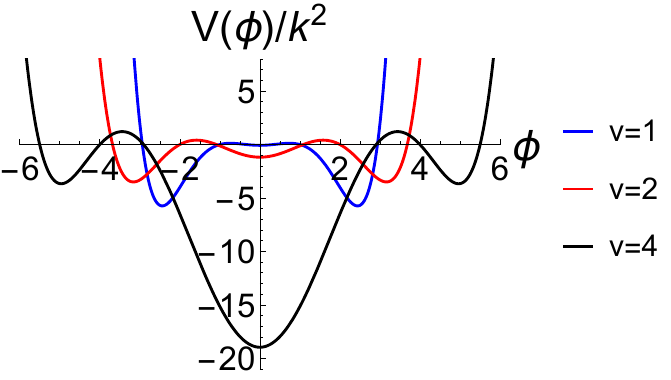}\label{15b}}
	\subfigure[the effective potential]{\includegraphics[width=6cm]{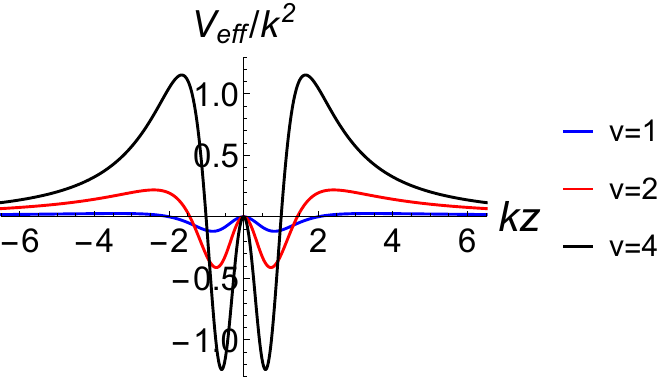}}
	\subfigure[the supersymmetric potential]{\includegraphics[width=6cm]{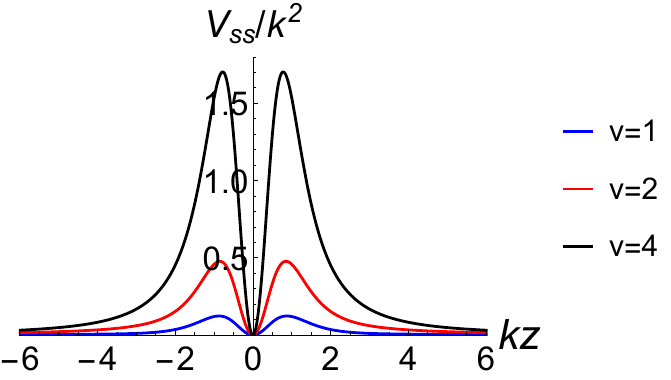}}
	\subfigure[the evolution of a Gaussian wave packet ($v=1$)]{\includegraphics[width=5cm]{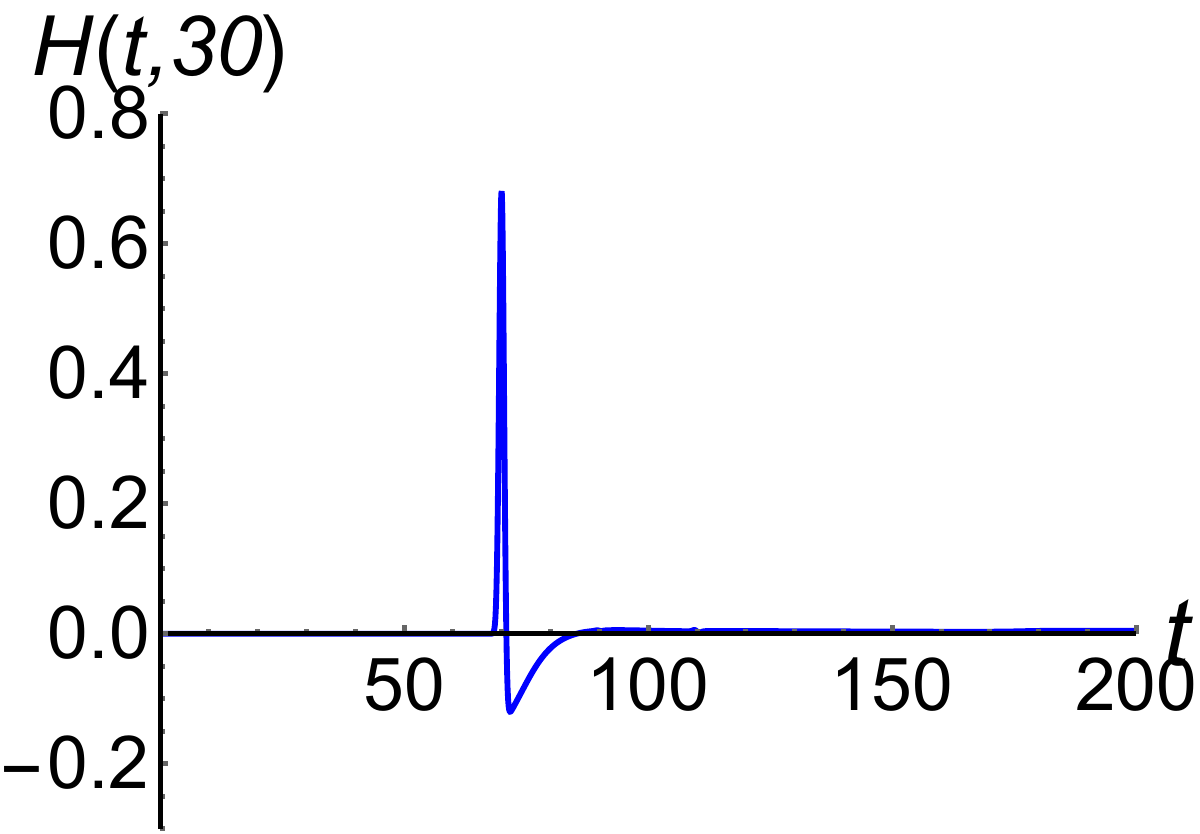}}
	\subfigure[the evolution of a Gaussian wave packet ($v=2$)]{\includegraphics[width=5cm]{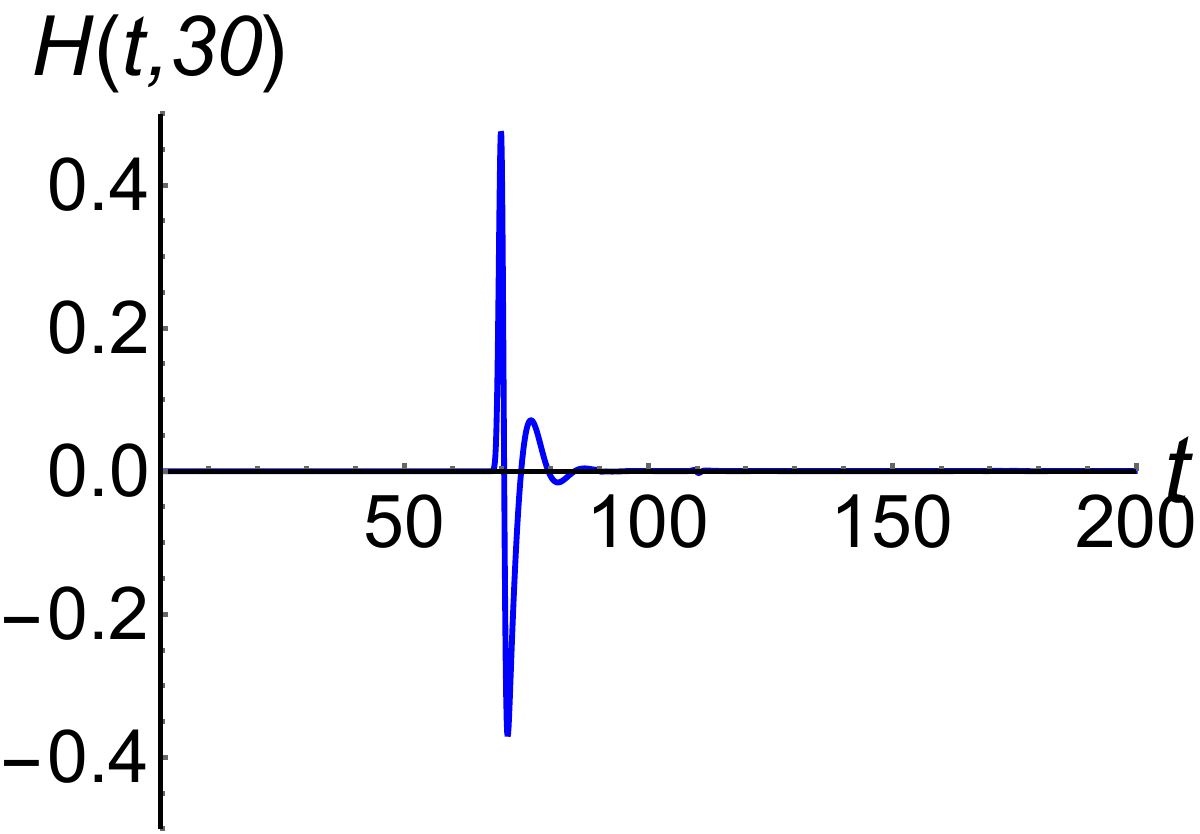}}
	\subfigure[the evolution of a Gaussian wave packet ($v=4$)]{\includegraphics[width=5cm]{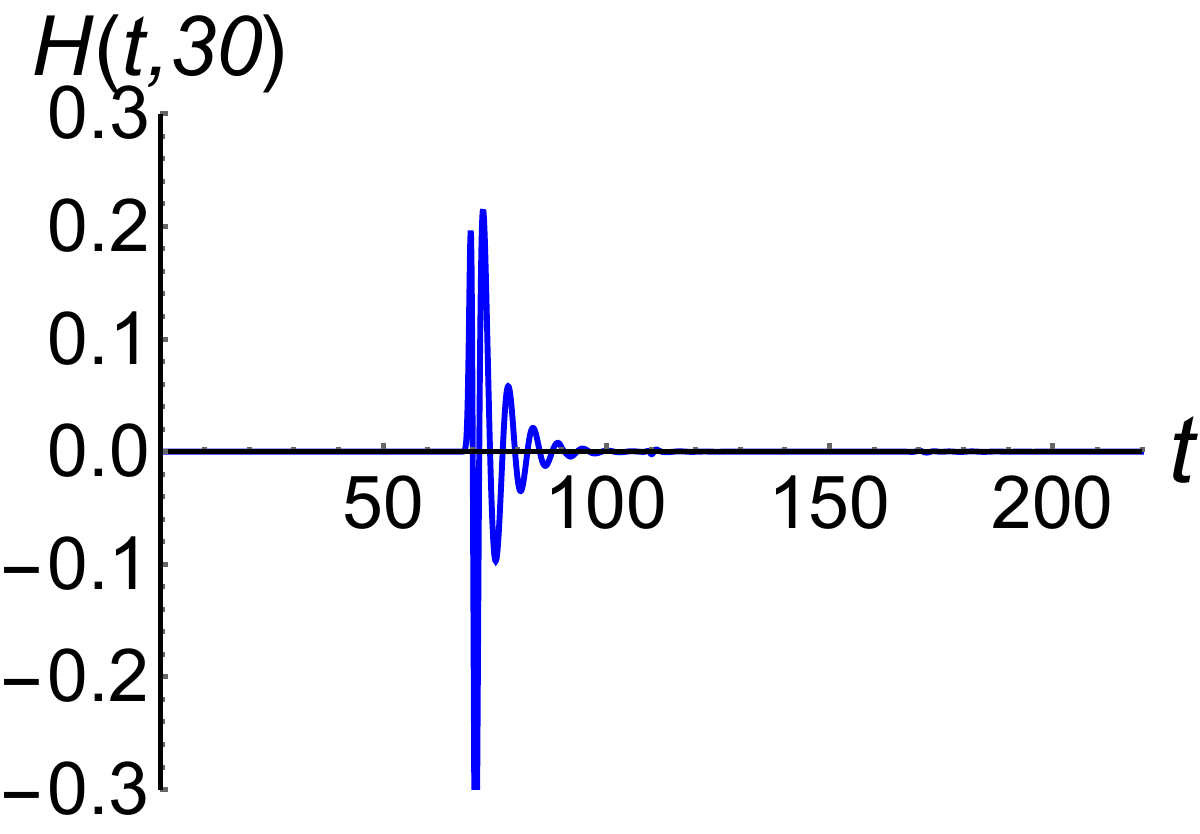}}
	\vskip -4mm \caption{Plots of the soliton kink scalar $\phi(y)$, the scalar potential $V(\phi)$, the effective potential $V_{\text{eff}}(z)$ of gravitational perturbation, the supersymmetric potential $V_{\text{ss}}(z)$, and the evolution $H(t, 30)$ of a Gaussian wave packet under corresponding spacetime with different $v$}.
	\label{Figsolitonscalarsystem}
\end{figure*}

The soliton scalar field $\phi(y)$, the scalar potential $V(\phi)$, the effective potential $V_{\text{eff}}(z)$ of gravitational perturbation, the supersymmetric potential $V_{\text{ss}}(z)$, and the evolution $H(t, 30)$ of a Gaussian wave packet are  shown in Fig.~\ref{Figsolitonscalarsystem}. The soliton solution not only requires a negative minimum values at $\phi=0$, but also requires positive maximum values on both sides which can be seen in Fig.~\ref{15b}. The peak value $v$ of a soliton increases with the absolute value of the scalar potential at $\phi=0$, $|V(0)|=2k^2 v^4/27$. Although both the effective potential and its supersymmetric potential of gravitational fluctuation exhibit two barriers, the distance between these barriers is too small. Consequently, gravitational echoes are not observed.

\begin{figure*}[htbp]
	\centering
	\subfigure[the scalar field ($b=4$)]{\includegraphics[width=6cm]{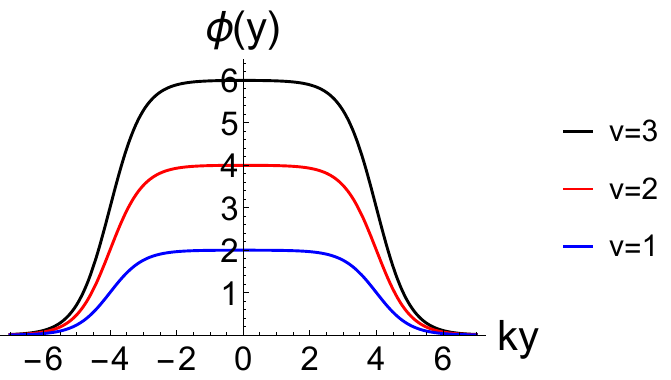}}
	\subfigure[the scalar field ($v=2$)]{\includegraphics[width=6cm]{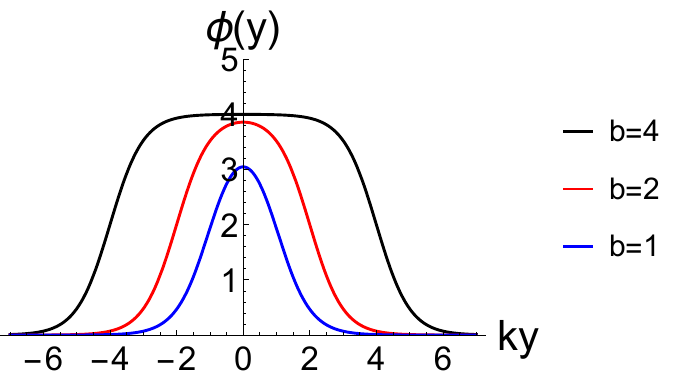}}
	\subfigure[the scalar potential ($b=4$)]{\includegraphics[width=6cm]{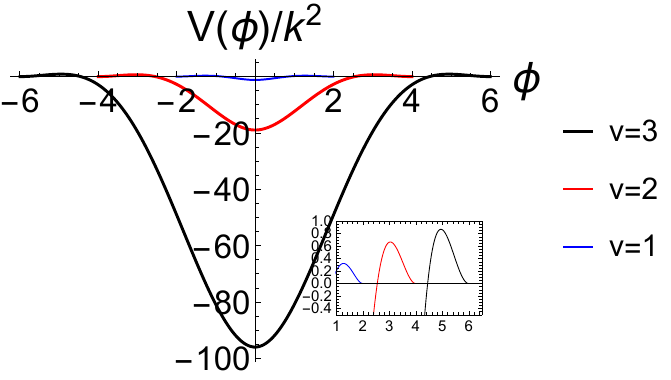}}
	\subfigure[the scalar potential ($v=2$)]{\includegraphics[width=6cm]{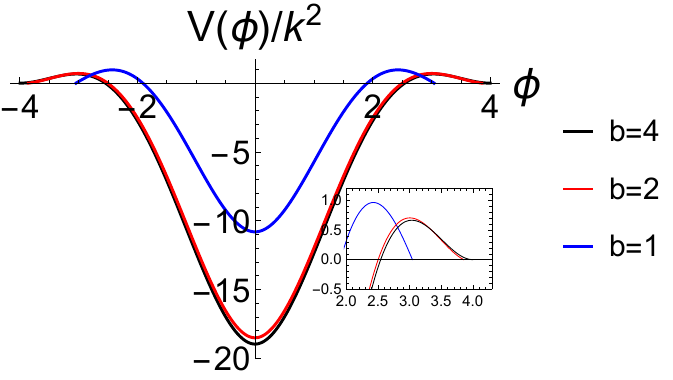}}
	\subfigure[the effective potential ($b=4$)]{\includegraphics[width=6cm]{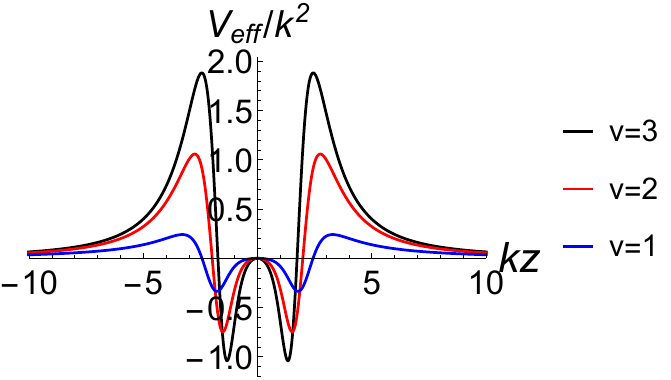}}
	\subfigure[the effective potential ($v=2$)]{\includegraphics[width=6cm]{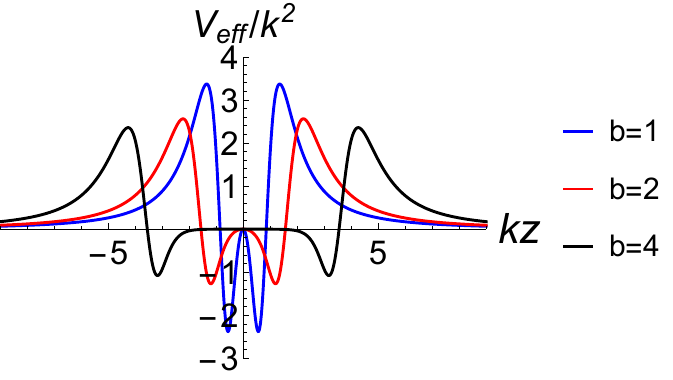}}
	\vskip -4mm \caption{Plots of the  platform-type soliton scalar $\phi(y)$, the scalar potential $V(\phi)$, the effective potential $V_{\text{eff}}(z)$ of gravitational perturbation with different values of $v$ and $b$.}
	\label{Fig2solitonscalarsystem}
\end{figure*}

At last, we consider a platform-type soliton solution,
\begin{eqnarray}	
	\phi(y)=v \left(\tanh(k y+b)-\tanh(k y-b)\right),
\end{eqnarray}	
for which we can only give numerical solution of the scalar potential. The shape of this scalar field $\phi(y)$, the scalar potential $V(\phi)$, the effective potential $V_{\text{eff}}(z)$ of gravitational perturbation are shown in Fig.~\ref{Fig2solitonscalarsystem}. The effective potentials of gravitational perturbation in this case are similar to that of the double kink solution. The effect of parameters $b$ and $v$ on the effective potential is also similar to that of the double kink solution. So we can observe gravitational echoes in this case. For fixed $b$ and varying $v$, the smaller the value of scalar potential at $\phi=0$, the lager the peak value of the scalar field, the higher the proportion of high-frequency components in the gravitational echoes. Conversely, for fixed $v$ and varying $b$, the smaller the absolute value of the slope of the scalar potential at $\phi=2v\tanh(b)$ on both sides, the larger the width of the scalar field platform, the higher the proportion of low-frequency components in the gravitational echoes.

 Now, we give a brief  summary and discussion. The effective potential of gravitational perturbation for the double kink solution or the platform-type soliton solution  only contains two barriers, resulting that there are equal time intervals between echoes, and each echoes will uniformly decay. In contrast, the effective potential of a multiple kink solution has multiple barriers. Therefore, the time intervals between echoes are generally different, and different barrier heights lead to non-uniform decay of echoes. For a $\phi^6$ model, we can have a double-kink solution at most.  A $\phi^8$ model can support a three-kink solution. In general, an $n$-kink scalar field requires a $\phi^{2n+2}$ model. Gravitational echoes may be observed in these multi-kink models. If the extra-dimensional gravitational echoes are detected, their properties can be used to infer the characteristics of the thick brane.

\subsection{Two-dimensional evolution}
In the previous part, we only focus on the extra-dimensional component of the gravitational waves. However, the matter of our universe is localized on the brane, and we can not observe the signal out of the brane. Gravitational waves propagating along the extra dimensions are  KK modes, and not all  KK modes propagate at the speed of light. Consequently, we simulate the evolution of a Gaussian wave packet propagating in the bulk and explore the phenomenon of the gravitational echoes when the gravitational wave propagates on the brane. The coordinate on the brane is chosen to be three-dimensional spherically symmetric coordinates $(r,\theta,\varphi)$. By separating the angle part from the other parts $\tilde{h}_{\mu\nu}=\sum_{lm}\frac{1}{r}R_{lm}(t,r,z)(Y_{lm})_{\mu\nu}(\theta,\varphi)$, we can get the evolution equation of the extra and radial components from Eq.~\eqref{perturbation eq},
\begin{eqnarray}
	\partial_t^2R(t,r,z)-\partial_{r}^2R(t,r,z)-\partial_z^2R(t,r,z)+V_{\text{eff}}(z)R(t,r,z)=-\frac{l(l+1)}{r^2}R(t,r,z).\label{2Dperturbation eq}
\end{eqnarray}
In the following, we consider the spherical wave $l=0$. Similarly, we also consider a Gaussian wave packet $R(0,r,z)=\e^{-\frac{z^2+r^2}{\sigma}}$ as the initial waveform, and take the radiative boundary conditions. In the numerical program of two-dimensional evolution, the radiation boundary condition ($\pd_t R=\pd_{\rho}R$ for $\rho\rightarrow\infty$, where $\rho=\sqrt{r^2+z^2}$ is the four-dimensional spherical coordinate) cannot completely suppress the waves reflected from the boundary due to numerical errors, so some modifications listed as follows are adopted~\cite{Alcubierre:2002kk}:
\begin{eqnarray}
	\frac{z}{\sqrt{r^2+z^2}} \partial_t R\pm\partial_z R\pm\frac{z}{(r^2+z^2)}R \simeq \frac{z~ h^{\prime}(t)}{(r^2+z^2)^{\frac{N+1}{2}}}, ~~ z\rightarrow \pm\infty,\nonumber \\
	\frac{r}{\sqrt{r^2+z^2}} \partial_t R\pm\partial_r R\pm\frac{r}{(r^2+z^2)}R \simeq \frac{r~ h^{\prime}(t)}{(r^2+z^2)^{\frac{N+1}{2}}}, ~~ r\rightarrow \pm\infty,\nonumber			
	\label{2Dboundarycondition}		
\end{eqnarray}
where $h(t)$ is a time function that resolves transient errors and $N$ is an optional parameter. In general, the value of $h(t)$ is calculated at the non-boundary position and subsequently applied at the boundaries. When $N=3$, the error is best eliminated.

\begin{figure*}[htbp]
	\centering
	\subfigure[~$t=3$]{\includegraphics[width=5.5cm]{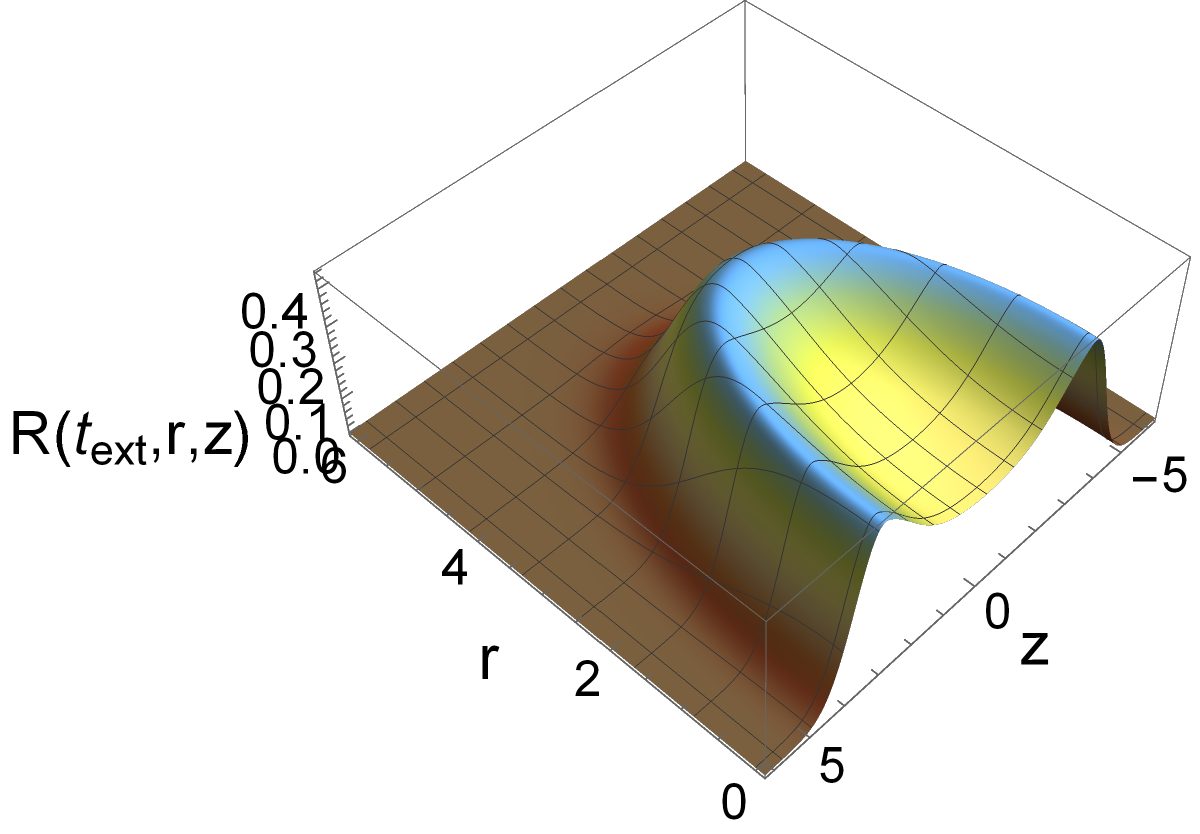}\qquad
		\includegraphics[width=5.5cm]{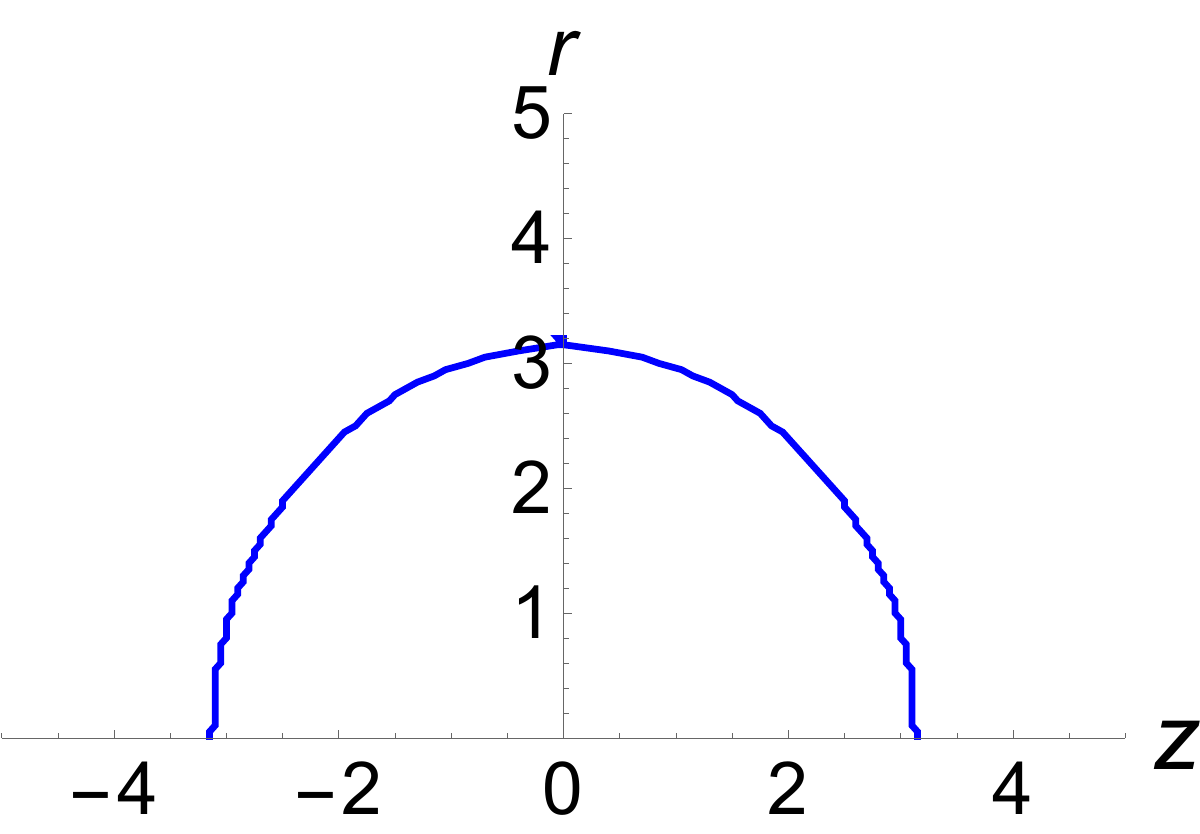}\label{Fig2Devolution1}}
	\subfigure[~$t=5$]{\includegraphics[width=5.5cm]{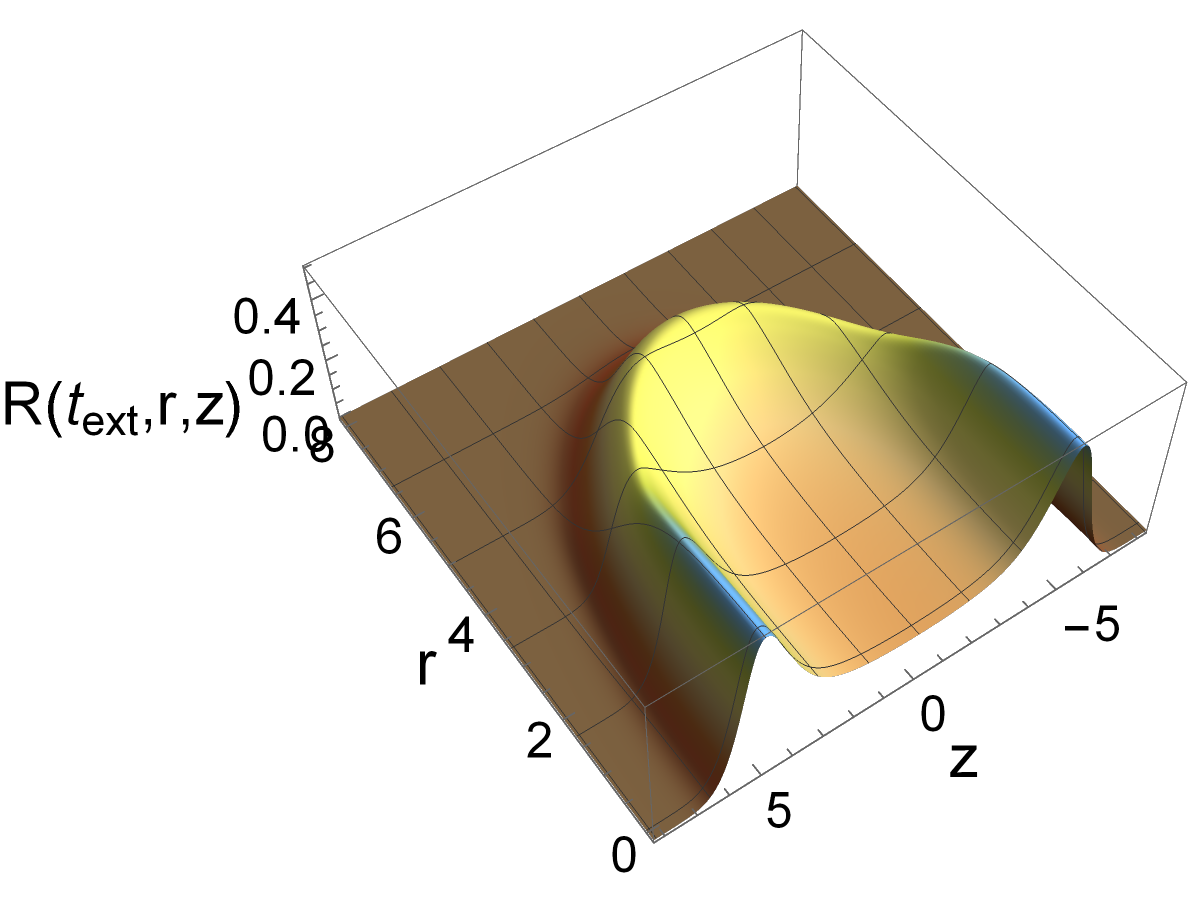}\qquad
		\includegraphics[width=5.5cm]{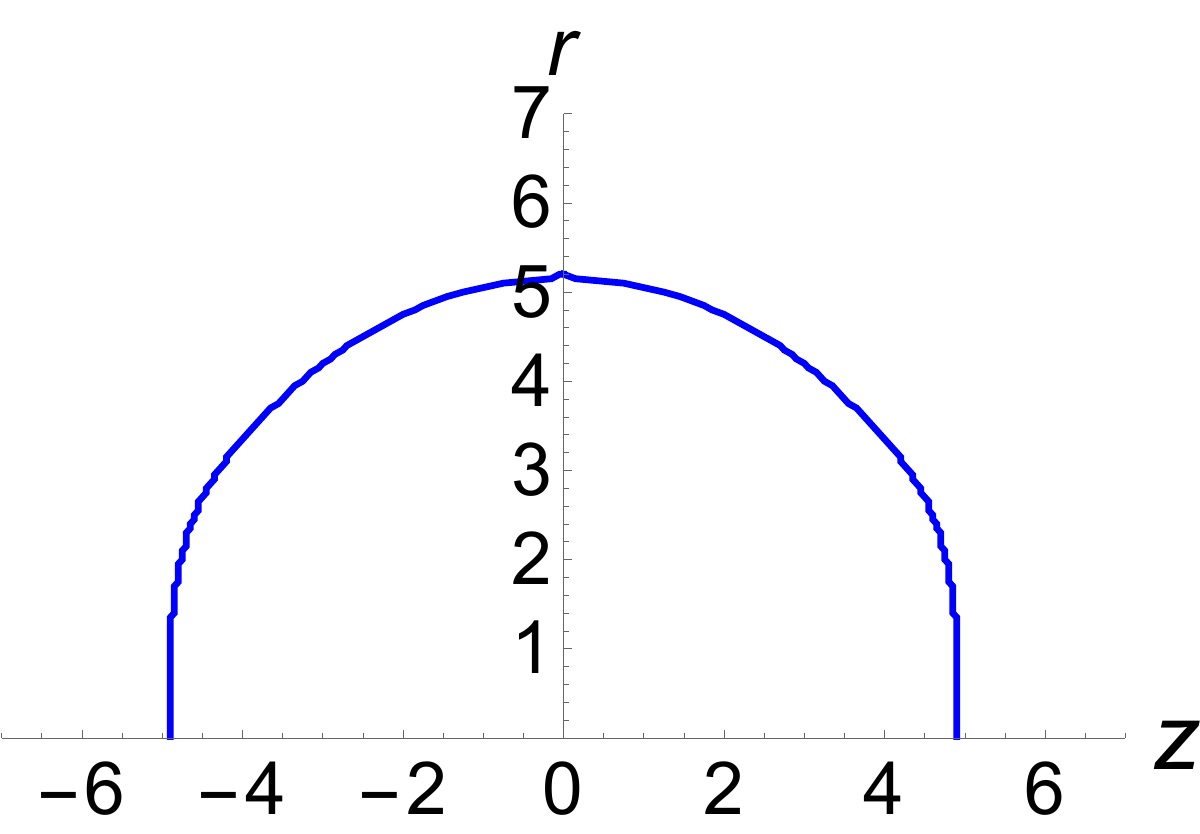}\label{Fig2Devolution2}}
	\subfigure[~$t=10$]{\includegraphics[width=5.5cm]{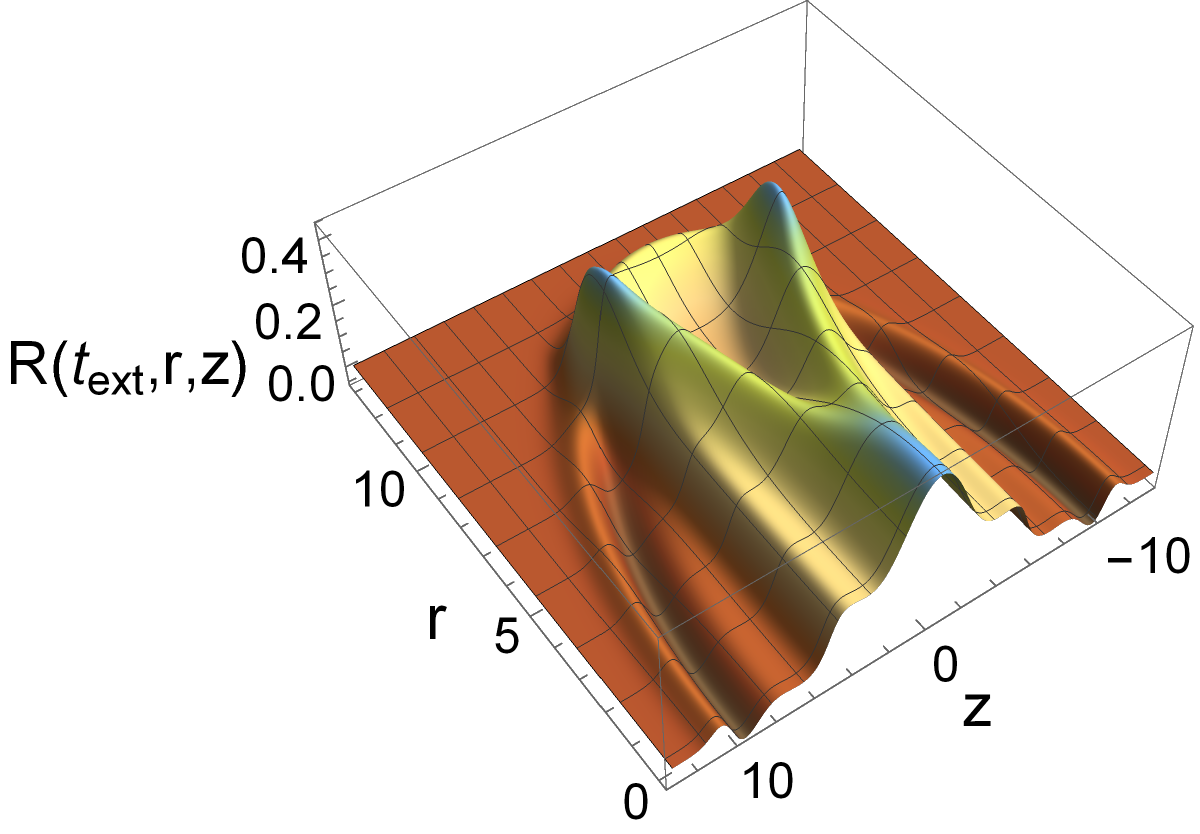}\qquad
		\includegraphics[width=5.5cm]{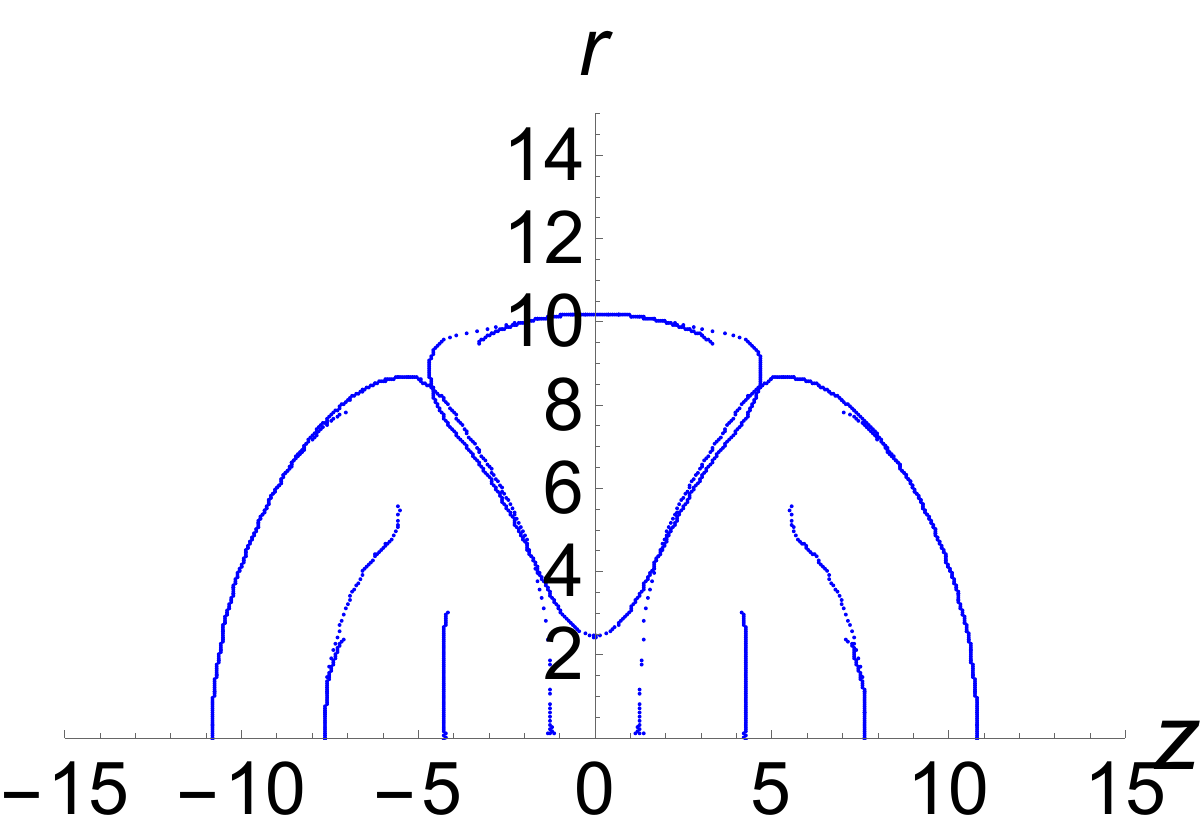}\label{Fig2Devolution3}}
	\subfigure[~$t=20$ ]{\includegraphics[width=5.5cm]{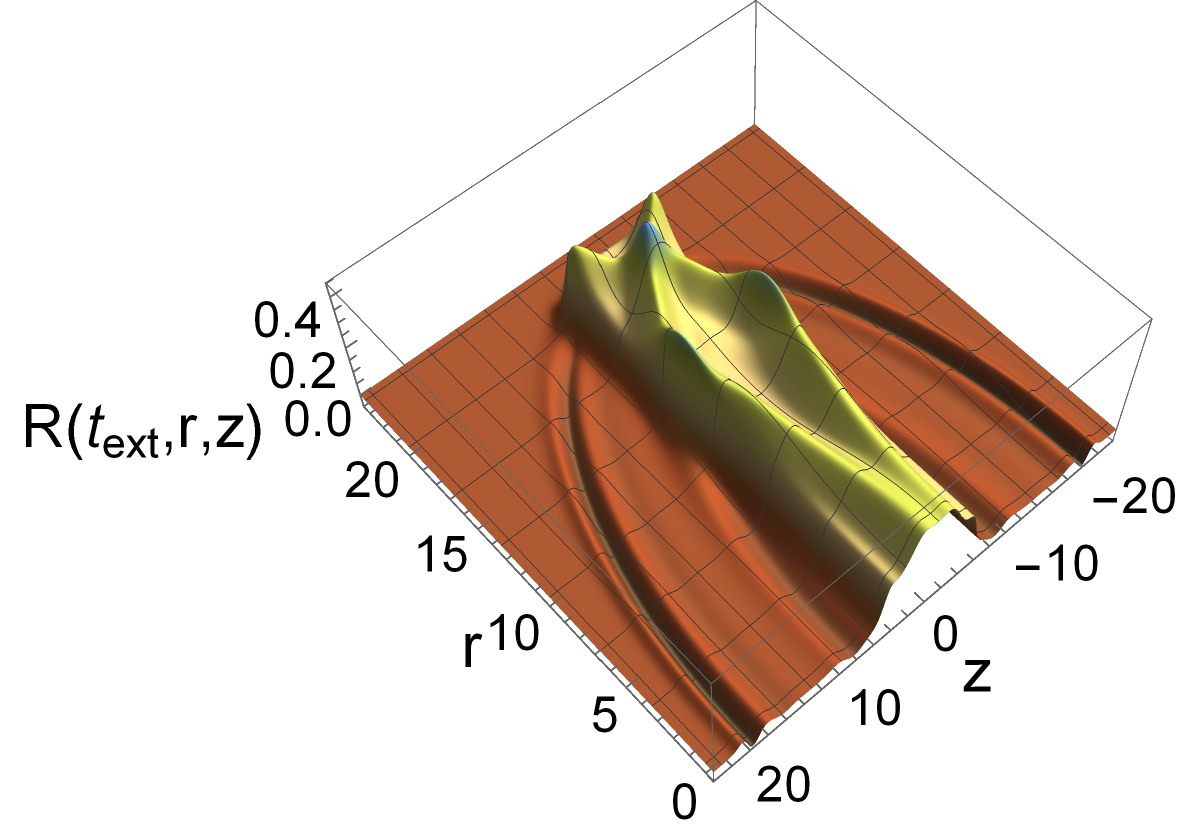}\qquad
		\includegraphics[width=5.5cm]{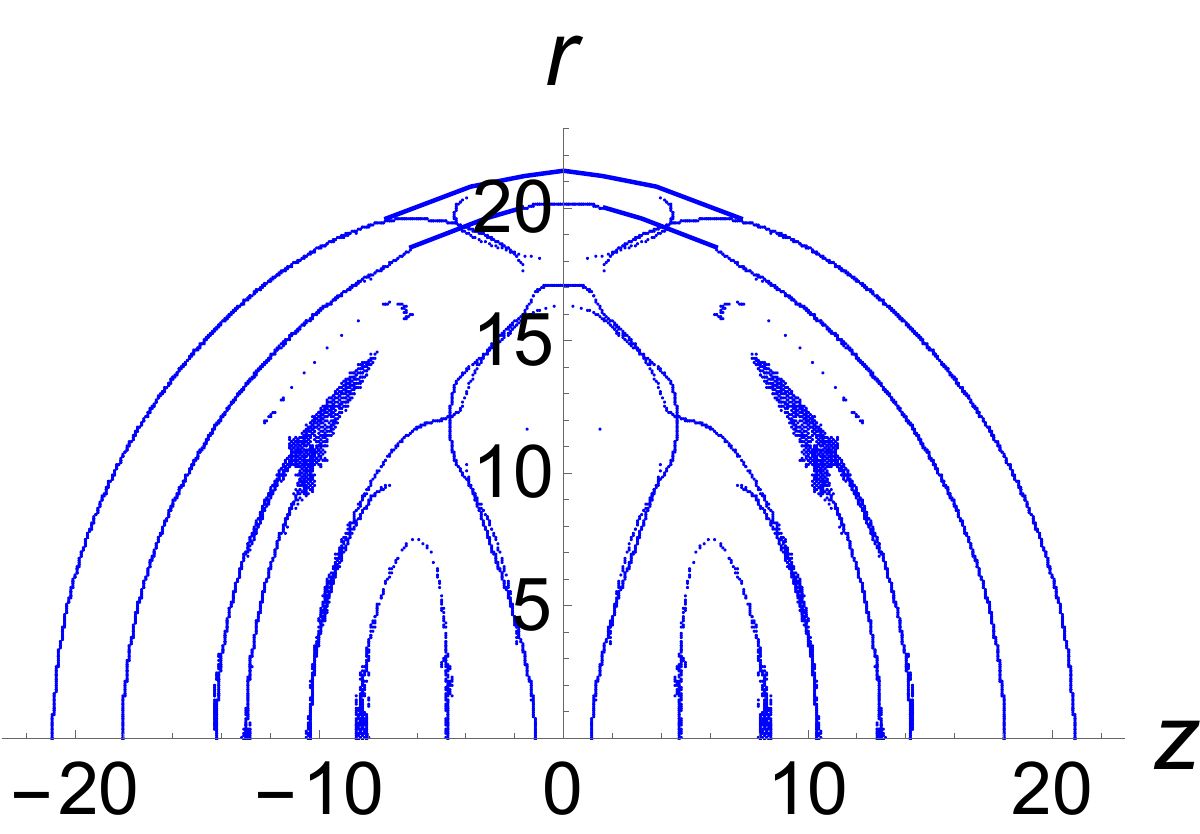}\label{Fig2Devolution4}}
	\vskip -4mm \caption{Plots of the evolution $R(t,r,z)$ of a two-dimensional Gaussian wave packet and the wave front of gravitational waves at different time. The parameters are set to $b=6$, $v=6$, and $\sigma=1$.}
	\label{Fig2Devolution}
\end{figure*}	

The left panels of Fig.~\ref{Fig2Devolution} show the two-dimensional distributions of the gravitational waveforms at different times. Additionally, we also use the peak point of the pulse signal to depict the position of the wave front at different times, as shown in the right panels of Fig.~\ref{Fig2Devolution}. The waveform at early time $t=3$ only has a single peak in Fig.~\ref{Fig2Devolution1}. Subsequently, at $t=5$, the waveform at $z=\pm5$  exhibits a slight deformation from a spherical wave which can be seen in Fig.~\ref{Fig2Devolution2}. Furthermore,  at $t=10$, as  Fig.~\ref{Fig2Devolution3} shows, in addition to the outermost primary wave $(r\sim10)$ on the brane, there is also a wave peak (echo) on the brane due to the reflection of the wave packet on the barrier. Lastly, at $t=20$, the echoes also appear outside the brane, and more complex echo waves appear on the brane due to the barrier reflection, which can be seen in Fig.~\ref{Fig2Devolution4}. We can clearly observe the multiple pulses outside the barrier. There are also echoes produced by the reflection of the Gaussian wave packet in the potential well. Due to the high reflectivity under the effective potential~\eqref{effectivepotential}, the signal inside the potential well is much stronger than the signal outside the potential well.
\begin{figure*}[htbp]
	\centering
	\includegraphics[width=7cm]{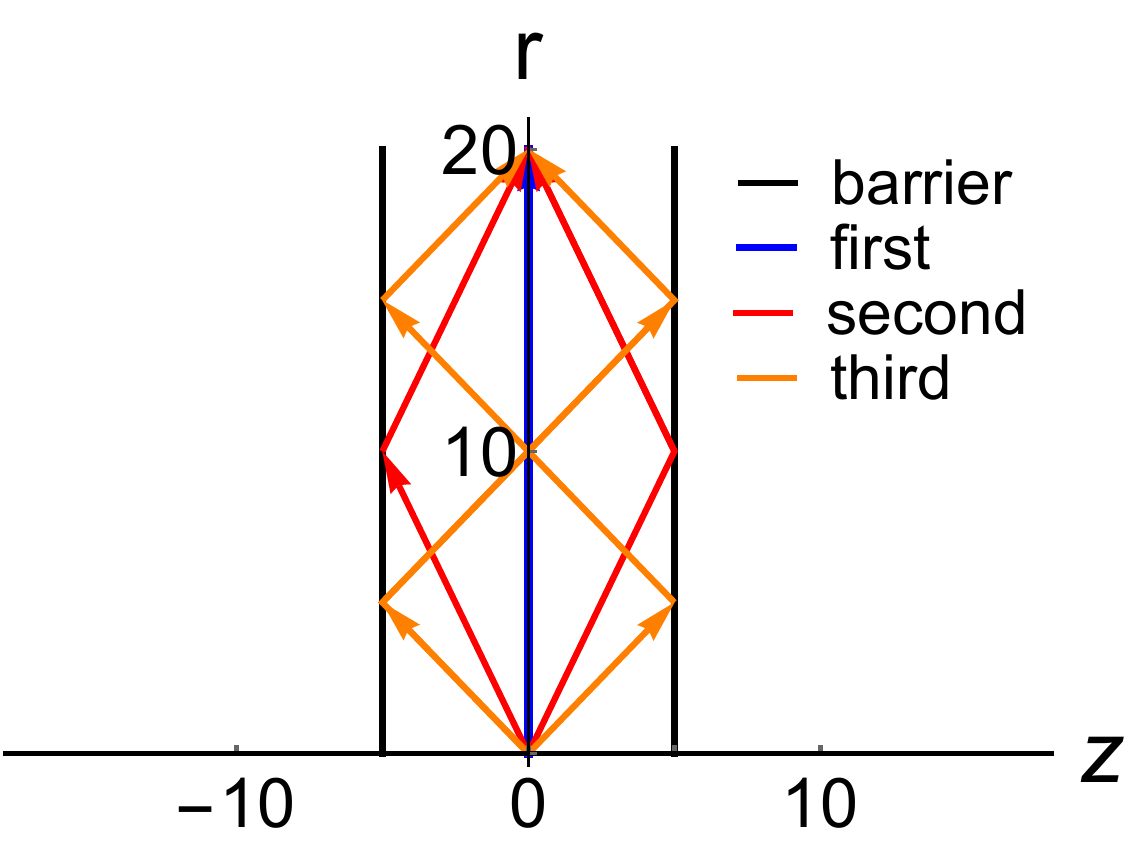}
	\vskip -4mm \caption{Plot of the propagation path of the gravitational waves. The blue, red, and orange lines represent the propagation path of the first three pulses, respectively.}
	\label{Figwavefront}
\end{figure*}

Figure \ref{Figwavefront} depicts a schematic diagram of the propagation path of the gravitational waves along the brane. It also illustrates the propagation of the first three pulses from the wave source to the observation point. It should be noted that, the generation of gravitational echoes in the extra-dimensional model occurs during the propagation path due to reflection by the potential barrier and is independent of the wave source. The gravitational echoes for compact stars are generated near the stars, which depends on the structure of the stars. The orange line represents the wave traveling in a straight line, the red lines represent the wave reflected once by the barrier, which constitutes the first echo signal, and the blue lines represent the wave reflected twice by the barrier, which constitutes the second echo signal. From this we can calculate the time interval between the $n$-th echo signal and the primary wave signal.
If the distance between the barriers is $2d$ and the distance between the source and the observer is $l$, then the time interval between the $n$-th gravitational echo and the primary wave is
\begin{eqnarray}
	t_n=\frac{2n\sqrt{\left(\frac{l}{2n}\right)^2+d^2}-l}{c}.
	\label{timedifference}
\end{eqnarray}

In Fig.~\ref{Fig2Devolution}, we can also observe the wavefront at different positions simultaneously. When $t=20$, the primary wave propagating along $r$ is roughly at $r=20$, and the position of the first echo is roughly at $r=17$, which is fundamentally consistent with the formula~\eqref{timedifference}. Note that the position of the reflecting surface is approximately at $z=4.6$. When $l\gg d$, Eq.~\eqref{timedifference} becomes $t_n(l\gg d)\approx4n^2d^2/(lc)$. 
Since no echo signal was detected in the first gravitational wave event, GW150914, there are two possibilities for this model. One is that the distance $d$ between the two barriers is too large, so the arrival time interval between the secondary pulse and the primary wave is too large. The binary black hole signal from GW150914, is approximately $410$Mpc away and the detectors operated for about 10 seconds~\cite{LIGOScientific:2016aoc}. From Eq.~\eqref{timedifference}, the distance between the barriers $2d$ should be more than 20 light year. The second possibility is that the reflectivity is too low, and the signal strength of the secondary pulse is not enough to be detected. For LIGO, the minimum signal sensitivity of its detectors is approximately 1/10 of the signal amplitude of GW150914~\cite{LIGOScientific:2016aoc}. Therefore, for the frequency range of the GW150914 signal (35-250Hz)~\cite{LIGOScientific:2016aoc}, the reflectivity should be less than 1/10. Roughly speaking, the reflectivity decreases with the frequency, which can be seen from Fig.~\ref{Figfourier}. (The transmittance increases with the frequency.) Thus, when $f\leqslant35\text{Hz}$, if the reflectivity is less than 1/10, then the amplitude of the gravitational echoes can be beyond the detection accuracy of the detector. With this criterion, we can obtain the constraints on the  five-dimensional fundamental mass. The relationship between the five-dimensional fundamental mass $M_5$ and the four-dimensional Planck mass $M_4$ is presented as follows:
\begin{eqnarray}
	M_4^2=\frac{M_5^3}{k}\int_{-\infty}^{+\infty}\e^{2A(ky)}d(ky).\label{5Dplankmass}
\end{eqnarray}
Through the transfer matrix method, we can calculate the dimensionless critical frequency $f_0/k$, such that, when $f>f_0$, the reflectivity is less than 1/10.  The maximum value of $f_0$ should be 35Hz, thus we can get the corresponding $k$. Through Eq.~\eqref{5Dplankmass}, we can get the maximum value of the five-dimensional fundamental mass.
\begin{table*}[!htb]
	\begin{center}
		\begin{tabular}{|| c | c | c | c | c | c | c  ||}
			\hline
			
			$kb$   & 6 & 6  & 6 & 4 & 6 & 8        \\ \cline{2-7}
			\hline
			$kv$   & 2 & 6  & 12 & 6 & 6 & 6        \\ \cline{2-7}
			\hline
			$f_0/k$  & 5.03  & 9.30 & 12.46 & 9.05  & 9.30 & 10.50        \\ \cline{2-7}
			\hline
			$M_5/\text{TeV}$ & $\,3.02\times10^4\, $  & $\,2.58\times10^4\, $ & $\,2.40\times10^4\, $ & $\,3.06\times10^4\, $ & $\,2.58\times10^4\, $ & $\,2.21\times10^4 \, $     \\ \cline{2-7}\cline{2-7}  \hline
			$m_1/\text{eV}$ & $2.06\times10^{-29} $ & $3.80\times10^{-29} $ & $5.11\times10^{-29} $ & $3.70\times10^{-29} $ & $2.06\times10^{-29} $ & $4.30\times10^{-29}  $     \\ \cline{2-7}\cline{2-7} 
			\hline
		\end{tabular}\\
		\caption{The maximum of the five-dimensional fundamental mass corresponding to the parameters $b$ and $v$ of the scalar field. $m_1$ is the minimum effective mass of the first echo on the brane. }
		\label{Tableplankparameter}
	\end{center}
\end{table*}

Table~\ref{Tableplankparameter} illustrates that the maximum value of the five-dimensional fundamental mass $M_5$  decreases with the vacuum expectation value $v$ and the distance between kinks $b$ of the scalar field. The values of $b$ and $v$ can be determined by the frequency of the late-time gravitational echoes by matching to the characteristic frequencies of the extra-dimensional model, provided that gravitational echoes can be detected in the future.

The frequency of the gravitational echo generated by compact stars depends on the effective potential felt by gravitational perturbations and the effective potential is determined by the compact stars' parameters, such as its mass and angular momentum. Consequently, the resonance frequencies of gravitational echoes of different gravitational wave events will vary according to the different parameters of the compact stars, resulting in different echo frequencies.
In contrast, gravitational echoes in extra-dimensional models arise from the reflection of gravitational waves by the effective potential along the extra-dimensional direction of the propagation path. Therefore, the echoes share the same resonance frequency spectrum for all gravitational wave events. For different gravitational wave events with similar initial frequency band, the resonance frequencies of the echoes will be close, and the decay rates of these echoes will also be similar. Furthermore, since the gravitational echo in extra-dimensional model corresponds to the KK modes, they would exhibit an effective mass if they can be detected on the brane, and their speed would be slightly less than the speed of light. For the gravitational wave event  GW150914, the speed of the first echo is not less than $(1-1.11\times10^{-16})c$. And the effective mass of the gravitational echo on the brane is also much smaller, about $10^{-29}$ eV. Besides, the time intervals of the gravitational echoes depend on the position of the detector, which can be seen from Eq.~\eqref{timedifference}. This is different from the gravitational echoes generated from compact stars. However, it seems that it is hardly to detect this phenomenon. In conclusion, if we can observe the echos that share the same frequency spectrum and echos with speed less than light speed, this will be the smoking gun of infinite extra dimensions.

\section{Echoes in Other Extra-Dimensional Theory}
There are also multiple barriers in the effective potential of gravitational perturbation in other thick brane theories. In this section, we investigate gravitational echo phenomena in two different thick brane theories that exhibit the obvious multiple barriers. 

\subsection{Echoes in Five-Dimensional $f(R)$ Theory}~\label{5FRT}
In this subsection, we investigate the echoes of the thick brane in $f(R)$ gravity. First, the action of the $f(R)$ brane is given by~\cite{Parry:2005eb,Zhong:2010ae,Yu:2015wma}
\begin{eqnarray}\label{actionfR}
	S=\int \mathrm{d}^5x\sqrt{-g}\left(\frac{M_5^3}{4}f(R)
	-\frac{1}{2}\partial^M\phi\partial_M\phi- \frac{1}{2}\partial^M\varphi\partial_M\varphi - V(\phi,\varphi) \right).
\end{eqnarray}
Varying the action~\eqref{actionfR} with respect to the metric $g_{MN}$ and using the flat brane metric~\eqref{metric}, we get the equations of motion:
\begin{eqnarray}
	\label{a26}
	f(R)\!+\!2f_R\left(4A'^2 \!+\! A''\right)
	\!-\! 6f'_RA'\!-\!2f''_R\!\!&=&\!\!2(\phi'^2\!+\varphi'^2\!+\!2V),~~~~~~   \\
	\label{b26}
	-8f_R\left(A''\!+\!A'^2\right)\!+\!8f'_RA'
	-f(R)\!\!&=&\!\!2(\phi'^2\!+\varphi'^2\!-\!2V),\\
	\phi^{\prime\prime}+4A^{\prime}\phi^{\prime}-\frac{\pd V(\phi)}{\pd \phi}&=&0,\\
	\varphi^{\prime\prime}+4A^{\prime}\varphi^{\prime}-\frac{\pd V(\phi)}{\pd \varphi}&=&0. \label{eqmotionfR}
\end{eqnarray}
We consider the degenerate Bloch-brane solution~\cite{deSouzaDutra:2008gm}:
\begin{eqnarray}
	\label{DBbrane1phi}
	{\phi}(y)
	\!\!&=&\!\!  v\frac{\sqrt{c_{0}^{2}-4}\sinh (2dvy)}
	{\sqrt{c_{0}^{2}-4}\cosh (2dvy)-c_{0}},
	\\
	\label{DBbrane1chi}
	{\varphi}(y)
	\!\!&=&\!\! \frac{2v}{\sqrt{c_{0}^{2}-4}\cosh (2dvy)-c_{0}},
	\\
	\label{DBbrane1E2A}
	e^{2A(y)}
	\!\!&=&\!\! \left( \frac{\sqrt{c_{0}^{2}-4}-c_{0}}
	{\sqrt{c_{0}^{2}-4}\cosh (2dvy)-c_{0}}\right)
	^{4v^{2}/9} \nonumber \\
	&\times&\!\!\exp \left[ -\frac{4v^2\left( c_{0}^{2}-4-c_{0}
		\sqrt{c_{0}^{2}-4}\right) }
	{9\left( \sqrt{c_{0}^{2}-4}
		-c_{0}\right) ^{2}}\right]\nonumber \\
	&\times&\!\! \exp \left[ \frac{4v^2\left( c_{0}^{2}\!-\!4\!-\!c_{0}
		\sqrt{c_{0}^{2}\!-\!4}\cosh(2dvy)\right) }
	{9\left( \sqrt{c_{0}^{2}\!-\!4}\cosh(2dvy)
		\!-\!c_{0}\right) ^{2}}\right].~~~~~~~
\end{eqnarray}\label{DBbrane1}
In our calculations,  the parameters are set as $a=d=1,~v=2$, and $c_0=-2-7.0\times 10^{-16}$~\cite{Yu:2015wma}.
\begin{figure*}[htbp]
	\centering
	\includegraphics[width=4.5cm]{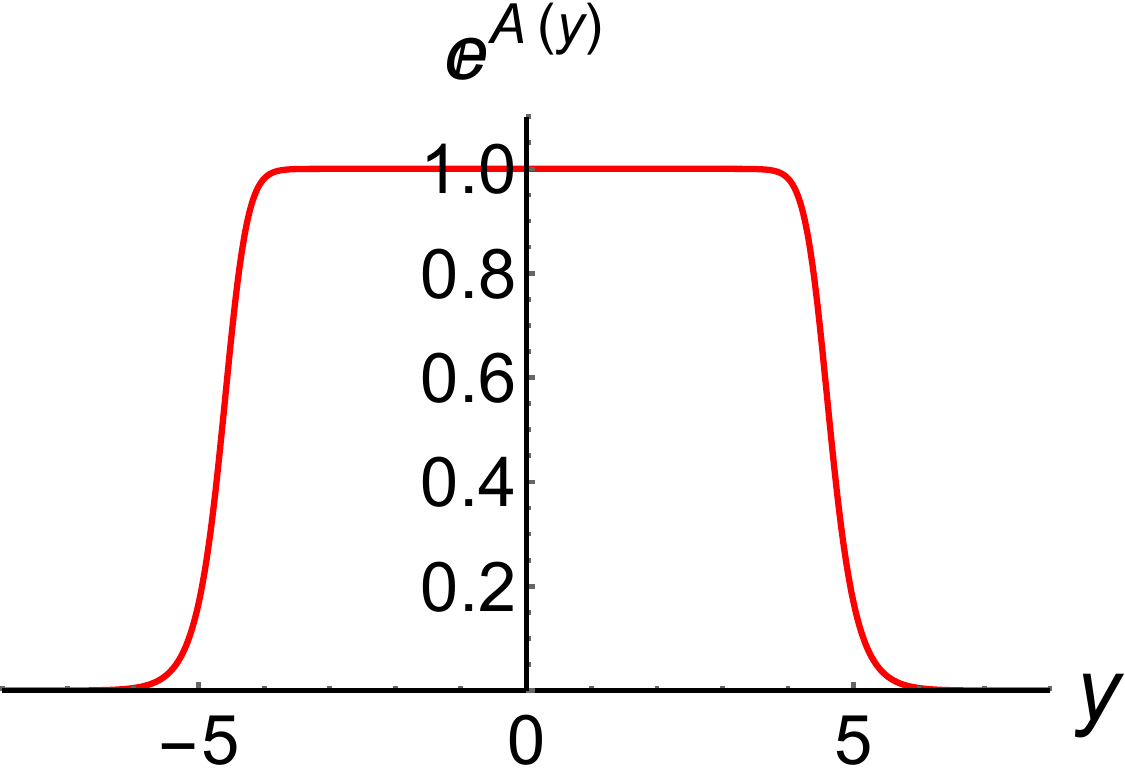}
	\includegraphics[width=4.5cm]{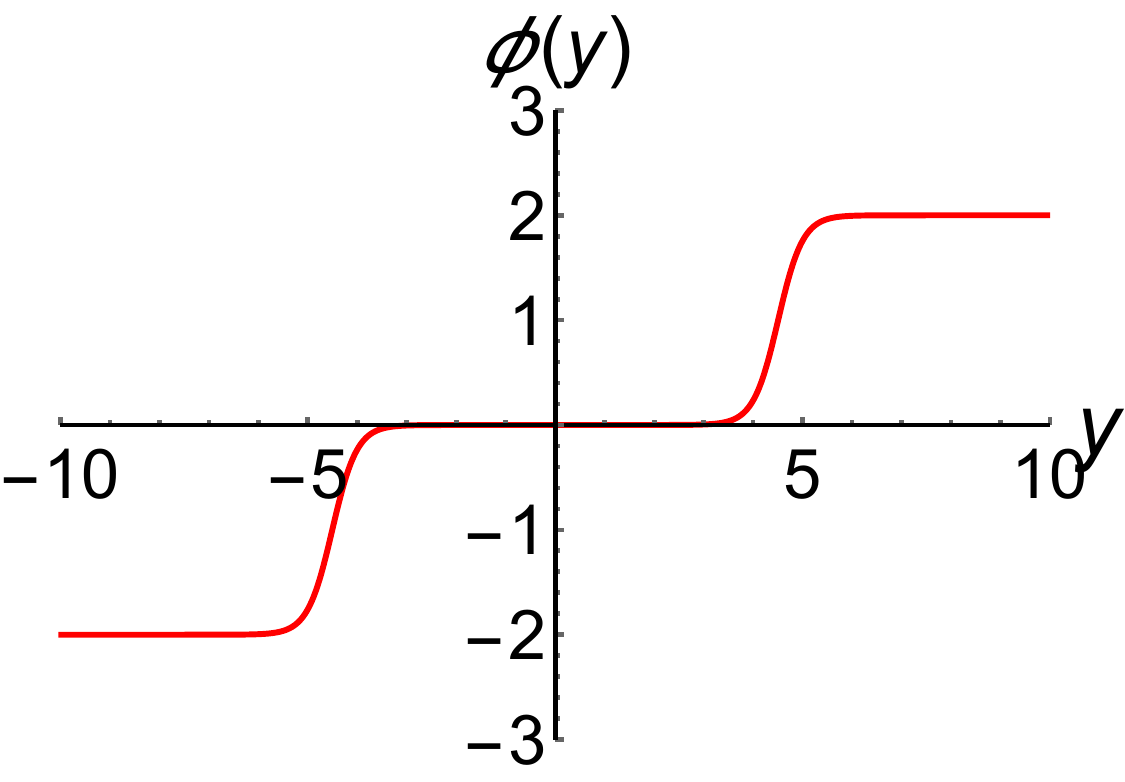}
	\includegraphics[width=4.5cm]{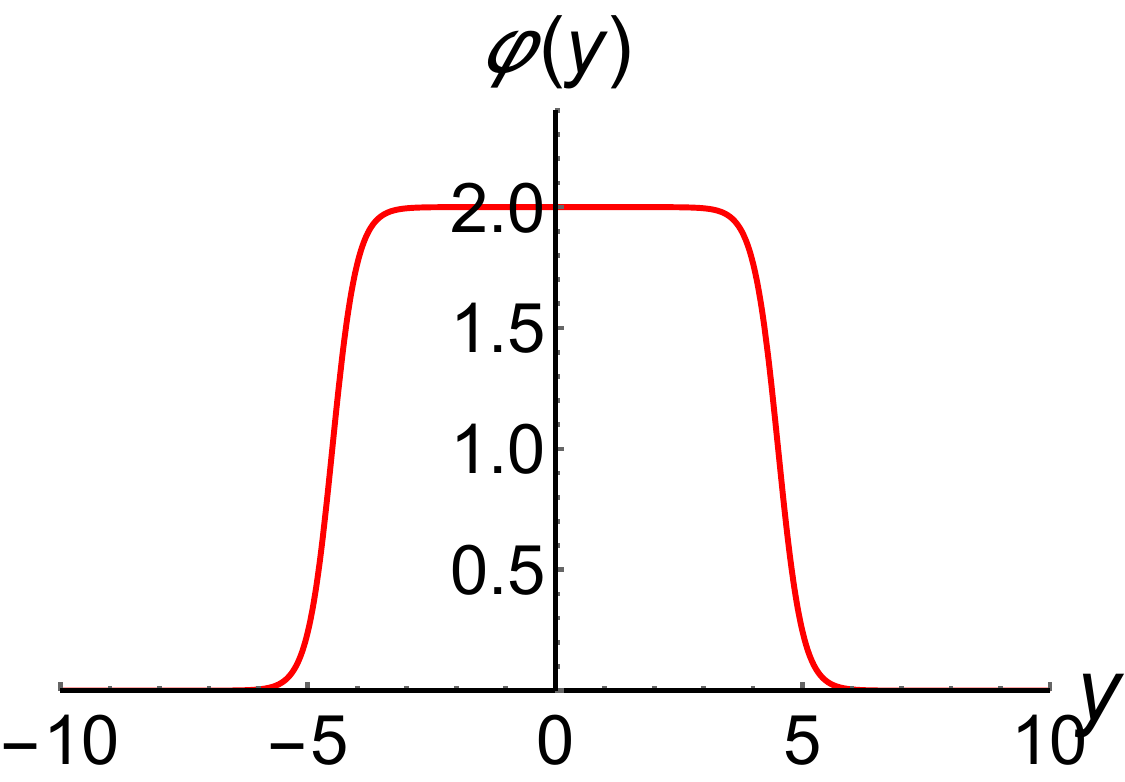}
	\vskip -4mm \caption{Plots of the warp factor and the two scalar fields. }
	\label{FigwarpedfactorfR}
\end{figure*}

Next we consider gravitational perturbations on the brane with RS gauge $h_{M0}=0$. The metric describing  the tensor perturbation is
\begin{eqnarray}
	ds^2=\text{e}^{2A(y)}(\eta_{\mu\nu}+h_{\mu\nu})dx^{\mu}dx^{\nu}+dy^2. \label{5DfRperturbationmetric}
\end{eqnarray}
With the coordinate transformation $dz=\e^{-A(y)}dy$, the perturbed Einstein equation reads
\begin{eqnarray}
	\label{SE}
	\left[\partial_z^2+ \Big(3\frac{\partial_z a}{a}+\frac{\partial_z f_R}{f_R}\Big)\partial_z+\Box^{(4)}\right]
	h_{\mu\nu}=0.
\end{eqnarray}
where $a(z)=\e^{A(z)}$. Then considering the following decomposition:
\begin{eqnarray}
	\label{decomposition}
	h_{\mu\nu}(t,x^i,z)=\epsilon_{\mu\nu}(x^i) a^{-3/2}f_{R}^{-1/2}H(t,z),
\end{eqnarray}
we obtain the evolution equation for the extra-dimensional component of the perturbation:
\begin{eqnarray}
	\label{Schrodinger equation fR}
	\left[\partial_t^2-\partial_{z}^2+V_\text{eff}(z)\right]H(t,z)=-p^2H(t,z),
\end{eqnarray}
where 
\begin{eqnarray}
	V_{\text{eff}}(z)=\frac{3}{4}\frac{(\partial_z a)^2}{a^2} + \frac{3}{2}\frac{\partial_z^2a}{a^2} + \frac{3}{2}\frac{\partial_z a \partial_z f_R}{a f_R} - \frac{1}{4}(\frac{\partial_z f_R}{f_R})^2 + \frac{1}{2}\frac{\partial_z^2 f_R}{f_R^2}.\label{fReffectivepotential} 
\end{eqnarray}
\begin{figure*}[htbp]
	\centering
	\includegraphics[width=7cm]{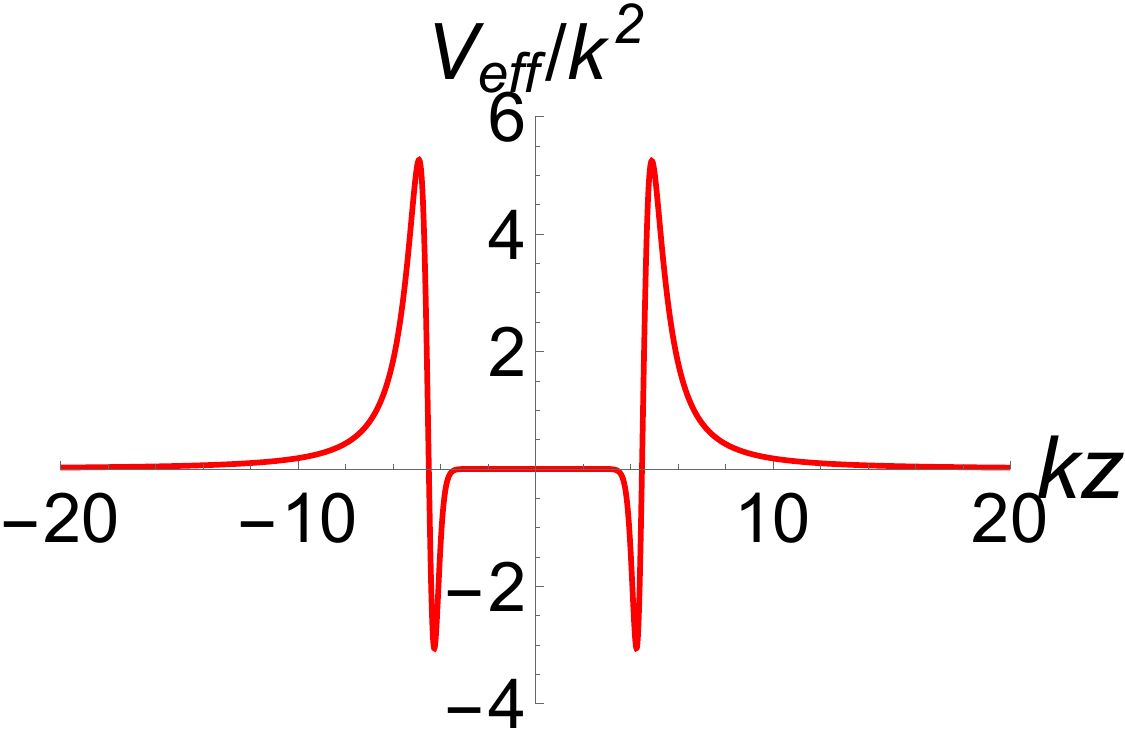}
	\vskip -4mm \caption{Plots of the effective potential~\eqref{fReffectivepotential} of the $f(R)$ brane. }
	\label{FigeffectivepotentialfR}
\end{figure*}
We still take  the same radiative boundary conditions as in Sec.~\ref{5fieldeq}, and use a Gaussian wave packet  $H(0,z)=\e^{-\frac{(z-z_0)^2}{\sigma}}$ for evolution. In Fig.~\ref{FigfRWFevolution}, it can be seen that the evolution of the Gaussian wave packet is basically similar to that in Sec.~\ref{5fieldeq}. The reason is that the warp factor is also a platform-type and the effective potential for gravitational perturbations is basically same as in Sec.~\ref{5fieldeq}.
\begin{figure*}[htbp]
	\centering
	\subfigure[~The echoes in the $f(R)$ braneworld]{
		\includegraphics[width=6cm]{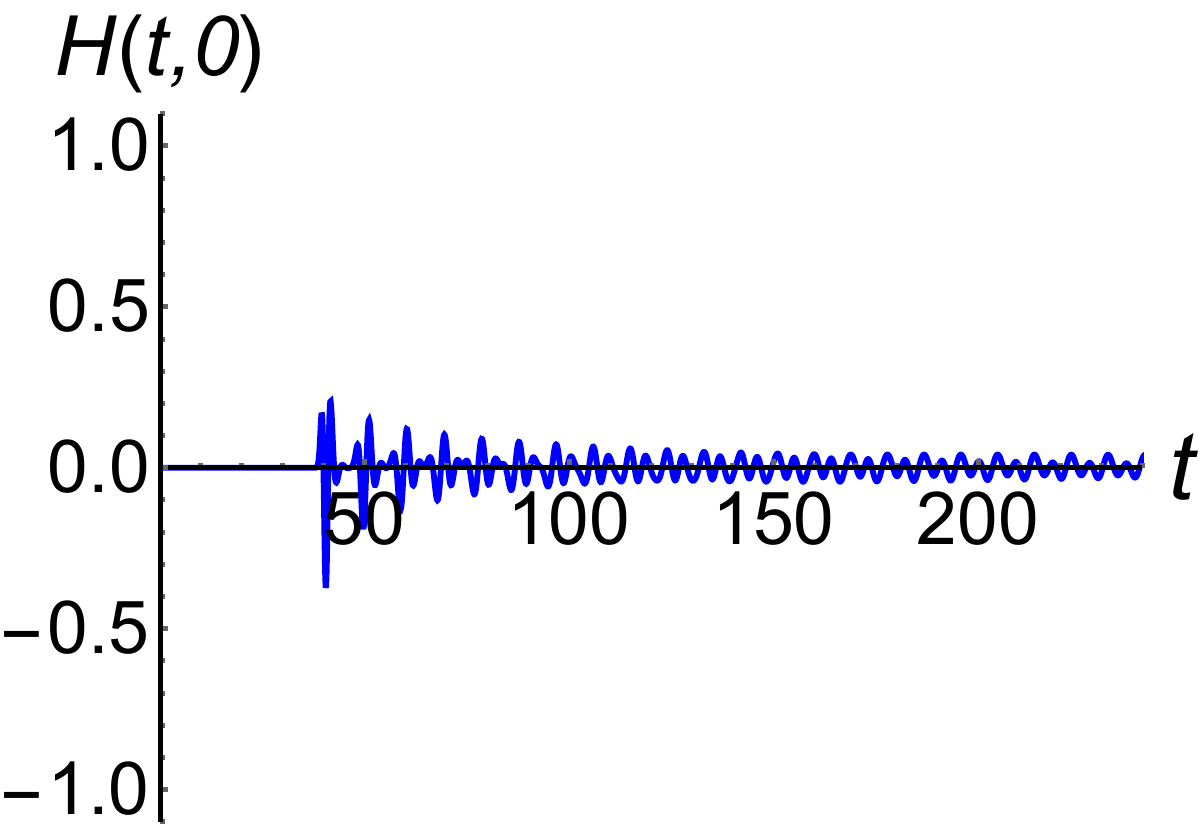}
		\includegraphics[width=6cm]{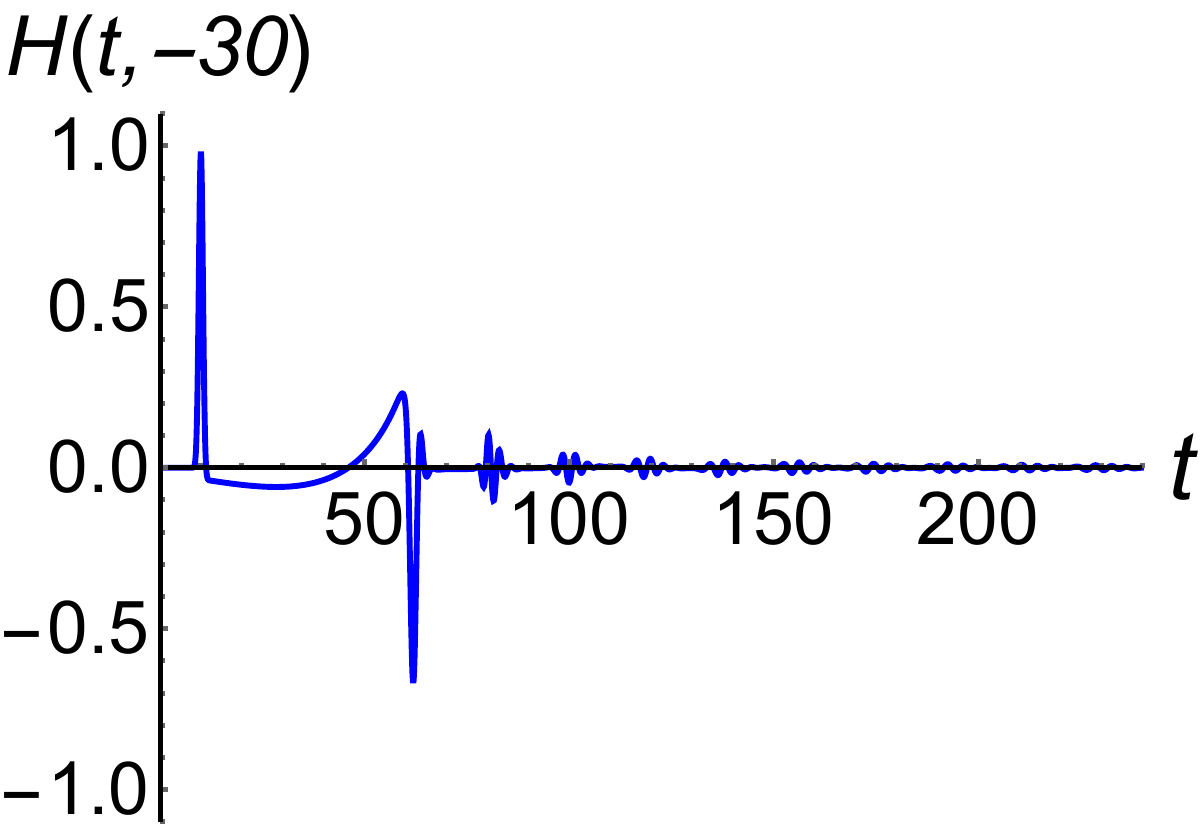}}
	\subfigure[~The echoes in the GR braneworld]{
		\includegraphics[width=6cm]{phi_tanhyWFy=40_0.pdf}
		\includegraphics[width=6cm]{phi_tanhyWFy=40_30.pdf}}
	\vskip -4mm \caption{Plots of the evolution $H(t,z)$ of a Gaussian wave packet. The parameters of the Gaussian wave packet are $\sigma=0.5$ and $z_0=-40$.}
	\label{FigfRWFevolution}
\end{figure*}

\subsection{Echoes in Five-Dimensional Non-minimally Derivative Coupling Scalar-Tensor Theory}~\label{5FDNDT}	
In this section, we investigate the echoes of the thick brane in a five-dimensional non-minimally derivative coupling scalar-tensor theory. First, the action of the five-dimensional non-minimally derivative coupling scalar-tensor theory is given by~\cite{Fu:2019xtx,Brito:2018pwe}
\begin{eqnarray}
	S=\int d^5x \sqrt{-g}\Big[\frac{1}{2 }F(\phi)R-b~G_{MN}\nabla^{M}\phi\nabla^{N}\phi-\frac{1}{2}\partial^M\phi\partial_M\phi-V(\phi)\Big]. ~\label{actionf}
\end{eqnarray}	
Varying the action~\eqref{actionf} with respect to the metric $g_{MN}$, we obtain the gravitational field equation:
\begin{eqnarray}
	F(\phi)G_{MN}&-&\nabla_M\nabla_N F(\phi)+g_{MN}\Box^{(5)} F(\phi)= T_{MN}+2b\Theta_{MN},~\label{EoMofEinstein}\\
	T_{MN}&=&\nabla_M\phi\nabla_N\phi-\frac{1}{2}g_{MN}(\nabla\phi)^2-g_{MN}V(\phi),\qquad \\
	\Theta_{MN}&=&-\frac12\nabla_M\phi\nabla_N\phi R+2\nabla_K\phi\nabla_{(M}\phi R_{N)}^K-\frac{1}{2}\left(\nabla\phi\right)^2G_{MN}\nonumber\\
	&+&\nabla^K\phi\nabla^L\phi R_{MKNL}+\nabla_M\nabla^K\phi\nabla_N\nabla_K\phi-\nabla_M\nabla_N\phi\Box^{(5)}\phi\nonumber\\
	&+&g_{MN}\left[-\frac{1}{2}\nabla^K\nabla^L\phi\nabla_K\nabla_L\phi+\frac{1}{2}\left(\Box^{(5)}\phi\right)^2
	-\nabla_K\phi\nabla_L\phi R^{KL}\right],\label{NMDCTmotion}
\end{eqnarray}	
where $\square^{(5)}=g_{MN}\nabla^{M}\nabla^{N}$. Unlike Eq.~\eqref{motionGRbrane} in Sec.~\ref{5fieldeq}, for Eq.~\eqref{EoMofEinstein}, we find that the warp factor $\e^{A(y)}$ can have a multi-peak shape solution. We can have a set of solutions given by
\begin{eqnarray}
	A(y)&=&\text{ln}\left(\frac{\text{cosh}\,d}{2}(\text{sech}(ky+d)+\text{sech}(ky-d))\right),\\
	\phi(y)&=&\text{tanh}\,(ky).
	\label{changedwarp}
\end{eqnarray}
\begin{figure}[!htb]
	\centering
	\includegraphics[width=5cm]{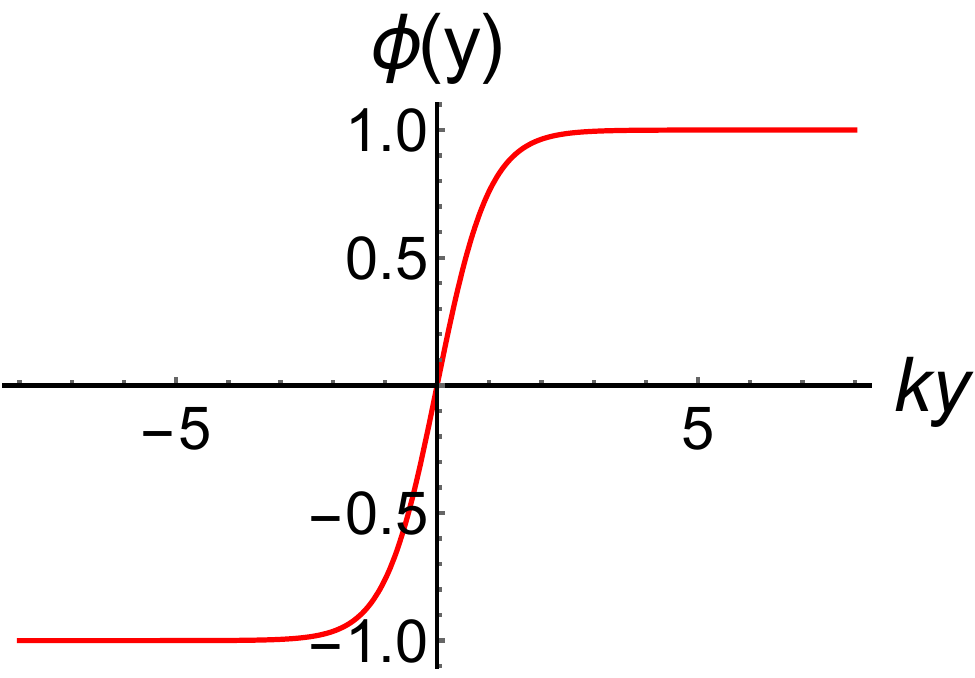}\qquad\qquad\quad
	\includegraphics[width=6cm]{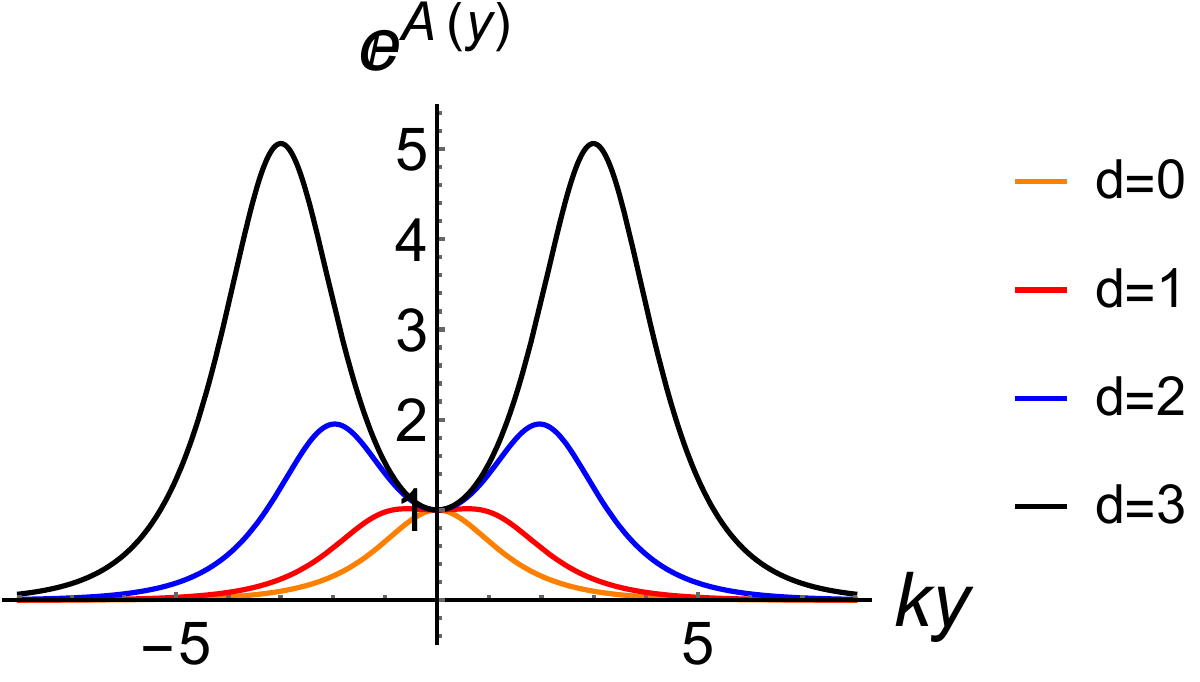}
		
	\includegraphics[width=6cm]{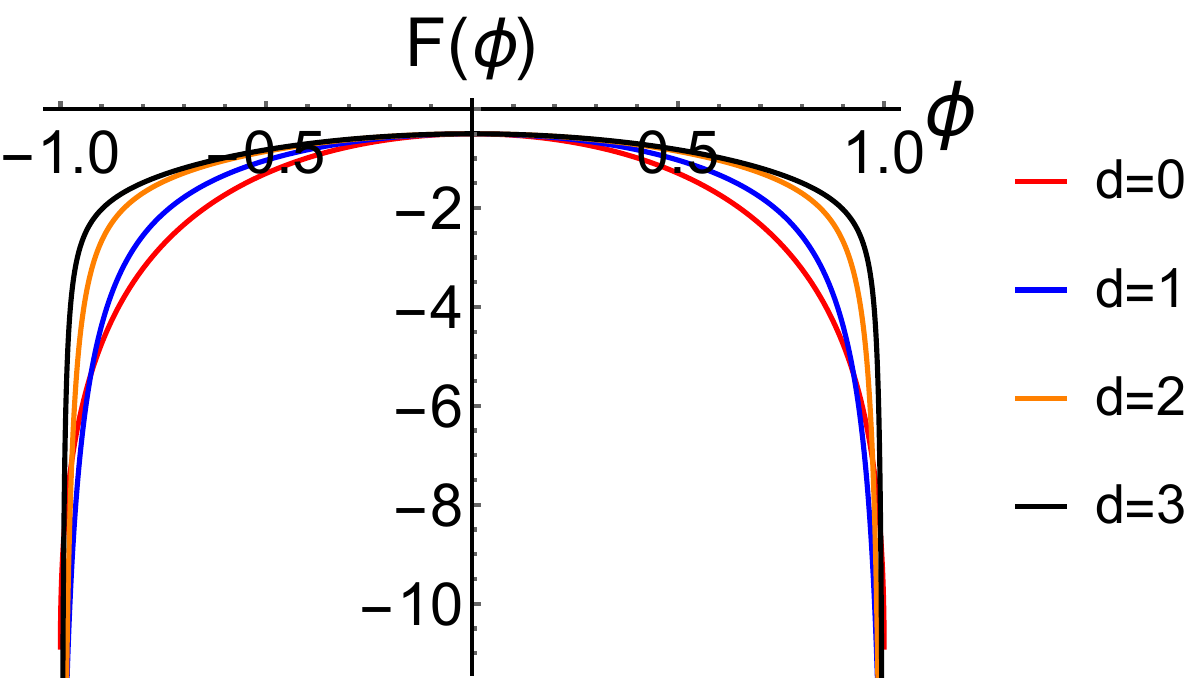}\qquad
	\includegraphics[width=6cm]{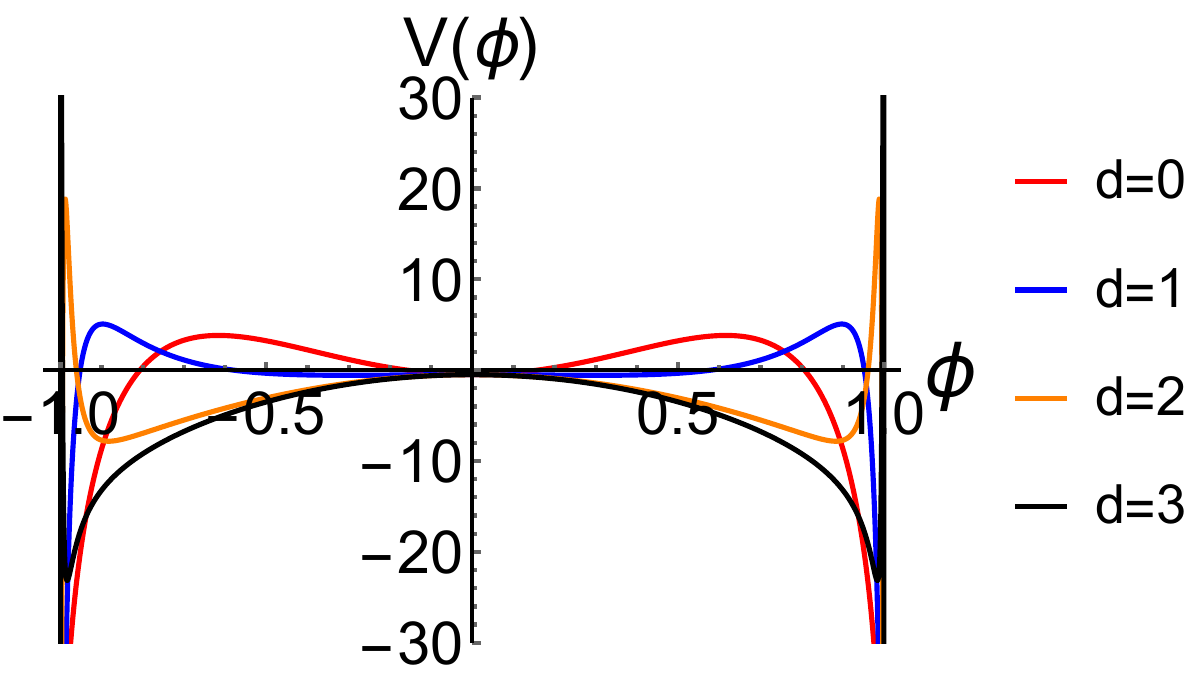}
	\vskip -4mm \caption{Plots of the warp factor $\e^{A}$, the scalar field $\phi$, the function $F(\phi)$ of the scalar field, and the scalar potential $V(\phi)$ with different peak distance $d$ of the warp factor for the theory\eqref{actionf}.}
	\label{Fighorndeskifunction}
\end{figure}
Figure~\ref{Fighorndeskifunction} shows that the influence of the distance $d$ between the two peaks of the warp factor on the function $F(\phi)$ and the scalar potential $V(\phi)$.  As $d$ increases, the shape of $F(\phi)$ becomes flatter near $\phi=0$ and the sides become steeper, and the two peaks of the scalar potential $V(\phi)$ become higher and closer to $\phi=1$. 
Under the conformal coordinate $z$, the equation of the gravitational tensor perturbation is
\begin{eqnarray}
	K(z)\partial_z^2 h_{\mu\nu}+L(z)\partial_z h_{\mu\nu}+\Box^{(4)}h_{\mu\nu}=0, 
	\label{perturbetioneq}
\end{eqnarray}
where
\begin{eqnarray}
	K(z)&=&\frac{F-be^{-2A}(\partial_{z}\phi)^2}{F+be^{-2A}(\partial_{z}\phi)^2}, \\
	L(z)&=&\frac{3\partial_{z}A(F-be^{-2A}(\partial_{z}\phi)^2)+\partial_{z}(F-be^{-2A}(\partial_{z}\phi)^2)}{F+be^{-2A}(\partial_{z}\phi)^2}.
\end{eqnarray}
Then making another coordinate transformations $dz=\sqrt{K(z)}dw$ and using the following decomposition:
\begin{eqnarray}
	h_{\mu\nu}(x^i,t,w)=\varepsilon_{\mu\nu}(x^i)T(w,t), \label{txzdecomposition}
\end{eqnarray}
we can simplify the perturbation equation as
\begin{eqnarray}
	\partial_{t}^2T-\partial_{w}^2T+Q(w)\partial_w T+m^2T=0.
\end{eqnarray}
where $Q(w)=\frac{L}{\sqrt{K}}-\frac{\partial_w K}{2K}$. Redefining $T(t,w)=\text{exp}(-\frac{1}{2}\int Q(w)dw)H(t,w)$, we further have 
\begin{eqnarray}
	\partial_t^2 H-\partial_w^2 H + V_\text{eff}(w)H = -p^2 H,
	\label{evolutionequation}
\end{eqnarray}
where 
\begin{eqnarray}
	V_\text{eff}(w)=\frac{1}{2}\partial_w Q+\frac{1}{4}Q^2. \label{horndeskieffectivepotential}
\end{eqnarray}	

\begin{figure*}[htbp]
	\centering
	\includegraphics[width=7cm]{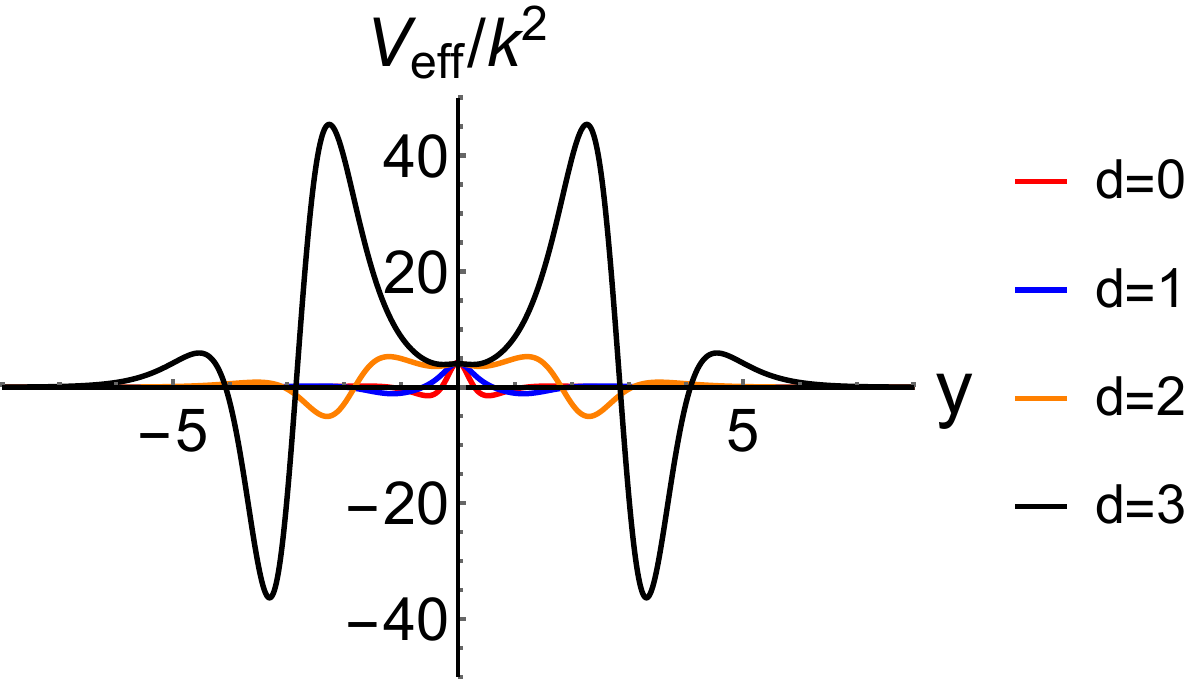}
	\vskip -4mm \caption{Plots of the effective potential~\eqref{horndeskieffectivepotential} of the gravitational perturbation in five dimensional non-minimally derivative coupling theory. }
	\label{Fighorndeskieffectivepotential}
\end{figure*}
In Fig.~\ref{Fighorndeskieffectivepotential}, it can be seen that the distance between the effective potential barriers, and the height of the effective potential barriers increase with $d$. Taking the same radiative boundary conditions as in Sec.~\ref{5fieldeq}, we evaluate a Gaussian wave packet  $H(0,w)=\e^{-\frac{(w-w_0)^2}{\sigma}}$.
\begin{figure*}[htbp]
	\centering
	\subfigure[~The echoes in the GR braneworld]{
		\includegraphics[width=6cm]{phi_tanhyWFy=40_0.pdf}
		\includegraphics[width=6cm]{phi_tanhyWFy=40_30.pdf}}
	\subfigure[~The echoes in the five-dimensional non-minimally derivative coupling theory]{\includegraphics[width=6cm]{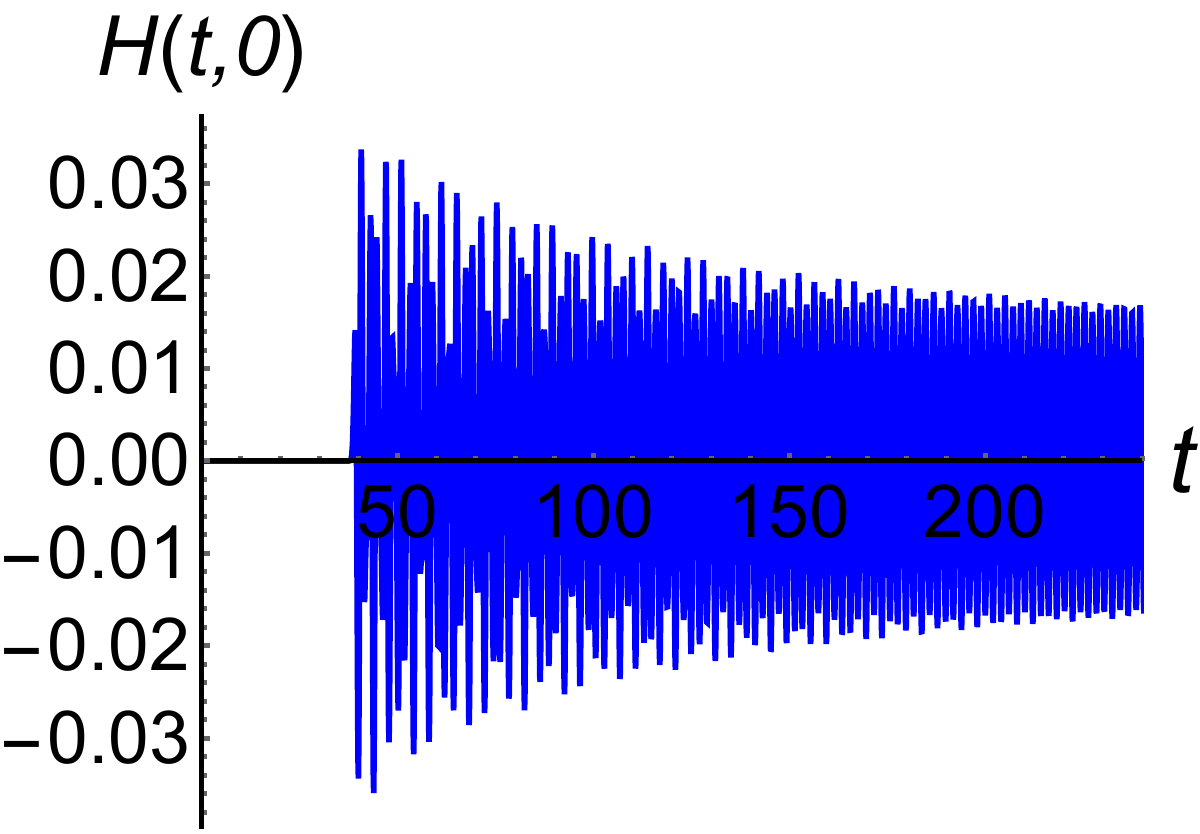}
		\includegraphics[width=6cm]{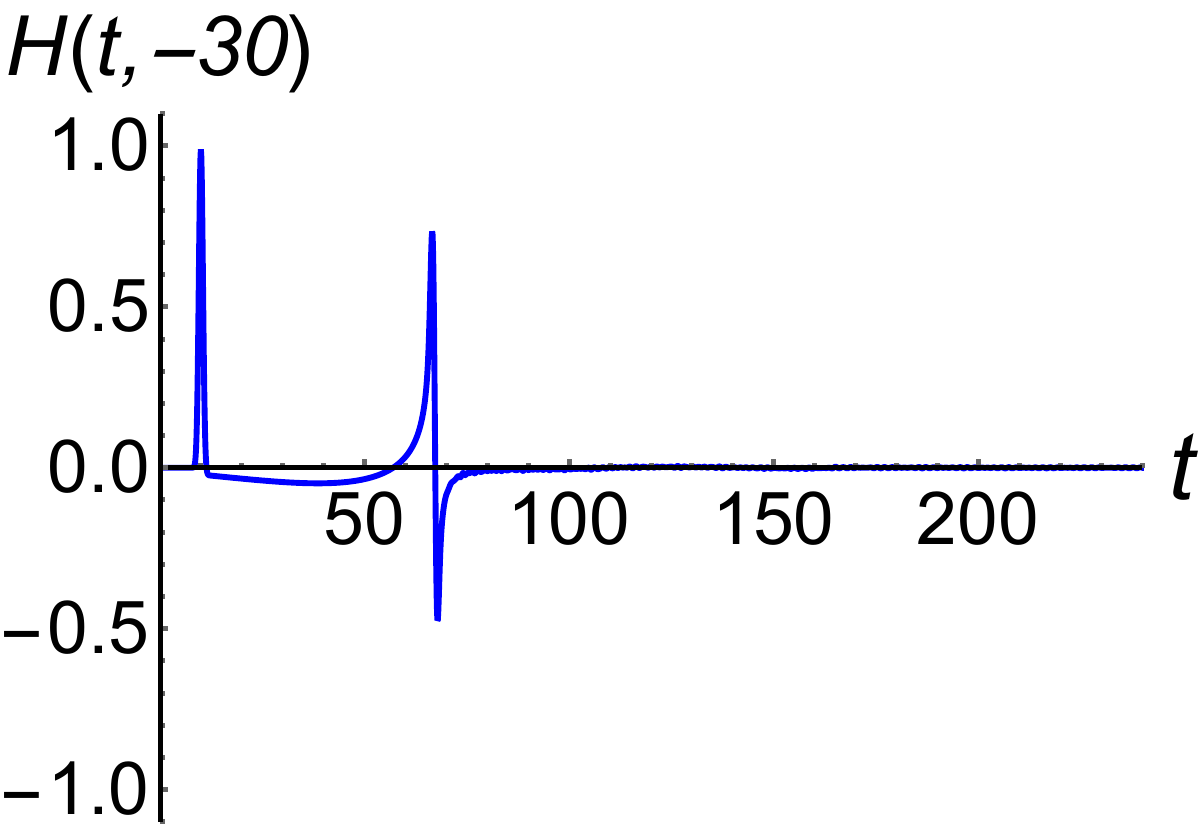}\label{FighorndeskiWFevolution2}}
	\vskip -4mm \caption{Plots of the evolution $H(t,w)$ of a Gaussian wave packet at fixed point. The parameters of the Gaussian wave packet and the effective potential are set to $\sigma=0.5$, $w_0=-40$, and $d=3$.}
	\label{FighorndeskiWFevolution}
\end{figure*}

The results of the evolution are shown in Fig.~\ref{FighorndeskiWFevolution2}.	 After the initial pulse, there are still several pulses, and the first echo is obvious but the amplitude of the later echoes are too small. The gravitational echoes observed at the outer regions of the brane are also more obvious than those inside the brane. Due to the small distance between the two barriers, the waveform reflects back and forth in a short time, leading to a relatively compact waveforms at points on the brane. The wave evolution on the brane gradually stabilizes, and exhibits a relatively slow decay. However, the height of the barrier varies greatly with distance $d$, which can be seen from Fig.~\ref{Fighorndeskieffectivepotential}. This variation results in higher reflectivity of the barrier for larger values of $d$, thereby making the first echo more pronounced. If it is a gravitational wave propagating on the brane, the energy outflow rate is extremely low, and the echoes are more obvious.

\section{Conclusion}~\label{conclusion}
In this paper, we investigated the gravitational echoes in various extra-dimensional theories.In thick brane models, the effective potential of gravitational perturbations typically displays a multi-barrier form. Consequently, the propagation of the gravitational waves in most thick brane models may lead to the gravitational echoes. We studied the gravitational echoes in three different thick brane models. Additionally, we examined the impact of various parameters on the echoes in these models. Based on these investigations, we explored the propagation of the gravitational echoes on the brane, which allows us to make some constraints on the parameters of the  extra-dimensional theories.

We investigated the thick brane generated by a background scalar field within the framework of general relativity. When the background scalar field is a double kink configuration, the warp factor of the metric exhibits a platform shape, thus enabling gravitational perturbations to possess an effective potential characterized by a distinct double barrier. The distance $b$ between two kinks of the scalar field and the vacuum expectation value $v$ affect both the height and width of the effective potential barriers. We used Gaussian wave packets as initial data to study the time evolution of the gravitational waves in Eq.~\eqref{EDevolution eq}. It becomes evident that the distance between barriers influences the time interval between the two pulses of the echoes. Furthermore, the height of the barriers affects the reflectivity of the waves and the amplitude of the echo pulse. Different frequency components of Gaussian wave packets exhibit different reflectivities, which will affect the amplitude of the echoes. Gravitational echoes may be produced in the brane model with a platform-type soliton scalar field or a multi-kink scalar field. The multi-kink scalar solution requires that the scalar potential has multiple minima, with the intermediate minima being higher than the minima on either side. The platform-type soliton scalar solution requires that the scalar potential has a negative minimum and two positive maxima.

We also explored the propagation of a Gaussian wave packet on the brane. The two-dimensional evolution equation is given by Eq.~\eqref{2Dperturbation eq}. If the gravitational waves are originated from a wave source on the brane, multiple gravitational wave signals can be detected at different locations on the brane due to reflections by barriers. The strength of these signals depends on the reflectivity. We also compute the time interval between the $n$-th echo signal and the primary wave and the speed of the first gravitational echo. 

We also conducted a brief calculation of the gravitational echoes in other two thick brane models. For the $f(R)$ thick brane, there is also the gravitational echo phenomenon. However, for the non-minimally derivative coupled scalar-tensor theory, the reflectivity of waves markedly increases due to the significant impact of the parameter $d$ on the barriers. Therefore the gravitational waves inside the brane are notably challenging to escape, while those external to the brane become highly detectable upon their initial reflection.

Based on the existence of the gravitational echoes in the evolution of Gaussian wave packets in the three models, it can be deduced that gravitational echoes may be a common phenomenon in most thick brane models. The gravitational echoes in the thick brane occurs during propagation, and the frequency and amplitude of the echoes only depend on the initial wave and the structure of the thick brane models. In contrast, echoes from black holes or other compact stars are generated near the stars, with their frequency and amplitude being influenced by the initial wave and the spacetime structure of the stars themselves. Consequently, the attenuation rate of the same frequency in gravitational echoes remains consistent for different events' gravitational wave signals in the thick brane model. And the speed of the gravitational echo in the thick brane would be slightly below the speed of light relative to the primary wave. We expect to observe gravitational echoes in future gravitational wave detectors and to detect same behavior in different gravitational echo signals and different propagation speeds among the different pulses. This would help us understand more deeply on higher-dimensional spacetime theories. Next, we aim to investigate the gravitational echoes of non-flat brane and brane black hole models.

\section*{Acknowledgments}
We are thankful to Bin~Guo, Yu-Peng~Zhang, and Qin~Tan for useful discussions. This work was supported by the National Key Research and Development Program of China (Grant No. 2020YFC2201503), the National Natural Science Foundation of China (Grants No.~12475056, No.~12247101, and No.12205129),  the 111 Project under (Grant No. B20063), and Gansu Province's Top Quality Leadership Talent Project.

\bibliographystyle{JHEP}
\bibliography{ref}

\end{document}